\documentclass[journal=jpcbfk,manuscript=article]{achemso}
\usepackage{longtable}
\usepackage{multirow}
\usepackage{amsmath}
\usepackage[usenames,dvipsnames]{color}
\usepackage{soul}
\usepackage{threeparttable}
\usepackage{xr}
\usepackage{graphicx}
\usepackage{amsmath,amssymb}
\usepackage{caption}
\usepackage{color}
\usepackage{dcolumn}
\usepackage{bm}
\usepackage{float}
\usepackage{subcaption}
\usepackage{textcomp}
\usepackage{url}
\usepackage[normalem]{ulem}
\usepackage{microtype}

\usepackage[unicode=true,
            bookmarks=true,
            bookmarksnumbered=true,
            bookmarksopen=true,
            bookmarksopenlevel=2,
            breaklinks=false,
            pdfborder={0 0 1},
            backref=false,
            colorlinks=true,
            hidelinks
            ]{hyperref}
            
\makeatletter

\author{Kai T\"opfer} \affiliation[University of Basel]{Department of
  Chemistry, University of Basel, Klingelbergstrasse 80, CH-4056
  Basel, Switzerland.}  \author{Silvan K\"aser}
\affiliation[University of Basel]{Department of Chemistry, University
  of Basel, Klingelbergstrasse 80, CH-4056 Basel, Switzerland.}
\author{Markus Meuwly} \affiliation[University of Basel]{Department of
  Chemistry, University of Basel, Klingelbergstrasse 80, CH-4056
  Basel, Switzerland.}  \email{m.meuwly@unibas.ch}

\title{Double Proton Transfer in Hydrated Formic Acid Dimer: Interplay
  of Spatial Symmetry and Solvent-Generated Force on Reactivity}

\begin{document}

\begin{abstract}
The double proton transfer (DPT) reaction in hydrated formic acid
dimer (FAD) is investigated at molecular-level detail. For this, a
global and reactive machine learned (ML) potential energy surface
(PES) is developed to run extensive (more than 100\,ns) mixed ML/MM
molecular dynamics (MD) simulations in explicit molecular mechanics
(MM) solvent at MP2-quality for the solute. Simulations with fixed -
as in a conventional empirical force field - and conformationally
fluctuating - as available from the ML-based PES - charge models for
FAD shows significant impact on the competition between DPT and
dissociation of FAD into two formic acid monomers. With increasing
temperature the barrier height for DPT in solution changes by about 10\,\% ($\sim 1$ kcal/mol) between 300\,K and 600\,K. The rate for 
DPT is
largest, $\sim 1$\,ns$^{-1}$, at 350\,K and decreases for higher
temperatures due to destabilisation and increased probability for
dissociation of FAD. The water solvent is found to promote the first
proton transfer by exerting a favourable solvent-induced Coulomb force
along the O--H$\cdots$O hydrogen bond whereas the second proton
transfer is significantly controlled by the O--O separation and other
conformational degrees of freedom. Double proton transfer in hydrated
FAD is found to involve a subtle interplay and balance between
structural and electrostatic factors.
\end{abstract}

\section{Introduction}
Understanding the molecular details of chemical reactions in solution
is one of the challenges in physical and general chemistry at
large.\cite{warshel:1980,warshel:1993,gajewski:1994,MM.claisen:2019,MM.reactions:2021,kamerlin:2009,anfinrud:2004,yang:2008,ihee:2015,Meuwly2019}
These processes are at the center of interest of organic, inorganic and 
biological
chemistry as the environment in which they take place can determine
reaction products, their rates and propensities. However, the direct
and atomistically resolved experimental characterization by which the
solvent influences reactions is challenging because simultaneous
control over temporal (i.e. ``rates'') and spatial (``collocation''
and local solvation) aspects of the process along the progression
coordinate is rarely possible. On the other hand, computer simulations
have matured to a degree that allows the study of realistic processes
at a sufficiently high level of theory including the
environment.\cite{HeadGordon2018,KonerMeuwly2020}\\

\noindent
More specifically, polar solvents such as water, can stabilize the
transitions state (TS) of the
reactant.\cite{Guest1997,Jung2007,MM.claisen:2019} For the Claisen
rearrangement the origin of the catalytic effect of the water solvent
is the differentially stronger hydrogen bonding to the TS structure
(allyl phenyl ether) compared with that to the reactant structure
because the TS is more polar which lowers the reaction
barrier.\cite{White1970,Acevedo2010} Further stabilization of the TS
and lowering of the reaction barrier by the water solvent can be
affected by solvent polarizability and dipole-dipole interaction
between solute in its TS structure and the solvent.\cite{Acevedo2010}
The explicit description of solvent effects by full \emph{ab initio}
MD simulations for reactive systems in solution is computationally
very demanding and can be typically only followed on the picosecond
time scale.\cite{Rothlisberger2017,Hage2017} To avoid this, the
(reactive) solute can be treated quantum mechanically (QM - ranging
from tight-binding to post-Hartree-Fock methods) whereas the remaining
degrees of freedom are described by an empirical force
field.\cite{Karplus2000,Senn2009,Groenhof2013} Alternatively, reactive
force fields have been developed which allow to follow chemical
reactions in solution on extended time
scales.\cite{warshel:1980,goddard01reaxff,MM.armd:2014} More recently,
another possibility has emerged by developing a machine learned
(ML) representation for the solute that can be used together with an
implicit\cite{Ang2021,Riniker2021} or explicit description of the
environment at an empirical level. The present work develops such a
scheme and applies it to formic acid dimer in solution.\\

\noindent
Formic acid dimer (FAD) has been subject of several
computational\cite{Pan1997,Lim1997,Klein1998,Kohanoff2000,Ushiyama2001,kalescky2013local,marx:2015,Miliordos2015,tew2016ab,bowman.fad:2016,MM.fad:2016,richardson:2017,qu2018high,qu2018quantum,qu2018ir,MM.fad:2022}
and experimental
studies.\cite{ito2000jet,freytes2002overtone,georges2004jet,zielke:2007,xue2009,suhm:2012,MM.fad:2016,suhm:2020}
In the gas phase, formic acid exists as hydrogen bonded
dimers\cite{reutemann2011formic} making it a prototype for complexes
with hydrogen bonds such as enzymes or DNA base
pairs.\cite{balabin2009} Such conformations allow double proton
transfer (DPT) to occur which leads to broadening of the line shape
for the O-H frequency in IR-spectra.\cite{MM.fad:2016} The
dissociation energy of FAD into two formic acid monomers (FAMs) has
been determined from spectroscopy and statistical thermodynamics to be
$D_0 = -59.5(5)$\,kJ/mol ($\sim -14.22$\,kcal/mol)\cite{suhm:2012}
which compares with $-14.23 \pm 0.08$\,kcal/mol from recent
computations.\cite{MM.fad:2022} \\

\noindent
Another quantity that has been accurately measured experimentally is
the tunneling splitting for which a recent value of 331.2\,MHz is
available from microwave spectroscopy.\cite{caminati:2019} From this,
a barrier for DPT of 2559\,cm$^{-1}$ (30.6\,kJ/mol or 7.3\,kcal/mol)
from analysis of a 3D model was determined.\cite{caminati:2019}. The
tunneling splitting from the microwave data compares with values of
0.011\,cm$^{-1}$ (340.8\,MHz) \cite{duan:2017} and 0.016\,cm$^{-1}$
(473.7\,MHz)\cite{ortlieb2007proton,goroya2014high} from infrared
spectroscopy. Computationally, a high-quality (CCSD(T)-F12a/haTZ),
full-dimensional PES for FAD has been determined and represented as
permutationally invariant polynomials which features a DPT barrier of
2853\,cm$^{-1}$ (8.16\,kcal/mol).\cite{bowman.fad:2016} The computed
splitting on this PES from using a ring-polymer instanton approach was
0.014\,cm$^{-1}$ (420\,MHz).\cite{richardson:2017}\\

\noindent
Given its thorough characterization in the gas phase, FAD is also an
interesting system to study reactions in solution. In particular the
balance between DPT and the dissociation of FAD into two monomers
provides an attractive aspect of the system to be explored also as a
proxy for DNA base-pair dimers.\cite{MM.dna:2004,arabi:2018} The
homologous series of carboxylic acids (formic, acetic, propionic,
butyric) in water have also served as model systems for studying the
balance and role of hydrogen bonding, hydrophobicity and entropy
changes which is of particular relevance in the context of protein
folding and stability.\cite{brooks:2008}\\

\noindent
For FAD in pure water detailed experimental results are scarce. Early
work reported that dimerization is most relevant for the
low-concentration range for all carboxylic acids investigated,
including formic acid.\cite{katchalsky:1951} This was subsequently
challenged and measurements in 3\,m NaCl by potentiometric titrations
were interpreted as leading to singly H-bonded, extended formic acid
oligomers.\cite{scheraga:1964} The hydration of formic acid in pure
water was also investigated with neutron diffraction experiments for
higher formic acid concentrations.\cite{imberti:2015} At all
conditions studied, a pronounced peak for the O$\cdots$O separation
with a maximum at 2.7\,\AA\/ was reported which, together with a peak
for the OH$\cdots$O angle at $180^\circ$, supports formation of singly
H-bonded or cyclic FAD. It is of interest to note that the extended
structures proposed by Scheraga are incompatible with the neutron
scattering results.\cite{imberti:2015} Surface scattering experiments
found that FAD colliding with the liquid water surface almost
completely inserts in its cyclic form into bulk water even at low
collision energies.\cite{sobyra:2017} This indicates that H-bond
breaking in FAD is not essential for uptake. Whether or not and for
how long FAD remains in its cyclic form inside water was, however, not
determined in these studies.\cite{sobyra:2017} Accompanying {\it ab
  initio} MD simulations reported that the cyclic dimer has fewer
FAD-water hydrogen bonds (i.e. is ``more hydrophobic'') than the
branched, singly H-bonded FAD isomer.\cite{hanninen:2018} Finally, a
study considering clathrate formation involving formic acid and water
interpreted the measured infrared spectra as partially originating
from water-mediated cyclic formic acid dimers in which the two formic
acid monomers are H-bonded by a bridging water
molecule.\cite{tarakanova:2019} An independent MD simulation for FAD
in water reported\cite{brooks:2008} a negligible stabilization free
energy for cyclic FAD of $0.6$\,kcal/mol which is surprising given the
substantial gas phase stability of FAD of 
$-16.8$\,kcal/mol.\cite{Miliordos2015,MM.fad:2022} Very recent Raman
spectroscopy measurements reported formation of both, singly and
cyclic H-bonded FADs.\cite{men:2020} Taken together, the available
studies suggest that FAD on and in pure water can exist as an
equilibrium between cyclic and singly H-bonded dimers. Therefore, the
present work will focus on the dynamics of cyclic FAD in pure water.\\

\noindent
The aim of the present work is to characterize DPT dynamics in FAD in
aqueous solution. 
  First, the methods and the ML/MM embedding are
  described.  Next, the quality of the PES is reported and the
  structural dynamics, free energies, and rates together with
  conformational coordinates relevant for DPT of FAD in solution are
  discussed. This is followed by an analysis of the solvent
  distributions for reactant and transition state structures, and the
  effect of the solvent-generated electrical field. Finally, the
  results are discussed in a broader context.\\

\section{Methods}
This section describes the generation of the potential energy surface
for the simulation of FAD in water and the molecular dynamics (MD)
simulations performed within the Atomic Simulation Environment (ASE).
\cite{HjorthLarsen2017} In the following the potential energy of FAD
is represented as a Neural Network (NN),\cite{Unke2019} referred
to as ``ML'' or ``PhysNet'' whereas solvent water is described with
the TIP3P model (MM).\cite{Jorgensen1983} First, the reference data
generation, the training of the NN-based PES and the embedding of the
ML (FAD) part in the MM environment (water) is described. This is
followed by a description of the MD simulations and their analysis.\\

\subsection{Potential Energy Surface}
All electronic structure calculations were carried out at the
MP2/aug-cc-pVTZ level of theory using the MOLPRO software
package.\citep{Werner2020} 
  The reference data included
  structures, energies, forces, and dipole moments for FAM (5000
  structures), FAD (47069), and a number (6000) of related fragments
  (``amons'')\cite{Huang2020} including H$_2$, CO, H$_2$O, CH$_4$,
  CH$_2$O and CH$_3$OH. Structures were generated from MD simulations,
  carried out at the PM7 level of theory using ASE. The ``amons'' were
  only used for training the NN and help in generalizing the ML model.
  Additional FAD structures were generated from simulations in
  solution (akin to adaptive sampling) by using PhysNet trained only
  on gas-phase structures.\cite{behler2015constructing, MM.fad:2022}
  The energies of the FAD structures cover a range up to $\sim
  100$\,kcal/mol above the global minimum compared with the TS for DPT
  which is $\sim 8$ kcal/mol above the global minimum.  
For the final
training the 58069 reference structures were split into training
(49300), validation (5800) and testing (2969) sets. The parameters of
PhysNet were fitted to reproduce {\it ab initio} energies, forces and
dipole moments of the training set.\cite{Unke2019} Including FAD
conformations exhibiting two, one or no H-bond in the training
set has been shown essential to ensure the correct propensity
towards the global minimum conformation of FAD with two formed
H-bonds.\\

\noindent
PhysNet computes the total potential energy $E$ for given coordinates
$\boldsymbol{\vec{x}}$ and nuclear charges $\boldsymbol{Z}$ of $N$
atoms from atomic energies $E_i$ and pairwise electrostatic and
dispersion interactions according to
\begin{equation}
    E = \sum_{i=1}^{N} E_i + k_e \sum_{i=1}^{N} \sum_{j > i}^{N}
    \dfrac{q_i q_j}{r_{ij}} + E_\text{D3}.
    \label{eq1}
\end{equation}
The prediction of the atomic energies $E_i$ and charges $q_i$ is based
on feature vectors that encode the local chemical environment of each
atom $i \in N$ to all atoms $j$ within a cutoff radius that was
$r_\text{cut} = 10$\,\AA~in the present work.\cite{Unke2018} For
charge conservation, the partial charges $q_i$ are scaled to the
correct total charge of the system and used for the calculation of the
electrostatic interaction energy where $k_e$ is the Coulomb constant
and $r_{ij}$ is the distance between atoms $i$ and $j$. For small
distances the electrostatic energy is damped to avoid instabilities
due to the singularity at $r_{ij} = 0$.\cite{Unke2019} The initial
parameters for the dispersion correction $E_\text{D3}$ are the
standard values recommended for the revPBE level of
theory.\cite{Grimme2011} Since the atomic features, electrostatic and
dispersion interaction depend only on pairwise distances and are
combined by summation, the PES is invariant to translation, rotation
and permutation of equivalent atoms. The forces $\boldsymbol{\vec{F}}$
with respect to the atomic coordinates $\boldsymbol{\vec{x}}$ are
computed by reverse mode automatic differentiation provided by
Tensorflow.\cite{tensorflow2015,Baydin2018} For further details, the
reader is referred to Ref.\citenum{Unke2019}.\\

\noindent
For the water solvent the TIP3P model\cite{Jorgensen1983} as 
implemented in the GPAW program package\cite{Mortensen2005} is used 
with a cutoff range of $10$\,\AA~due to its improved performance 
compared to that of ASE. The interaction between formic acid and the
water solvent includes van-der-Waals and electrostatic terms which
were treated with a cutoff at $14$\,\AA~or half the simulation box
edge length at most. The van-der-Waals interactions are calculated
using the Lennard-Jones-Potential and the parameters from
CGenFF\cite{cgenff:2010} for formic acid.\\

\noindent
The electrostatic interactions between the TIP3P atomic charges of
water and the NN-predicted atomic charges of FAD are computed within
PhysNet. This allows exploitation of reverse mode automatic
differentiation to calculate the necessary derivatives of the
fluctuating atomic charges from the atomic coordinates and ensures
energy conservation. For the van-der-Waals and electrostatic
interactions between the ``high level'' treatment (here PhysNet) and
MM atoms a switch function $s(r)$
\begin{align}
	s(r) &= 
	\begin{cases}
		1 & \text{for } r \leq r_1 \\
		1 - 6 y^5 + 15 y^4 - 10 y^3 & \text{for } r_1 < r < r_2 \\ 
		0 & \text{for } r \geq r_2
	\end{cases} \\[1em]
	y &= \dfrac{r - r_1}{r_2 - r_1}
\end{align}
was used to set the interactions zero in the range from $r_1 =
13$\,\AA\/ to $r_2 = 14$\,\AA\/, where $r$ is the separation between
ML and MM atoms, respectively.  Additionally, the electrostatic
Coulomb potential $V_\text{C}$ is shifted to zero at the cutoff
distance of $r_{\rm cut} = 14$\,\AA\/ following the shifted force
method to prevent the inconsistent gradient at the cutoff
point.\cite{Spohr1997} While the van-der-Waals interaction between
both species become negligible at that cutoff distances, this is not
the case for the Coulomb forces between two charges $q_i$ and $q_j$ of
atoms $i$ from the ML part and $j$ from the MM part separated by
$r_{ij}$.
\begin{align}    
    V_\text{C} &= \sum_{i}^{N_\text{ML}} \sum_{j}^{N_\text{MM}}
    V_{\text{C}, ij}^\text{shifted}(r_{ij}) \\[1em] V_{\text{C},
      ij}^\text{shifted}(r_{ij}) &= s(r_{ij}) \cdot \left(
    V_{\text{C}, ij}(r_{ij}) - V_{\text{C}, ij}(r_\text{cut}) -
    \dfrac{\partial V_{\text{C}, ij}}{\partial r} \mid_{r_\text{cut}}
    \cdot (r_{ij} - r_\text{cut}) \right)\\[1em] V_{\text{C},
      ij}(r_{ij}) &= \dfrac{1}{4 \pi \epsilon} \dfrac{q_{i} \cdot
      q_{j}}{r_{ij}}
\end{align}

\noindent
An ASE calculator class for the PhysNet model was written as interface
to the ASE program package. Further extensions are implemented to
enable constant atomic charges in formic acid only for the calculation
of the electrostatic potential between the ML and MM part of the
simulation setup. In addition, custom energy functions were added to
the potential energy to perform biased
simulations.\cite{torrie1977nonphysical}\\

\subsection{Molecular Dynamics Simulations}
All MD simulations were performed using Python packages of a modified
version of ASE ($3.20.1$).
\cite{HjorthLarsen2017} Simulations are initialized with FAD in its
minimum conformation in a cubic simulation box with edge length
$28.036$\,\AA\/ containing 729 water molecules which corresponds to a
density of $0.997$\,g/cm$^3$. The density is not adjusted for
simulations at different temperatures.  All simulations were run with
periodic boundary conditions. To maintain the solute near the center
of the simulation box a center-of-mass harmonic constraint with a
force constant of $0.23$\,kcal/mol/\AA$^{2}$ was applied on formic
acid molecules when moving further than 10\,\AA\/ from the center of
the simulation box. RATTLE-type holonomic
constraints\cite{Andersen1983} as implemented in the GPAW program
package\cite{Enkovaara2010} were applied to the oxygen-hydrogen and
hydrogen-hydrogen distances of the water molecules.  The simulations
are carried out in the $NVT$ ensemble using the Langevin propagator
with a time step of $\Delta t = 0.2$\,fs and a friction coefficient of
$10$\,ps$^{-1}$.\\

\noindent
First, the structure of the system was minimized to release strain.  A
heating run is performed for $10$\,ps to reach the respective target
temperatures of 300\,K, 350\,K, 400\,K, 450\,K, 500\,K and 600\,K.  As
no rigorous implementation of a barostat is available in ASE,
simulations at constant volume but increased temperature were carried
out to qualitatively assess the influence of increased pressure on the
DPT dynamics. The system equilibrates within $5$\,ps during the
heating runs. Equilibration is followed by production simulations of
$100$\,ps each, accumulating to a total of $16$\,ns simulation time at
each temperature. Positions and momenta are stored (1) every $10$\,fs
for the complete trajectory and (2) every $1$\,fs in a time window of
$\Delta t = \left[ -20, +50 \right]$\,fs around every attempt of a
reaction registered during the simulation (see Analysis).\\

\noindent
The implementation is validated by examining energy conservation for a
100 ps $NVE$ MD simulation at 600\,K. The total energies at each time
step are within a standard deviation of $6.1 \cdot 10^{-3}$\,kcal/mol
from the average total energy with a maximum deviation of $\pm 25.2
\cdot 10^{-3}$\,kcal/mol and no drift is found. This confirms that the
forces are correctly implemented.\\

\noindent
To determine the impact of the fluctuating atomic charges available
from PhysNet on the dynamics and energetics, MD simulations were also
performed by fixing the atomic charges to those of the TS structure of
FAD to retain the symmetry of the system before and after PT. These
fixed atomic charges were used for the solute-solvent interaction
only. For the internal interaction of the ML part (FAD) fluctuating
charges were retained in order to correctly represent the PES for the
isolated solute. 
  Finally, as FAD was found to be largely
  hydrophobic from earlier simulations\cite{hanninen:2018,brooks:2008}
  and to probe the effect of the electrostatic interactions between
  solute and solvent on DPT, simulations were also carried out with
  zeroed charges on all atoms of the solute.\\

\subsection{Analysis}
Proton transfer (PT) is a transient process. To characterize a PT
event it is often useful to use a geometric
criterion.\cite{MM.amm:2002} Here, the criterion was the distance
between the hydrogen and the oxygen atoms O$_\mathrm{A}$ and
O$_\mathrm{B}$ of the respective H-bond. A PT in one H-bond was
identified by a sign change of the progression (or reaction)
coordinate $\xi =d(\mathrm{H-O}_\mathrm{A}) -
d(\mathrm{H-O}_\mathrm{B})$. Further, the oxygen atom bonded to the
hydrogen atom is the donor oxygen (O$_\mathrm{don}$) and the second
oxygen atom in the H-bond is the acceptor oxygen (O$_\mathrm{acc}$),
see Figure \ref{fig:1}. 
  Successful and attempted DPT always
  involve two consecutive PTs, see Figure S1. For
  ``successful'' DPT the two PTs include a ``first'' and a ``second''
  hydrogen atom H$_{\rm A}$ and H$_{\rm B}$ in both H-bonds,
  respectively, whereas for an ``attempted'' DPT both PTs involve only
  hydrogen atom H$_{\rm A}$.  DPT is considered to be successful when
  a sign change in both progression coordinates $\xi_1$ and $\xi_2$
  occurs. Attempted - as opposed to successful - DPT events are those
  for which only one forth and back PT occurs along the first H-bond
  without PT along the second H-bond.\\

\noindent
In addition to unbiased simulations, biased simulations were
carried out at different temperatures for FAD in the gas phase and in
solution. For this, an artificial harmonic potential $V_\text{bias}
(\xi_i) = \frac{1}{2} \sum_{i=1}^2 \left( \,k_\text{bias} \cdot (\xi_i
- \xi_{i,e})^2 \right)$ involving both progression coordinates $\xi_1$
and $\xi_2$ was added to the total potential energy. For FAD in
solution the force constant was $k_\text{bias} = 16.1$\,
kcal/mol/\AA$^2$ and $5 \cdot 10^6$ snapshots were recorded from 1\,ns
trajectories at each temperature. 
  Both constraining harmonic
  potentials are centered at the TS coordinate, $\xi_\mathrm{1,TS} =
  \xi_\mathrm{2,TS} = 0$, and the magnitude of the force constant was
  chosen to lift the potential of the equilibrium conformation ($
  \xi_\mathrm{eq} = \pm 0.66$\,\AA) of FAD roughly to the level of the
  TS conformation. Additional tests were run with lower values of
  $k_\text{bias}$. The resulting potentials of mean force (PMFs) were
  found to be insensitive of this choice, see Figure S2.\\

\section{Results}
\subsection{Quality of the Potential Energy Surface}
\begin{figure}[htb!]
\centering
\includegraphics[width=0.75\linewidth]{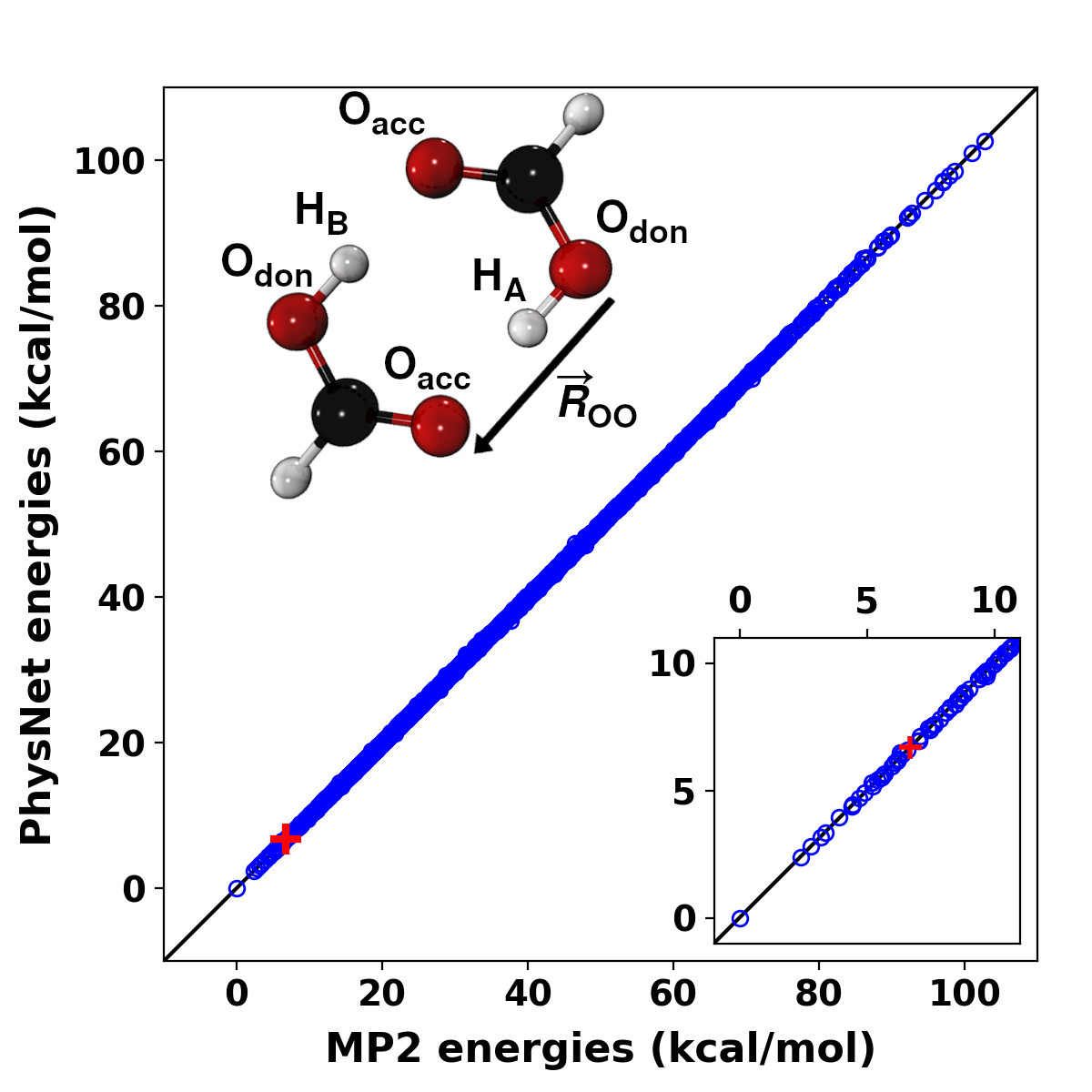}
\caption{Correlation between the MP2/aug-cc-pVTZ reference and the
  NN-predicted energies from PhysNet for the test set. The 2969
  randomly chosen structures which were not used during training
  contain samples of FAM, FAD, and their ``amons''. The RMSE between
  reference and NN-predicted energies is $0.336$\,kcal/mol with a
  Pearson coefficient of $R^2 = 1 - 2\cdot 10^{-6}$. The red cross
  indicates the energy for the TS. A sketch of FAD with atom labels is
  shown in the upper left corner. The vector pointing from
  O$_\mathrm{don}$ towards O$_\mathrm{acc}$, $\vec{R} _\mathrm{OO}$,
  is also reported.}
\label{fig:1}
\end{figure}

Figure \ref{fig:1} shows the quality of the ML PES by comparing
reference MP2 energies with those predicted from the trained PhysNet
model for formic acid dimer for 2969 test structures. The energies of
the smaller fragments (``amons'') are outside the energy range of the
graph but are included in the root mean squared error (RMSE) of
$0.336$\,kcal/mol and the Pearson coefficient is $R^2 = 1 - 2\cdot
10^{-6}$. The energy deviation between the MP2 energy and the PhysNet
model prediction is smaller than $3 \cdot 10^{-3}$\,kcal/mol for the
minimum conformation of FAD and the TS for DPT.\\

\noindent
The MP2 method is a good compromise between speed and accuracy. For
example, the barrier for DPT of $6.7$\,kcal/mol compares with a value
of $7.2$\,kcal/mol from a morphed MMPT-MP2 force field that is
consistent with infrared spectroscopy\cite{MM.fad:2016} and a barrier
height of $7.3$\,kcal/mol from analysis of microwave spectroscopy
data.\cite{caminati:2019} Calculations at the CCSD(T) level of theory
report somewhat higher barriers of $7.9$\,kcal/mol and
$8.2$\,kcal/mol, depending on the basis set and additional F12
corrections used.\cite{MM.fad:2016,bowman.fad:2016} Furthermore, the
dissociation of the dimer into two monomers is of interest. 
  In
  the gas phase the PhysNet model yields $D_{\rm e} = -16.8$\,kcal/mol
  which is also the value from the reference MP2/aug-cc-pVTZ
  calculations. {\it Ab initio} calculations at the higher
  CCSD(T)/aug-cc-pVTZ level of theory find $D_{\rm e} =
  -16.8$\,kcal/mol\cite{Miliordos2015,MM.fad:2022} and
  $-16.0$\,kcal/mol at the basis set limit.\cite{Miliordos2015}\\

\noindent
Next, the energetics of increasingly hydrated FAD is determined from
the present ML/MM energy function, from the CGenFF\cite{cgenff:2010}
parametrization of FAD together with the TIP3P water model, and from
electronic structure calculations. For this, 50 snapshots from a 2\,ns
ML/MD simulation with FAD in its dimeric structure were extracted. The
15 water molecules closest to the center of mass of FAD were retained
for each snapshot. 
  Electronic structure calculations for this
  analysis were carried out at the B3LYP+D3/aug-cc-pVDZ level of
  theory\cite{becke:1993,lyp:1988,grimme:2010} because MP2/aug-cc-pVTZ
  calculations for FAD surrounded by up to 15 water molecules are
  computationally too demanding. 
The total interaction energies for
FAD--(H$_2$O)$_n$ $(n = 1,\dots ,15)$ complexes were determined. Water
molecules were retained in increasing order of their distance from
FAD. The results are reported in Figures S3 and
S4 and the analysis shows that the interaction energies
with 15 water molecules from the MM/ML energy function differ on
average by less than 1\,kcal/mol from the reference B3LYP+D3
calculations whereas CGenFF underestimates the reference results by
about 4\,kcal/mol.\\

\noindent
The 50 snapshots were also analyzed from retaining the 4, 8, and 12
closest water molecules and correlating the MM/ML and CGenFF 
interaction energies with those from the reference B3LYP+D3
calculations (Figures S5 to S7).  
  Mean
  absolute errors between MM/ML and DFT reference calculations are
  $1.13$, $2.06$ and $2.85$\,kcal/mol compared with $2.42$, $3.36$ and
  $3.62$\,kcal/mol for CGenFF. It is also worth to point out that
  CGenFF finds a dimer stabilization energy of only $9$\,kcal/mol in
  the gas phase compared with high-level electronic structure
  calculations that yield
  $-16.8$\,kcal/mol\cite{Miliordos2015,MM.fad:2022} and
  $-17.4$\,kcal/mol at the B3LYP+D3/aug-cc-pVDZ level which is
  $-0.6$\,kcal/mol lower than for the MP2 and CCSD(T) levels of theory
  with the aug-cc-pVTZ basis set.
Overall, the comparison with the
B3LYP+D3 results validates the quality of the ML/MM energy function
whereas the CGenFF parametrization is found to considerably
underestimate the stability of FAD.\\

\subsection{Structural Dynamics in Solution}
\begin{figure}[htb!]
\centering
\includegraphics[width=\linewidth]{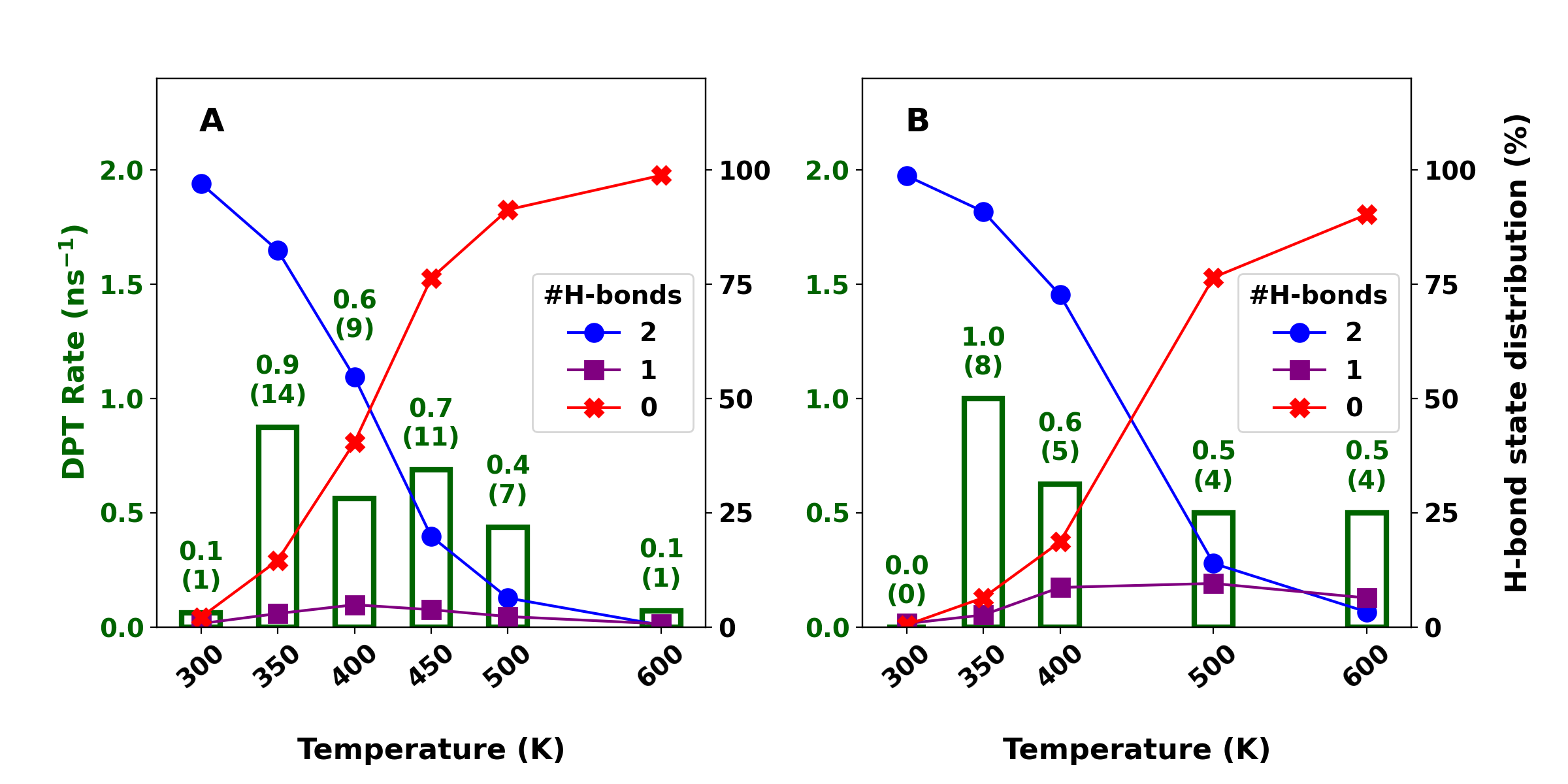}
\caption{DPT rates and distribution of the number of formed H-bonds in
  the simulation of FAD in water at different temperatures using
  fluctuating atomic charges (panel A) and fixed atomic charges (panel
  B). The green numbers atop the bars show the respective rate and
  absolute number of observed DPT events. The markers show the
  propensity of FAD to have two formed H-bonds (blue circle), to be
  partially dissociated by one (purple square) or completely separated
  into two formic acid monomers (red cross). The lines are to guide
  the eye.}
\label{fig:2}
\end{figure}

Figure \ref{fig:2} reports the propensity of FAD to exist as a doubly
(blue circle) or singly (violet square) H-bonded dimer or as two
separate monomers (red cross) in water. This is consistent with what
has been inferred from experiments.\cite{imberti:2015,sobyra:2017} For
the equilibrium conformation the O$_{\rm don}$--O$_{\rm acc}$ distance
is $\sim 2.66$\,\AA. An H-bond in FAD is considered ``broken'' if the
O$_{\rm don}$--O$_{\rm acc}$ distance exceeds 4\,\AA. MD simulations
with fluctuating atomic charges on FAD show (Figure \ref{fig:2}A) that
the propensity for a doubly H-bonded FAD decreases from $97.0$\,\% at
300\,K to $0.4$\,\% at 600\,K. The probability for the singly H-bonded
dimer increases from $0.8$\,\% at 300\,K to a maximum of $4.9$\,\% at
400\,K and decreases again to $0.7$\,\% at 600\,K. At the temperature
with the highest DPT rate (350\,K) both H-bonds are formed for
$82.4$\,\% of the propagation time.\\

\noindent
  In conventional force fields the partial charges are
  fixed\cite{cgenff:2010} and do not change with conformation.
  PhysNet provides geometry-dependent charges, see
  Figure~S8, which fluctuate by $\pm 0.05$\,e to $\pm
  0.1$\,e around their mean. The average charges for the C, H$_{\rm
    C}$ and one of the oxygen atoms from PhysNet are similar to those
  in CGenFF whereas for the H$_{\rm O}$ and the second oxygen they
  differ by about $0.1$\,e. It is also found that the charge
  distributions from PhysNet differ for cyclic and branched FAD.
Given these differences it is of interest to determine the differences
between the simulations analyzed with conformationally fluctuating
charges and those with fixed partial charges on the FAD. The charges
assigned to the atoms are those of the TS structure to ensure a
symmetric atomic charge distribution before and after successful
DPT. The distribution of formed H-bonds for FAD in water with
fluctuating and fixed charges in Figure \ref{fig:2} shows that the
propensity for H-bond formation significantly depends on fixing the
charges. With the fixed charge model the probability for the doubly
H-bonded dimer to exist is larger at all temperatures from 300 K
($98.7$\,\%) to 600 K ($3.3$\,\%) compared with the fluctuating charge
model ($97.0$\,\% and $0.4$\,\% at 300\,K and 600\,K,
respectively). In addition, the probability for singly H-bonded FAD
and two separate monomers at 600\,K differs with the fixed charge
model with $6.4$\,\% and $90.3$\,\%, respectively, from the
fluctuating charge model with $0.7$\,\% and $98.8$\,\%. Thus, a fixed
charge model with charges from the TS structure predicts a higher
propensity for FAD to exist in solution compared to the fluctuating
charge model.\\

\noindent
The higher propensity for dimer dissociation in solution using
fluctuating rather than fixed charges correlates with the differences
in the magnitudes of the atomic charges between the two charge
models. From the equilibrium conformation of FAD towards separated
FAMs the polarization along the atoms in the H-bond increases. The
fluctuating charges along the O$_{\rm don}$--H$\cdots$O$_{\rm acc}$
bond change from $[-0.39$, $ +0.32$, $-0.43]\,e$ in the equilibrium
dimer conformation to $[-0.42, +0.36,-0.45]\,e$ for separated
monomers, respectively, which compares with $[-0.40, +0.30, -0.40]\,e$
in the fixed charge models with charges taken from the TS
structure. Comparable amplitudes for charge transfer ($\delta q =
0.025\,e$) have been found for the HO$_{\rm don}$--H$\cdots$O$_{\rm
  acc}$H$_2$ hydrogen bond in bulk water when the energy function was
fit to match experimental frequencies and intensities of the infrared
and Raman bands.\cite{MM.water:2018} The larger absolute values of the
atomic charges in the fluctuating charge model (differences of up to
$20$\,\% and $13$\,\% for the hydrogen and oxygen atoms of the H-bonds
relative to the charges in the TS structure, respectively) lead to
stronger electrostatic interactions with the surrounding solvent. This
also impacts on the conformational sampling and the reaction barrier
height.\\

\noindent
To evaluate the effect of the water solvent on dissociation, $NVT$
simulations for FAD in the gas phase were performed using the same
setup as for the simulations in solution. The presence of both H-bonds
is a prerequisite for DPT to occur as the top of the barrier becomes
inaccessible if even one H-bond is broken. Thus, dissociation of the
FAD competes with DPT and therefore governs the DPT rate. In the gas
phase it is found that - similar to the situation in solution - the
propensity towards broken H-bonds increases for higher
temperatures. At 300\,K both H-bonds are formed for $99.4$\,\% of the
propagation time.  This fraction decreases at higher temperatures:
$27.3$\,\% and $2.1$\,\% at 500\,K and 600\,K, respectively.\\

\subsection{Free Energies of Activation and Rates for DPT}
\begin{figure}[htb!]
\centering
\includegraphics[width=0.75\textwidth]{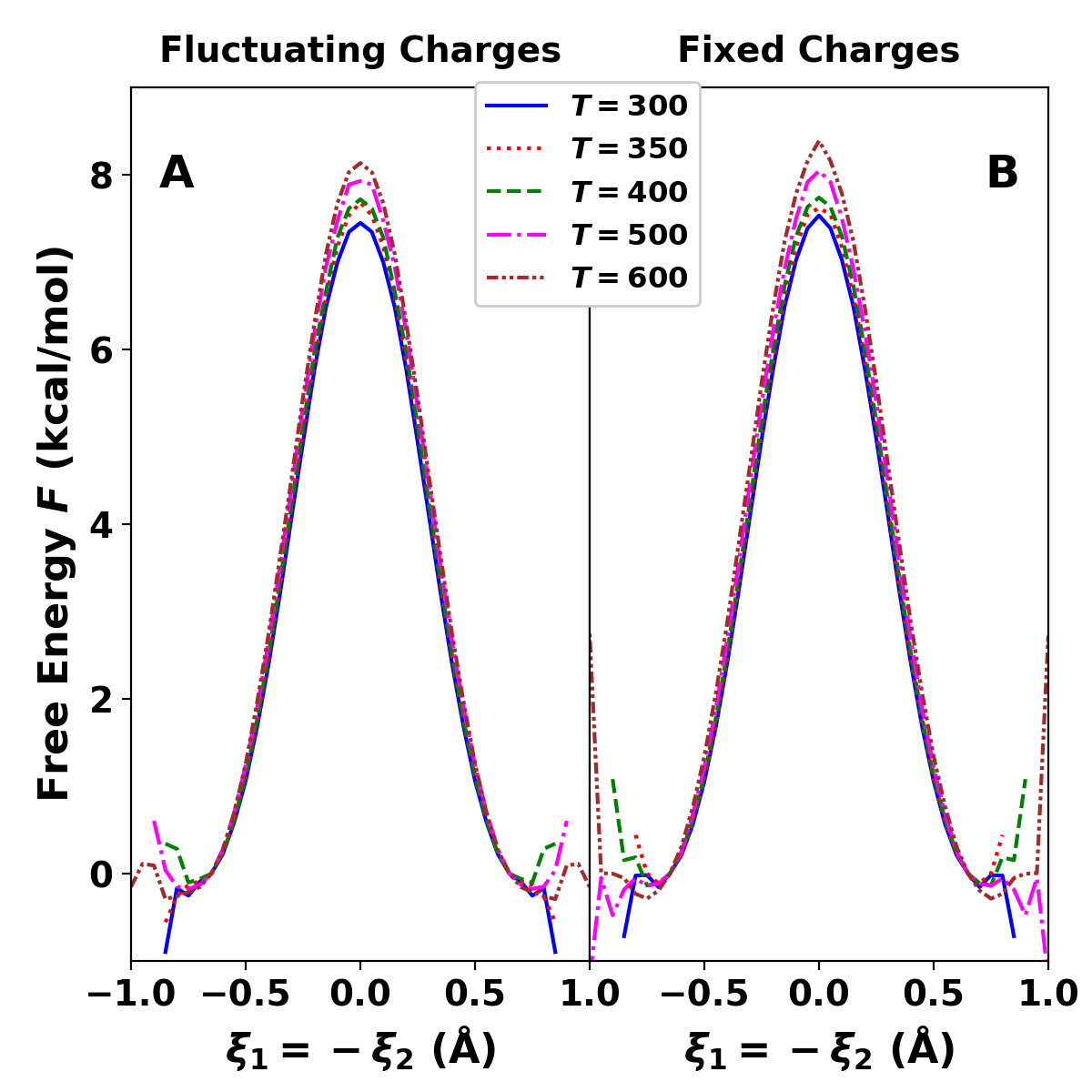}
\caption{ Free energy profile for DPT in FAD with fluctuating (panel A)
  and fixed charges (panel B) for different temperatures in solution.
  The free energy curve follows the minimum energy path of the
  2-dimensional free energy surface of antiparallel reactive
  coordinates $\xi_1 = -\xi_2$.  The average temperature of FAD in
  solution in the $NVT$ simulations are given in the legend and the
  standard deviation ranges from $ \pm6$\,K at 300\,K to $\pm13$\,K at
  600\,K.  The PMF is symmetrized around $\xi_1 = \xi_2 = 0$ to
  increase data sampling as the artifacts at large absolute values of
  $\xi$ result from insufficient sampling especially for biased
  simulations at low temperatures.}
\label{fig:3}
\end{figure}

Biased simulations were performed to investigate the free energy
barrier for DPT. Figure \ref{fig:3} shows the 1-dimensional cuts along
the constrained progression coordinate $\xi$.  
  The
  2-dimensional free energy surfaces $G(\xi_1,\xi_2)$ are given in
  Figure S9.  
The energies around the equilibrium
$\xi_\mathrm{eq} = \pm 0.66$\,\AA~are set to zero. Biased simulations
are carried out for fluctuating and fixed charges. Between 300\,K and
600\,K the free energy barrier increases from $7.45$\,kcal/mol to
$8.13$\,kcal/mol for simulations using the fluctuating charge
model. With fixed charges the barriers change from $7.53$\,kcal/mol to
$8.39$\,kcal/mol. Thus, as a function of temperature, the barrier
increases by $0.68$\,kcal/mol with fluctuating charges, compared with
$0.85$\,kcal/mol with fixed charges. In other words, the {\it change}
of the barrier height as a function of temperature differs by 25\,\%
from fluctuating to fixed charges on solution.\\

\noindent
Simulations were also carried out in the $NVE$ ensemble for FAD in the
gas phase and yield a free energy barrier for DPT of $8.25$\,kcal/mol
at a temperature of 310\,K which was determined from the kinetic
energy of all atoms in this biased simulation. Earlier classical
\emph{ab initio} metadynamics simulations with the same reaction
coordinates as those used in the present work at the BLYP level with a
plane wave basis for FAD in the gas phase reported a free energy
barrier height of $4.7$\,kcal/mol at 300\,K compared with the
potential barrier height of $5.0$\,kcal/mol.\cite{marx:2015} This
differs qualitatively from the present results for which the free
energy barrier in the gas phase increases from the potential barrier
of $6.7$\,kcal/mol.\\

\noindent 
To quantify the electrostatic contribution of the solute/solvent
interactions to the free energy barrier, biased simulations of FAD in
solution were performed with the atomic charges of FAD set to zero. At
300\,K the free energy of activation is $7.46$\,kcal/mol and increases
by $+0.87$ kcal/mol to $8.33$\,kcal/mol at 600\,K. This compares with
$7.45$ and $7.53$\,kcal/mol with fluctuating and fixed charges at
300\,K, respectively, which increase to $8.13$ and $8.39$\,kcal/mol at
600\,K. At 300\,K the contribution of fluctuating electrostatics is
thus $-0.01$\,kcal/mol as opposed to $+0.07$\,kcal/mol if the charges
are fixed and a reduction by $-0.20$\,kcal/mol and an increase of $
+0.06$\,kcal/mol at 600\,K for fluctuating and fixed charges,
respectively. This finding suggests that FAD behaves largely as a
hydrophobic solute for which direct electrostatic interactions with
the solvent are less relevant.\\

\noindent
The $T-$dependence to the free energy ($- T \Delta S$) suggests that
the change $\Delta S$ between the reactant and the TS state is
negative for all charge models used but the magnitude of $\Delta S$
differs. This difference is accommodated in the solvent ordering
around the solute. Linear interpolation of the free energy barriers at
the different temperatures yields an activation enthalpy and activation
entropy of $[\Delta U, \Delta S] =$ [$6.87$\,kcal/mol,
  $-2.1$\,cal/mol/K] for fluctuating charges, [$6.62$\,kcal/mol,
  $-2.9$\,cal/mol/K] for fixed charges, and [$6.70$\,kcal/mol,
  $-2.7$\,cal/mol/K] for zero charges.\\

\noindent
Because the constraints in the biased simulations favour FAD over two
separated monomers (and the dimer has been found to dissociate at
higher temperatures for free dynamics) it is mandatory to also run and
analyze unbiased simulations that allow DPT to occur in solution. For
this, an aggregate of 16\,ns simulations was run and analyzed at each
temperature. As the barrier for DPT is relatively high, DPT is a rare
event. DPT rates in unbiased simulations are highest at 350\,K for
both fluctuating and fixed charge models, see Figures \ref{fig:2}A and
B. With the fluctuating charge model (Figure \ref{fig:2}A) the rate is
$0.9$\,ns$^{-1}$ at 350\,K and decreases to $0.4$\,ns$^{-1}$ at
500\,K, while the dimer is only present for $82.4$\,\% and $6.4$\,\%
of the propagation time, respectively. Hence, the true rates are $\sim
6.5$ times higher at 500\,K compared with those at 350\,K.
  Within transition state theory energy barriers of $\sim 8$\,kcal/mol
  from the biased simulations correspond to rates of $\sim
  0.1$\,ns$^{-1}$ which is consistent with results from unbiased
  simulations shown in Figure \ref{fig:2}.\\

\noindent
With the fixed charge model a similar increase of the DPT rates is
observed. At 300\,K, no DPT was observed during 8\,ns of propagation
time whereas for $T = 350$\,K and 500\,K they are $1.0$\,ns$^{-1}$ and
$0.5$\,ns$^{-1}$, respectively. As the simulation time with fixed
charges is 8\,ns - shorter than simulations with the fluctuating
charge model - the rates for DPT are based on fewer events with larger
uncertainties which are not quantified in this work.\\

\subsection{Conformational Coordinates Involved in DPT}
Conformationally relevant degrees of freedom to consider are the
O$_\mathrm{don}$--O$_\mathrm{acc}$ and the CO bond distances. The
involvement of the O$_\mathrm{don}$--O$_\mathrm{acc}$ separation
becomes apparent from noting that it contracts from
$d_\mathrm{eq}(\mathrm{OO}) = 2.66$\,\AA\/ in the global minimum to
$d_\mathrm{TS}(\mathrm{OO}) = 2.41$\,\AA\/ in the TS. In other words,
compressing the O--O separation towards the TS structure facilitates
(D)PT. The CO bond in the COH moiety has bond order 1 (${\rm BO} = 1$)
whereas the second CO bond is formally a double bond (${\rm BO} =
2$). This bonding pattern reverses after DPT and for the TS structure
all bond orders are $1.5$.\\

\noindent
The first PT leads to a short-lived, metastable ion-pair with one
protonated formic acid monomer and the corresponding anion, see Figure
S1. Subsequently, the transferred hydrogen atom returns
either to where it came from (attempted DPT) within an average delay
time of $4.8 \pm 2.9$\,fs at 350\,K or the hydrogen atom along the
second H-bond transfers (successful DPT) with an average delay time of
$5.9\pm 2.6$\,fs. At 450\,K the analysis of unbiased simulations find
delay times of $5.1 \pm 3.9$\,fs and $8.2 \pm 7.1$\,fs for attempted
and successful DPT, respectively. This is consistent with results from
earlier work that reported an average delay time of $\sim 8$\,fs for
successful DPT.\cite{Ushiyama2001}\\

\begin{figure}[htb!]
\centering \includegraphics[width=\linewidth]{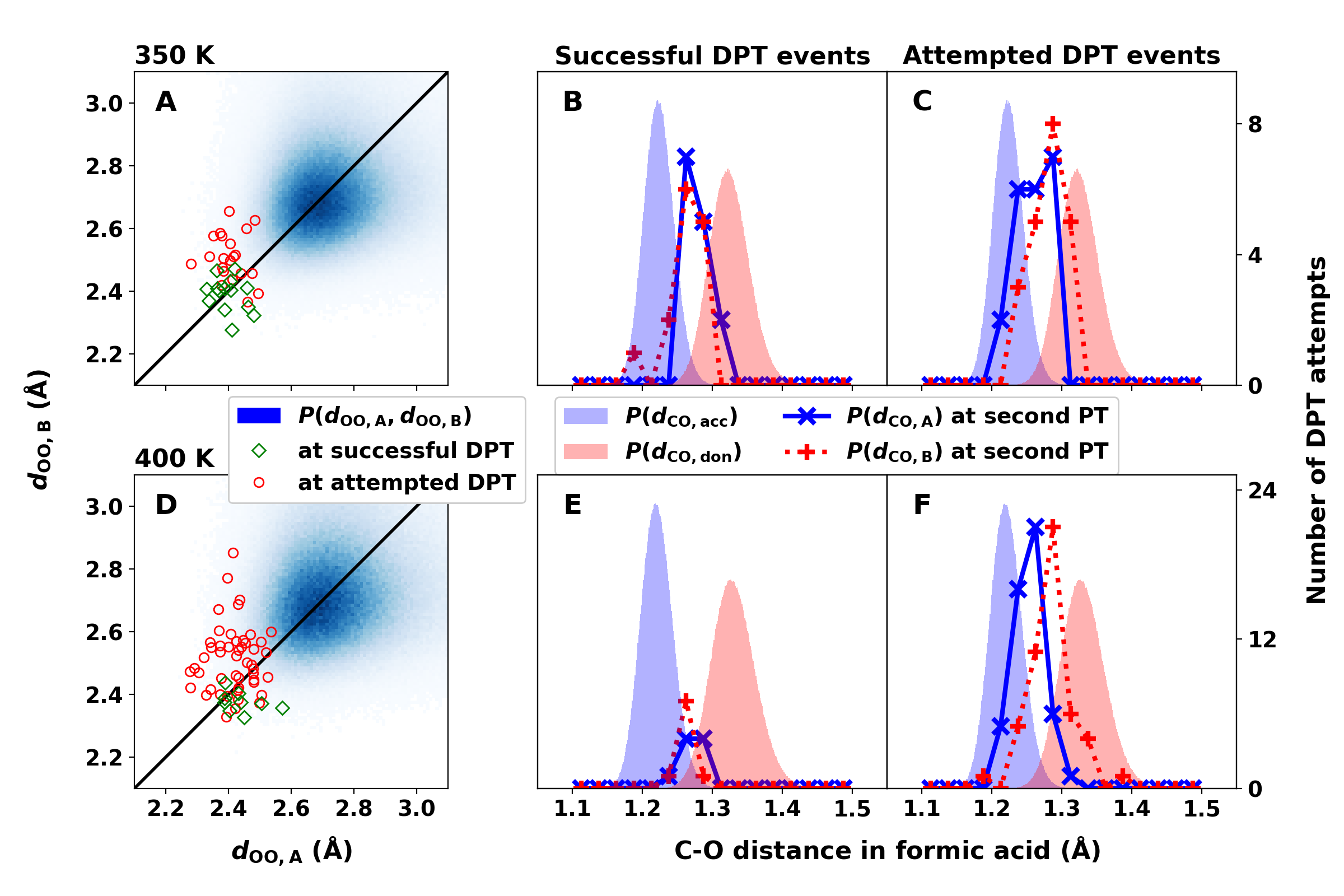}
\caption{Bond distance distributions in FAD from simulations at 350\,K
  (A-C) and 400\,K (D-F). The blue colormap (panels A and D) shows the
  probability distribution of the O$_\mathrm{don}$--O$_\mathrm{acc}$
  distances $P(d_{\rm OO,A},d_{\rm OO,B})$ in both H-bonds.
  Oxygen-oxygen separations $d_\mathrm{OO}$ for successful (green
  diamonds) and attempted DPT events (red circles) at the time of the
  second PT are shown for both hydrogen bonds. The blue and red
  distributions (panels B, C, E, and F) show the distribution
  $P(d_\mathrm{CO,acc})$ and $P(d_\mathrm{CO,don})$ for the complete
  simulations. 
    Distributions $P(d_\mathrm{CO,A})$ (blue cross)
    and $P(d_\mathrm{CO,B})$ (red plus) at the time of the second PT
    are reported for successful (B, E) or attempted DPT (C, F). For
    successful DPT they peak at the same CO separation (symmetric)
    whereas for attempted DPT they overlap less.}
\label{fig:4}
\end{figure}

\noindent
Figures \ref{fig:4}A and D compare the instantaneous
O$_\mathrm{don}$--O$_\mathrm{acc}$ separations at the time of the
second PT. The probability distribution functions $P(d_{\rm
OO,A},d_{\rm OO,B})$ for the O--O separations in unbiased
simulations at 350\,K and 400\,K peak at $(2.75 \pm 0.18)$\,\AA\/ and
$(2.78 \pm 0.22)$\,\AA, respectively. These distances shorten
significantly to $(2.40 \pm 0.05)$\,\AA\/ and $(2.42 \pm 0.06)$\,\AA\/
for successful DPT; see Figure \ref{fig:4}. For successful DPT both
O--O separations are between 2.4\,\AA\/ and 2.6\,\AA\/ whereas for
attempted DPT the first O--O separation ranges from 2.3\,\AA\/ to
2.5\,\AA\/ but the second oxygen donor-acceptor distance covers a
range between 2.4 and 2.9\,\AA\/. This suggests that a more
symmetrical geometry at the time of the first PT increases the
probability for successful DPT during the second step, i.e. if the
distances between {\it both} oxygen atoms in the H-bonds are
simultaneously contracted.\\

\noindent
Next, the CO bond lengths are analysed. Figure \ref{fig:4}B, C, E, and
F shows the distribution of the CO$_{\rm acc}$ double and CO$_{\rm
  don}$ single bond distances from all MD simulations at 350\,K and
400\,K. The maximum peak positions are clearly distinguishable and
demonstrate the single and double bond character of the CO bonds with
${\rm BO} = 1$ and ${\rm BO} = 2$, respectively. In addition, the
histograms for the CO distances at the second PT of a successful
(panels B and E) and attempted (panels C and F) DPTs are reported. For
successful DPT the distributions overlap for both temperatures and
${\rm BO} = 1.5$ whereas for attempted DPT they are clearly
non-overlapping and closer to ${\rm BO} = 1$ and ${\rm BO} = 2$,
respectively.\\

\begin{figure}[htb!]
\centering
\includegraphics[width=0.75\linewidth]{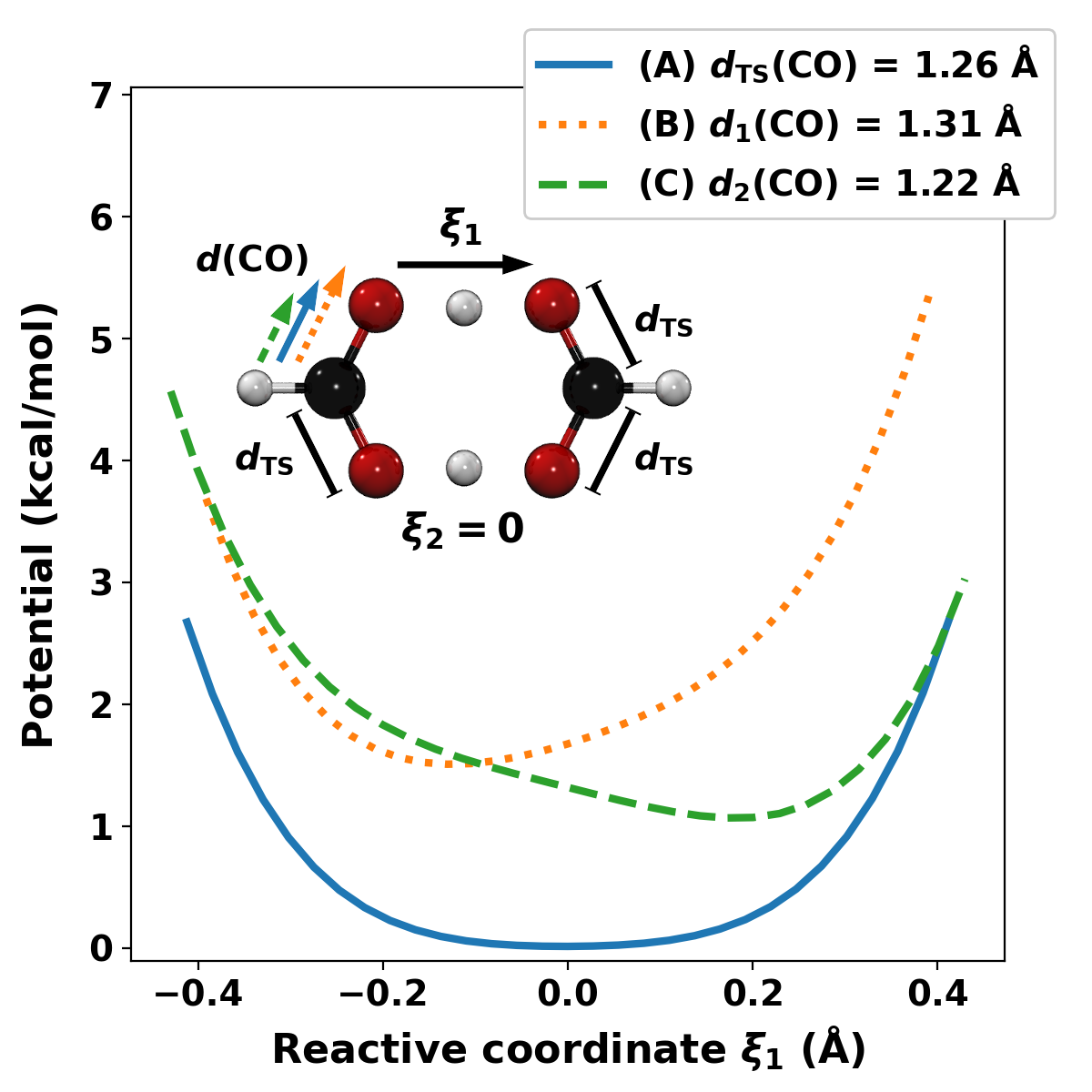}
\caption{Scan of the potential curve along the reactive coordinate
  $\xi_1$ of the hydrogen atom in the H-bond of FAD in the TS
  conformation. The three cases are for different elongations of the
  CO bond involving the first H-bond. CO distances are for the
  (A) TS conformation $d_\mathrm{TS}(\mathrm{CO})$ (solid line, blue),
  (B) average CO single bond separation $d_1(\mathrm{CO})$ (dotted
  line, orange) and (C) average CO double bond length
  $d_2(\mathrm{CO})$ (dashed line, green). The minimum of the
  potential for the TS conformation is shifted to 0 for reference.}
\label{fig:5}
\end{figure}

\noindent
The correlation between narrowly overlapping distance distributions 
of the CO single and double bond
and the higher probability for successful DPT
can be explained by the respective OH bond potential in the protonated
formic acid monomer. Figure \ref{fig:5} shows the PhysNet potential
along the progression coordinate $\xi_1$ of a hydrogen atom in the TS
of FAD. With all other bond lengths frozen in the TS conformation, a
scan along $\xi_1$ is performed for three different distances of the
CO bond involved in the first H-bond as in A) the TS conformation, B)
a CO single bond and C) CO double bond (see illustration in Figure
\ref{fig:5}). The remaining CO bonds are kept frozen at the TS bond
length of $1.26$\,\AA. For $d_\mathrm{TS}(\mathrm{CO})=1.26$\,\AA\/
(TS conformation with ${\rm BO} = 1.5$) the potential minimum along
the reactive coordinate is at $\xi_1 = 0$\,\AA\/, which changes if the
CO bond contracts (B) or extends (C). The OH-stretch potential is
energetically more attractive in the covalent bonding range if the CO
bond has ${\rm BO} = 1$ rather than ${\rm BO} = 2$.  Vice versa, the
OH-stretch potential becomes repulsive at shorter CO distances as in
CO double bonds. Thus, hydrogen abstraction from an oxygen donor atom
involved in a CO double bond is energetically more favourable compared
to that involved in a CO single bond.\\

\subsection{Solvent Distribution}
\begin{figure}[htb!]
\centering
\includegraphics[width=\linewidth]{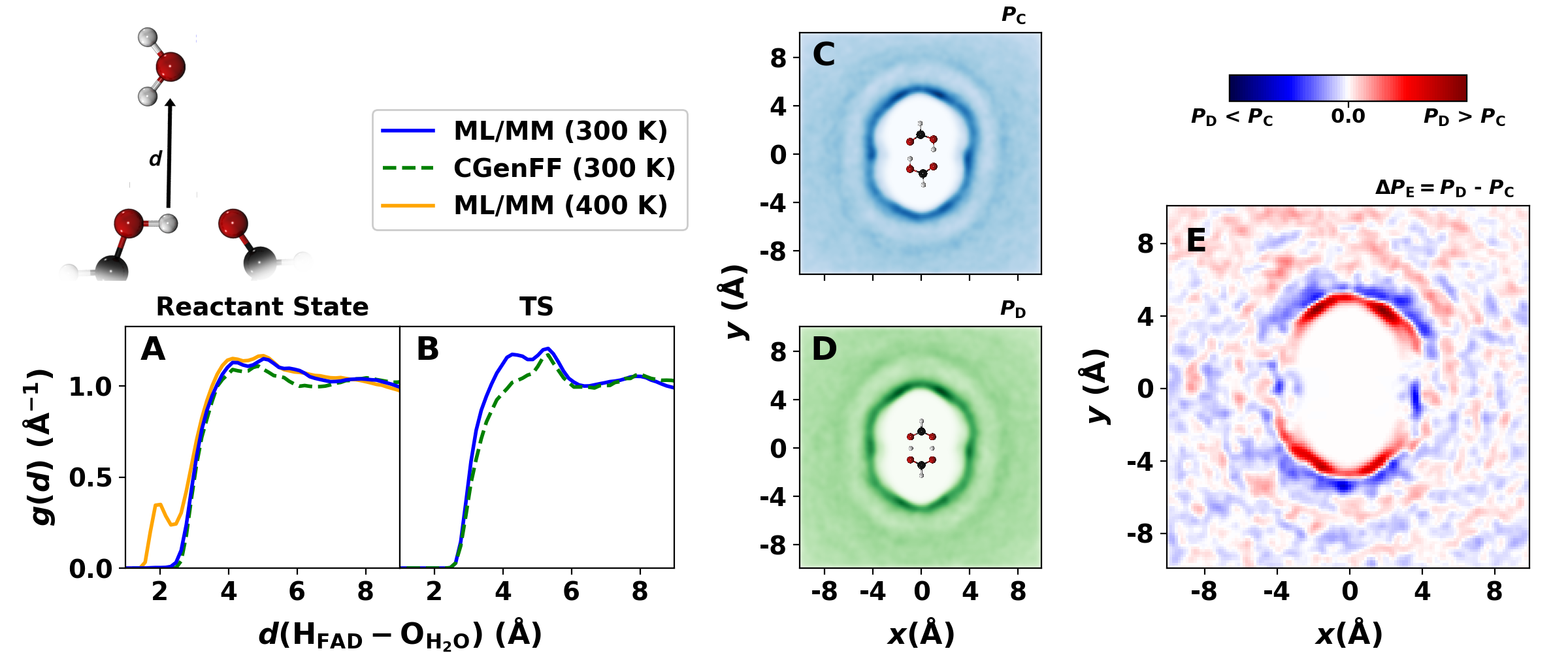}
\caption{The radial distribution function $g(d)$ for the water-oxygen
  atoms around hydrogen atoms H$_{\rm A}$ and H$_{\rm B}$ of FAD in
  the reactant (A) and TS (B) structure from the from ML/MM (blue
  line) and CGenFF (dashed green line) simulations at 300\,K. Panel A
  also shows $g(d)$ from one entire ML/MM simulation at 400\,K (orange
  line) including formic acid dominantly as cyclic dimer and separated
  monomers (see Figure~\ref{fig:2}A). Panels C and D report the
  solvent distribution functions $P(x,y)$ around FAD from ML/MM
  MD simulations in its reactant (C) and TS (D) structures,
  respectively, and panel E reports the difference $\Delta P(x,y)
    = P_{\rm D} - P_{\rm C}$ between the two.}
\label{fig:6}
\end{figure}

MD simulations also provide information about the solvent distribution
along the reaction path. Figure \ref{fig:6} reports the radial and
distribution of the water oxygen atoms in the vicinity of the hydrogen
atoms in the H-bonds of FAD from simulations at 300\,K and 400\,K. The
solvent distributions are obtained either from unbiased MD simulations
with FAD primarily in its reactant state and from biased simulations
with FAD constrained close to its TS conformation by applying harmonic
constraints along $\xi_1$ and $\xi_2$ with a force constant of $k=115$
\,kcal/mol/\AA$^2$. Analysis of the solvent distribution in the TS
compared with the reactant structure is of interest to assess the
amount of solvent reorganization that is required to reach the TS.\\

\noindent
The solvent radial distribution functions around the hydrogen atoms of
the H-bonds in Figures \ref{fig:6}A and B exhibit a weak double peak
between 4 and 6\,\AA. The maxima indicate the first solvation shell
around FAD and the double peak structure is more pronounced for
simulations with FAD in its TS structure. In the biased simulations
FAD is more rigid which leads to a more structured solvent
distribution. The radial distribution also shows a broad maximum
around 8\,\AA\/ that can be associated with a second solvation
shell. The solvent radial distribution from simulation at 400\,K
(orange line in Figure \ref{fig:6}A) features an additional peak
around 2\,\AA\/ that represents H-bond formation between branched,
i.e. non-cyclic FAD, separated formic acid monomers and water
molecules. Such H-bonds become available in simulation at temperatures
higher than 300\,K as the probability for almost exclusively cyclic
FAD decreases significantly (see Figure \ref{fig:2}A).\\

\noindent
  For a more comprehensive characterization of the changes in the
  hydration of FAD between product/reactant geometries and the TS
  structure, 2-dimensional solvent distributions were determined.
  Figures \ref{fig:6}C and D show the actual distributions projected
  onto the $xy-$plane which is the average plane containing the
  solute. Figure \ref{fig:6}E reports the solvent density difference
  around FAD in its reactant and TS conformation, i.e. the difference
  density from panels C and D in Figure \ref{fig:6}. In the
  2d-histograms the first and second solvation shells are clearly
  visible as was also found from the radial distribution functions
  (Figure \ref{fig:6}A and B). 
There is also a dent for the first
solvation shell at $(-4, 0)$\,\AA\/ and $(4, 0)$\,\AA\/ along the
horizontal axis joining the hydrogen atoms along the H-bonds of FAD.\\

\noindent
The solvent distribution difference in Figure \ref{fig:6}E indicates a
shift of the first solvation shell closer to the side of the CH group
of FAD for the biased simulations due to the short monomer-monomer
distance in the TS structure compared with the reactant state. Hence,
the solvent probability in this region is higher for FAD in its TS
than in its reactant state. This is consistent with the shorter
monomer-monomer distance of FAD in the TS conformation. The
monomer-monomer contraction in the TS also flattens the dent towards
the hydrogen atoms in the H-bond as shown by the blue area that
indicates a lower solvent distribution density around FAD in the TS
than the reactant state conformation.  Finally, the solvent
distribution difference exhibits the symmetry of FAD which indicates
the convergence of the simulations.\\

\subsection{Influence of the Solvent-Generated Electrical Field}
Finally, DPT events are analyzed from the perspective of the
alignment of the H-transfer motif relative to the solvent-generated
electric field. 
  Although cyclic FAD was found to be rather
  hydrophobic, the long ranged nature of the solvent-generated
  electric field can still significantly impact H-transfer along the
  progression coordinate for DPT. The effect of external fields on
  chemical reactions has been demonstrated from both,
  experiments\cite{huang:2019} and computations.\cite{shaik:2004}\\

\begin{figure}[htb!]
\centering 
\includegraphics[width=\linewidth]{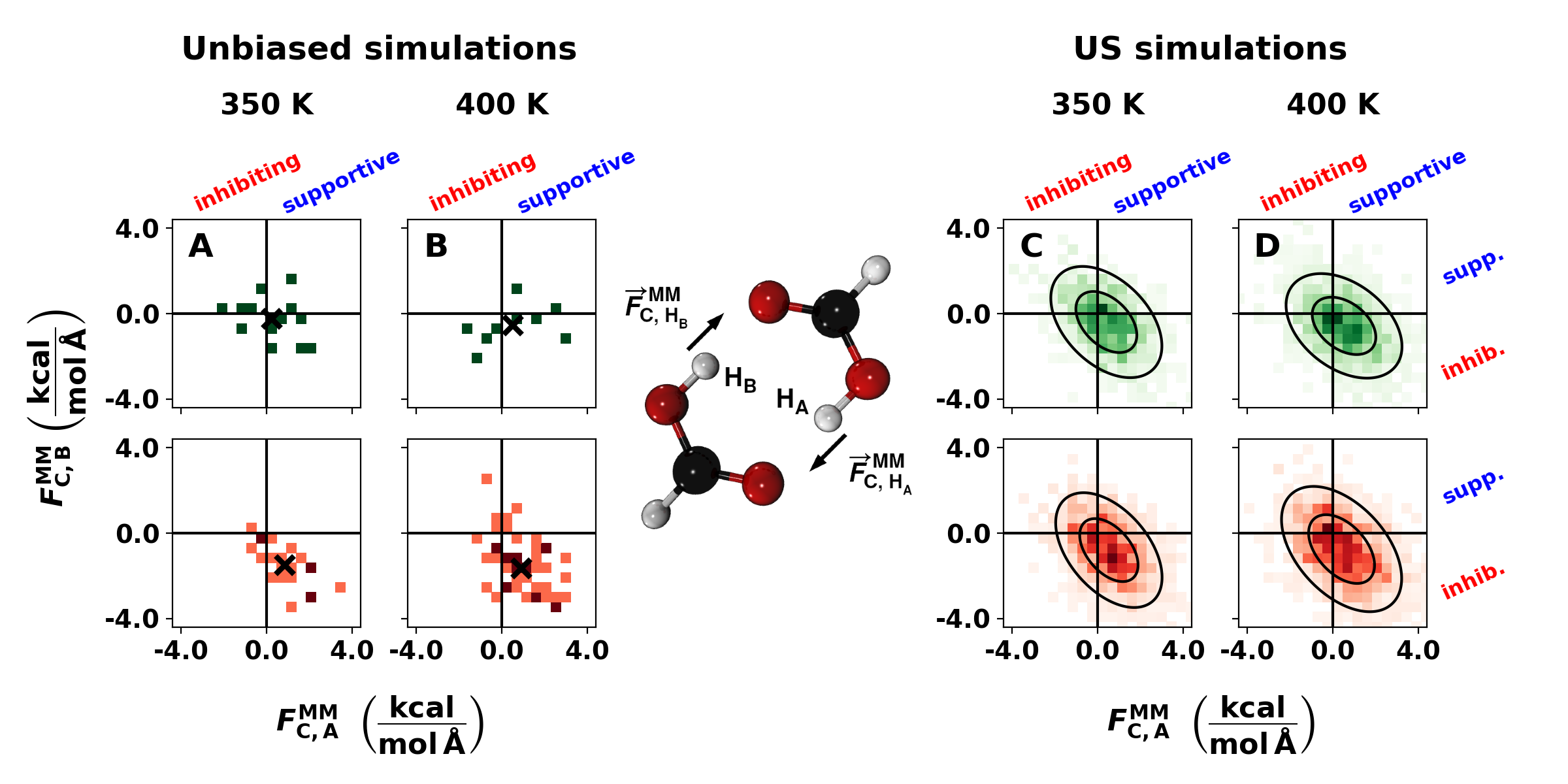}
\caption{Correlation between the magnitude and direction of the
  electrostatic component of the solvent-generated Coulomb force
  $\vec{F}_{\rm C}^{\rm MM}$ at the position of the transferring
  hydrogen atom during the first PT that leads to successful (green)
  or attempted (red) DPT; for definitions see Analysis. Results from
  unbiased (A and B) and biased simulations (C and D) are shown
  separately. The panels show the results from simulation at (A and C)
  350\,K and (B and D) 400\,K. The forces on the hydrogen atom
  H$_\mathrm{A}$ that has completed the first PT are given along the
  horizontal axis. On the vertical axis the forces on the hydrogen
  atom H$_\mathrm{B}$ are shown for either supporting or inhibiting
  the second PT (for successful DPT or attempted DPT,
  respectively). By definition, the force vector always points towards
  O$_{\rm acc}$ in the H-bond. Effective electrostatic force on the
  hydrogen atom for a PT towards O$_{\rm acc}$ in the respective
  H-bond is ``supportive'', and ``inhibiting'' otherwise. For the same
  analysis involving the second PT, see Figure S10.}
\label{fig:7}
\end{figure}

\noindent
In the following the electric field at the position of the
transferring hydrogen atoms along the hydrogen bond was analyzed. For
this, the force on the transferring hydrogen atom at the time of PT
was determined as the linear interpolation of the forces for the 
$1$\,fs-frame before and after PT. The ``time of PT'' was determined 
by the sign change of the linearly interpolated reactive coordinate 
of the respective H-bond along which PT occurs. In Figures \ref{fig:7}
and S10 the first PT - to which hydrogen atom H$_{\rm A}$ is
attributed - is associated with the horizontal $x- $axis. In a
successful DPT, the second PT is performed by hydrogen atom
H$_\mathrm{B}$ (vertical $y$-axis) whereas in an attempted DPT the
second PT is also performed by H$_\mathrm{A}$ (back-transfer).\\

\noindent
Figures \ref{fig:7} and S10 report the projection of the
solvent-generated (MM) Coulomb (C) force $F^\mathrm{MM}_\mathrm{C,A}$
along the O$_{\rm don}$--H$_\mathrm{A} \cdots$O$_{\rm acc}$ and
O$_{\rm don}$--H$_\mathrm{B}\cdots$O$_{\rm acc}$ hydrogen bonds for
successful and attempted DPTs during the first and second PT,
respectively. As per definition, a positive force always supports PT
of a hydrogen atom whereas a negative force has an inhibitory
effect. The analysis was carried out for the second PT independent of
whether transfer of H$_\mathrm{B}$ did (successful DPT) or did not
(attempted DPT with H$_\mathrm{A}$ transferring back) occur.\\

\noindent
In the unbiased simulations, the first PT is typically accompanied by
a supportive Coulomb force. This effect is more pronounced at higher
temperatures ($T = 400$\,K, see Figure \ref{fig:7}B) than at lower
ones (see Figure \ref{fig:7}A). Biased simulations increase the
probability for DPT and provide improved statistics which also allows
to fit 2-dimensional Gauss distributions to the data (Figures
\ref{fig:7}C and D). These results demonstrate that for the first PT
the field along the H$_\mathrm{A}$-bond is typically
supportive. Interestingly, for an attempted DPT the field along the
first H$_\mathrm{A}$-bond at the time of PT is supportive but
$F^\mathrm{MM}_\mathrm{C,B}$ along the second H$_\mathrm{B}$-bond is
more inhibiting. This applies to both, unbiased and biased
simulations, see red distributions in Figure \ref{fig:7}. Regarding
Figure \ref{fig:7} the gravity center of the average solvent-generated
Coulomb force is [$\bar{F}^\mathrm{MM}_\mathrm{C,A}$,
  $\bar{F}^\mathrm{MM}_\mathrm{C,B}$] = [$0.24$, $-0.23$] and [$0.86$,
  $-1.50$]\,kcal/mol/\AA\/ for successful and attempted DPT at 350\,K,
respectively, and shifts to [$0.51$, $-0.56$] and [$-0.93$,
  $-1.65$]\,kcal/mol/\AA\/ at 400\,K. In biased simulations the
gravity centers are at [$0.45$, $-0.43$] and [$0.56$,
  $-0.78$]\,kcal/mol/\AA, [$0.56$, $-0.55$] and [$-0.51$,
  $-0.81$]\,kcal/mol/\AA\/ accordingly. Hence, for successful DPT the
opposing electrostatic field along the second H-bond is considerably
smaller in magnitude than for attempted DPT in all cases.\\

\noindent
For the second PT the solvent-generated Coulomb force along both
H-bonds is in general inhibiting (red distributions in Figure
S10). In unbiased simulations at 350\,K the center of
gravity of the average solvent-generated Coulomb force is at
[$\bar{F}^\mathrm{MM}_\mathrm{C,A}$,
  $\bar{F}^\mathrm{MM}_\mathrm{C,B}$] = [$-0.21$, $-0.22$] and
[$-0.84$, $-1.52$]\,kcal/mol/\AA\/ for successful and attempted DPT,
respectively. This shifts to [$-0.52$, $-0.58$] and [$-0.92$,
  $-1.64$]\,kcal/mol/\AA\/ at 400\,K. In biased simulations the
gravity centers are at [$-0.45,-0.42$] and [$-0.56$,
  $-0.78$]\,kcal/mol/\AA\/ at 350\,K and [$-0.56$, $-0.54$] and
[$-0.51$, $-0.81$]\,kcal/mol/\AA\/ at 400\,K. Except for the sign
change in the projection of the Coulomb force on the hydrogen atoms of
the H$_\mathrm{A}$-bond the center of gravity does not change
significantly between the first and second PT in successful and
attempted DPT.\\

\noindent
Comparison of the solvent-generated Coulomb force for successful and
attempted DPT further establishes that the field of the water
molecules at the position of the transferring hydrogen atom does not
change appreciably within $\sim 14$\,fs, which is the maximum time
between first and second PT observed in the simulations at
350\,K. This is expected as the rotational reorientation time of water
in solution is on the order of several picoseconds.\cite{skinner:2001}
The time evolution of the Coulomb force along the H-bonds is shown in
Figures S11 and S12 for an example of a
successful and attempted DPT, respectively.\\

\noindent
  In summary, successful DPT for FAD in water is often accompanied
  by a mildly supporting solvent-generated Coulomb force along the
  first PT in the H$_\mathrm{A}$-bond whereas for the second PT in the
  H$_\mathrm{B}$-bond the force is inhibiting but smaller in magnitude
  than for the back transfer along the first H$_\mathrm{A}$-bond. As
  the inhibiting force along the H$_\mathrm{B}$-bond increases,
  back-transfer along the first H$_\mathrm{A}$-bond becomes favoured
  and results in attempted DPT with FAD returning to its initial
  reactant conformation, see Figure S1. The findings
complete the picture that the probability for successful DPT primarily
depends on the conformation of FAD itself at the first PT event. At
conformation of high symmetry and low reaction barriers for PT events
in both H-bonds, the solvent-generated Coulomb field impacts the
probability of performing one PT in both H-bonds each (successful DPT)
or a forth and back PT along one H-bond (attempted DPT).\\

\section{Discussion and Conclusions}
In this work double proton transfer in cyclic FAD in solution was
characterized from extensive ML/MM MD simulations (total of $\sim 100$
ns for all temperatures and all charge models considered) using an
energy function at MP2/aug-cc-pVTZ level quality for the solute akin
to a QM/MM treatment at this level of quantum chemical theory. 
  To
  put the present simulations in context it is worthwhile to note that
  100\,ps of a ML/MM MD simulation (with ML trained at the
  MP2/aug-cc-pVTZ level) on 1 CPU takes $\sim 5$ days for the $5 \cdot
  10^5$ energy and force evaluations. At the MP2/aug-cc-pVTZ level of
  theory the $5 \cdot 10^5$ energy evaluations alone, i.e. without
  calculating forces, on 1 CPU would take at least 4200 days. Clearly,
  QM/MM simulation at such a level of theory are not feasible at
  present and in the foreseeable future. It is further worth to
  mention that the PhysNet representation captures bond polarization
  through geometry-dependent, fluctuating charges. On the other hand,
  intramolecular polarization of FAD by the water solvent is not
  included but is expected to be small due to the hydrophobicity of
  cyclic FAD.\\

\noindent
Spontaneous DPT is observed on the nanosecond time scale at 300\,K to
350\,K with a maximum rate of 0.9\,ns$^{-1}$ at 350\,K. The total
ML/MM energy function used was validated vis-a-vis B3LYP+D3
calculations and agrees to within $\sim 1$\,kcal/mol for hydrated FAD
with up to 15 surrounding water molecules. One particularly
interesting aspect of the dynamics of FAD in solution is the fact that
dissociation into a branched, singly H-bonded structure or into two
separate formic acid monomers competes with DPT. The present results
(see Figure \ref{fig:2}) suggest that on the $\sim 14$\,ns time scale
the probability to form branched, singly H-bonded structures is
considerably smaller than for cyclic FAD to exist (which decreases as
$T$ increases) or that of forming two separated monomers (which
increases as $T$ increases). Simulations at higher temperature allow
to qualitatively assess conditions at higher pressure which has been
used to induce proton transfer in other
systems.\cite{martins:2009,ma:2017}\\

\noindent
  Results from previous simulation studies are somewhat
  ambiguous. Recent {\it ab initio} MD simulations using the BLYP
  functional including dispersion correction point towards pronounced
  hydrophobicity of cyclic FAD\cite{hanninen:2018} as was also found
  from earlier simulations using a fully empirical force
  field.\cite{brooks:2008} Both are consistent with the present
  work. On the other hand, simulations with this non-reactive, fully
  empirical force field report facile transitions from FAD to branched
  structures.\cite{brooks:2008} Such branched structures were also
  implied from experiments which were, however, carried out in 3 molal
  NaCl.\cite{scheraga:1964} 
Analysis of the $T-$dependent free
energies for DPT yields a change in solvation entropy of $\Delta S =
-2.1$\,cal/mol/K. It is of interest to juxtapose this value with
earlier work that rather focused on dissociation of FAD into two
separated monomers for which $\Delta S_{\rm D} = -3$\,cal/mol/K was
reported also using the TIP3P water model.\cite{brooks:2008} However,
it is quite likely that the two entropy values which differ by about
30\,\% are not directly comparable as dimerization is driven by the
different arrangements of solvent molecules around FAD and the FAMs
whereas for DPT the reorganization is less pronounced; see Figure
\ref{fig:6}E.\\

\noindent
As the calculated p$K_{\rm D}$ underestimates the experimentally
measured one by 1 to 2 units (0.25 vs. 1.2 to 2.1),\cite{brooks:2008}
it is expected that the energy function used in these earlier free
energy simulations for cyclic FAD\cite{brooks:2008} underestimate the
enthalpic contribution by about one order of magnitude. The reported
stabilization from umbrella sampling simulations using the CHARMM22
force field was $\Delta G \sim -0.5$\,kcal/mol. Hence, to be
consistent with the experimentally determined p$K_{\rm D}$, the
enthalpic contribution to the stabilization of FAD in solution should
rather be $-5$\,kcal/mol or larger. Adding this to the stabilization
energy for FAD in the gas phase using the CGenFF energy function
($-9$\,kcal/mol) yields an estimated stabilization of $-14$\,kcal/mol
or larger which is more consistent with results from recent high-level
electronic structure calculations which find a gas-phase stability for
cyclic FAD of $-16$\,kcal/mol.\cite{Miliordos2015,MM.fad:2022} In
other words, the unreactive, empirical CGenFF energy function
considerably underestimates the stabilization of cyclic FAD in the gas
phase and in solution which is also confirmed by the results in
Figures S3 to S7 and leads to ready
dissociation of FAD into two FAMs from such
simulations.\cite{brooks:2008}\\

\noindent
  Successful DPT is found to chiefly depend on the conformation of
  the cyclic FAD. In particular, increased symmetry with
  simultaneously contracted O--O distances in both H-bonds and CO
  bonds with ${\rm BO} \approx 1.5$ lowers the reaction barrier for PT
  along both H-bonds. Successful DPT requires the solvent induced
  electrostatic force to be only mildly inhibiting $\sim
  -0.5$\,kcal/mol/\AA\/ along the second H-bond. If the inhibiting
  force originating from the solvent is too strong along the second
  H-bond, attempted DPT dominates. For DPT, the MD simulations show an
  average delay between two PT events of $\sim 5$ and $\sim 6$\,fs for
  attempted and successful DPT at 350\,K, respectively, and the
  solvent-generated Coulomb field assists the first PT. A time
  difference of $\sim 6$\,fs between the successive PT events for
  successful DPT compares with a time scale of $\sim 155$ fs for the
  O$_{\rm don}$--O$_{\rm acc}$ vibration. This suggests that DPT in
  hydrated FAD is essentially concerted and not stepwise; both PTs
  occur within one O$_{\rm don}$--O$_{\rm acc}$ vibration period. For
  a stepwise DPT the intermediate state (both protons on one FAM)
  would need to stabilize, i.e. have a lifetime, which is, however,
  not what is found here.\\

\noindent
From a chemical perspective, one point of particular note is the
successful and correct description of the change between bond orders 1
and 2 for the CO bond in PhysNet. Conventional empirical force fields
encode the bond order in the equilibrium separation of the bonded term
which does not allow easily for changes in the bond order depending on
changes in the chemical environment. Although there are examples for
capturing such effects\cite{MM.oxalate:2017} making provisions for it
in the context of empirical force fields is cumbersome. As Figures
\ref{fig:4}B, C, E, and F demonstrate, a NN-trained energy function
based on reference electronic structure calculations successfully
captures such chemical effects.\\

\noindent
  In summary, a ML/MM MD scheme was implemented and applied to DPT
  for FAD in solution. The results from $\sim 100$\,ns of ML/MM MD
  simulations show that DPT and dissociation into two FAMs compete
  depending on temperature. FAD is predominantly hydrophobic which
  agrees with earlier findings, and the rate for DPT of $\sim
  0.1$\,ns$^{-1}$ is consistent between biased and unbiased
  simulations. Extending the present work to other reaction types and
protein-ligand binding will provide deeper chemical understanding and
improved models suitable for statistically significant sampling to
give molecular level insight into processes in the condensed
phase. 
  Contrary to straight AIMD simulations at the correlated
  level, an ML/MM MD ansatz is feasible even for nanosecond time scale
  simulations.\\

\section*{Supporting Information}
The supporting information reports additional graphics
S1-S14.

\section*{Data Availability Statement}
The data needed for the PhysNet representation of the MP2 PES is
available at \url{https://github.com/MMunibas/fad.git}.

\section*{Acknowledgments}
This work was supported by the Swiss National Science Foundation
grants 200021-117810, 200020-188724, NCCR MUST, and the University of
Basel which is gratefully acknowledged.  This project received funding
from the European Union's Horizon 2020 research and innovation program
under the Marie Sk{\l}odowska-Curie grant agreement No 801459 -
FP-RESOMUS.

\bibliography{refs.tidy}

\providecommand{\latin}[1]{#1}
\makeatletter
\providecommand{\doi}
  {\begingroup\let\do\@makeother\dospecials
  \catcode`\{=1 \catcode`\}=2 \doi@aux}
\providecommand{\doi@aux}[1]{\endgroup\texttt{#1}}
\makeatother
\providecommand*\mcitethebibliography{\thebibliography}
\csname @ifundefined\endcsname{endmcitethebibliography}
  {\let\endmcitethebibliography\endthebibliography}{}
\begin{mcitethebibliography}{92}
\providecommand*\natexlab[1]{#1}
\providecommand*\mciteSetBstSublistMode[1]{}
\providecommand*\mciteSetBstMaxWidthForm[2]{}
\providecommand*\mciteBstWouldAddEndPuncttrue
  {\def\EndOfBibitem{\unskip.}}
\providecommand*\mciteBstWouldAddEndPunctfalse
  {\let\EndOfBibitem\relax}
\providecommand*\mciteSetBstMidEndSepPunct[3]{}
\providecommand*\mciteSetBstSublistLabelBeginEnd[3]{}
\providecommand*\EndOfBibitem{}
\mciteSetBstSublistMode{f}
\mciteSetBstMaxWidthForm{subitem}{(\alph{mcitesubitemcount})}
\mciteSetBstSublistLabelBeginEnd
  {\mcitemaxwidthsubitemform\space}
  {\relax}
  {\relax}

\bibitem[Warshel and Weiss(1980)Warshel, and Weiss]{warshel:1980}
Warshel,~A.; Weiss,~R.~M. An Empirical Valence Bond Approach for Comparing
  Reactions in Solutions and in Enzymes. \emph{J. Am. Chem. Soc.}
  \textbf{1980}, \emph{102}, 6218--6226\relax
\mciteBstWouldAddEndPuncttrue
\mciteSetBstMidEndSepPunct{\mcitedefaultmidpunct}
{\mcitedefaultendpunct}{\mcitedefaultseppunct}\relax
\EndOfBibitem
\bibitem[{\AA}qvist and Warshel(1993){\AA}qvist, and Warshel]{warshel:1993}
{\AA}qvist,~J.; Warshel,~A. Simulation of Enzyme Reactions Using Valence Bond
  Force Fields and Other Hybrid Quantum/classical Approaches. \emph{Chem. Rev.}
  \textbf{1993}, \emph{93}, 2523--2544\relax
\mciteBstWouldAddEndPuncttrue
\mciteSetBstMidEndSepPunct{\mcitedefaultmidpunct}
{\mcitedefaultendpunct}{\mcitedefaultseppunct}\relax
\EndOfBibitem
\bibitem[Gajewski and Brichford(1994)Gajewski, and Brichford]{gajewski:1994}
Gajewski,~J.; Brichford,~N. Secondary Deuterium Kinetic Isotope Effects in the
  Aqueous Claisen Rearrangement: Evidence against an Ionic Transition State.
  \emph{J. Am. Chem. Soc.} \textbf{1994}, \emph{117}, 3165--3166\relax
\mciteBstWouldAddEndPuncttrue
\mciteSetBstMidEndSepPunct{\mcitedefaultmidpunct}
{\mcitedefaultendpunct}{\mcitedefaultseppunct}\relax
\EndOfBibitem
\bibitem[Brickel and Meuwly(2019)Brickel, and Meuwly]{MM.claisen:2019}
Brickel,~S.; Meuwly,~M. Molecular Determinants for Rate Acceleration in the
  Claisen Rearrangement Reaction. \emph{J. Phys. Chem. B} \textbf{2019},
  \emph{123}, 448--456\relax
\mciteBstWouldAddEndPuncttrue
\mciteSetBstMidEndSepPunct{\mcitedefaultmidpunct}
{\mcitedefaultendpunct}{\mcitedefaultseppunct}\relax
\EndOfBibitem
\bibitem[Meuwly(2021)]{MM.reactions:2021}
Meuwly,~M. Machine Learning for Chemical Reactions. \emph{Chem. Rev.}
  \textbf{2021}, \emph{121}, 10218--10239\relax
\mciteBstWouldAddEndPuncttrue
\mciteSetBstMidEndSepPunct{\mcitedefaultmidpunct}
{\mcitedefaultendpunct}{\mcitedefaultseppunct}\relax
\EndOfBibitem
\bibitem[Kamerlin \latin{et~al.}(2009)Kamerlin, Haranczyk, and
  Warshel]{kamerlin:2009}
Kamerlin,~S.~C.; Haranczyk,~M.; Warshel,~A. Are Mixed Explicit/implicit
  Solvation Models Reliable for Studying Phosphate Hydrolysis? A Comparative
  Study of Continuum, Explicit and Mixed Solvation Models. \emph{ChemPhysChem}
  \textbf{2009}, \emph{10}, 1125--1134\relax
\mciteBstWouldAddEndPuncttrue
\mciteSetBstMidEndSepPunct{\mcitedefaultmidpunct}
{\mcitedefaultendpunct}{\mcitedefaultseppunct}\relax
\EndOfBibitem
\bibitem[Plech \latin{et~al.}(2004)Plech, Wulff, Bratos, Mirloup, Vuilleumier,
  Schotte, and Anfinrud]{anfinrud:2004}
Plech,~A.; Wulff,~M.; Bratos,~S.; Mirloup,~F.; Vuilleumier,~R.; Schotte,~F.;
  Anfinrud,~P.~A. Visualizing Chemical Reactions in Solution by Picosecond
  X-ray Diffraction. \emph{Phys. Rev. Lett.} \textbf{2004}, \emph{92},
  125505\relax
\mciteBstWouldAddEndPuncttrue
\mciteSetBstMidEndSepPunct{\mcitedefaultmidpunct}
{\mcitedefaultendpunct}{\mcitedefaultseppunct}\relax
\EndOfBibitem
\bibitem[Hu and Yang(2008)Hu, and Yang]{yang:2008}
Hu,~H.; Yang,~W. Free Energies of Chemical Reactions in Solution and in Enzymes
  with Ab Initio Quantum Mechanics/molecular Mechanics Methods. \emph{{Ann.
  Rev. Phys. Chem.}} \textbf{2008}, \emph{59}, 573--601\relax
\mciteBstWouldAddEndPuncttrue
\mciteSetBstMidEndSepPunct{\mcitedefaultmidpunct}
{\mcitedefaultendpunct}{\mcitedefaultseppunct}\relax
\EndOfBibitem
\bibitem[Kim \latin{et~al.}(2015)Kim, Kim, Nozawa, Sato, Oang, Kim, Ki, Jo,
  Park, Song, Sato, Ogawa, Togashi, Tono, Yabashi, Ishikawa, Kim, Ryoo, Kim,
  Ihee, and Adachi]{ihee:2015}
Kim,~K.~H.; Kim,~J.~G.; Nozawa,~S.; Sato,~T.; Oang,~K.~Y.; Kim,~T.; Ki,~H.;
  Jo,~J.; Park,~S.; Song,~C. \latin{et~al.}  Direct Observation of Bond
  Formation in Solution with Femtosecond X-ray Scattering. \emph{Nature}
  \textbf{2015}, \emph{518}, 385--389\relax
\mciteBstWouldAddEndPuncttrue
\mciteSetBstMidEndSepPunct{\mcitedefaultmidpunct}
{\mcitedefaultendpunct}{\mcitedefaultseppunct}\relax
\EndOfBibitem
\bibitem[Meuwly(2019)]{Meuwly2019}
Meuwly,~M. Reactive Molecular Dynamics: From Small Molecules to Proteins.
  \emph{WIREs Comput. Mol. Sci.} \textbf{2019}, \emph{9}, e1386\relax
\mciteBstWouldAddEndPuncttrue
\mciteSetBstMidEndSepPunct{\mcitedefaultmidpunct}
{\mcitedefaultendpunct}{\mcitedefaultseppunct}\relax
\EndOfBibitem
\bibitem[Nerenberg and Head-Gordon(2018)Nerenberg, and
  Head-Gordon]{HeadGordon2018}
Nerenberg,~P.~S.; Head-Gordon,~T. New Developments in Force Fields for
  Biomolecular Simulations. \emph{Curr. Opin. Struct. Biol.} \textbf{2018},
  \emph{49}, 129--138\relax
\mciteBstWouldAddEndPuncttrue
\mciteSetBstMidEndSepPunct{\mcitedefaultmidpunct}
{\mcitedefaultendpunct}{\mcitedefaultseppunct}\relax
\EndOfBibitem
\bibitem[Koner \latin{et~al.}(2020)Koner, Salehi, Mondal, and
  Meuwly]{KonerMeuwly2020}
Koner,~D.; Salehi,~S.~M.; Mondal,~P.; Meuwly,~M. Non-conventional Force Fields
  for Applications in Spectroscopy and Chemical Reaction Dynamics. \emph{J.
  Chem. Phys.} \textbf{2020}, \emph{153}, 010901\relax
\mciteBstWouldAddEndPuncttrue
\mciteSetBstMidEndSepPunct{\mcitedefaultmidpunct}
{\mcitedefaultendpunct}{\mcitedefaultseppunct}\relax
\EndOfBibitem
\bibitem[Guest \latin{et~al.}(1997)Guest, Craw, Vincent, and
  Hillier]{Guest1997}
Guest,~J.~M.; Craw,~J.~S.; Vincent,~M.~A.; Hillier,~I.~H. The Effect of Water
  on the Claisen Rearrangement of Allyl Vinyl Ether: Theoretical Methods
  Including Explicit Solvent and Electron Correlation. \emph{J. Chem. Soc.,
  Perkin Trans. 2} \textbf{1997}, 71\relax
\mciteBstWouldAddEndPuncttrue
\mciteSetBstMidEndSepPunct{\mcitedefaultmidpunct}
{\mcitedefaultendpunct}{\mcitedefaultseppunct}\relax
\EndOfBibitem
\bibitem[Jung and Marcus(2007)Jung, and Marcus]{Jung2007}
Jung,~Y.; Marcus,~R.~A. On the Theory of Organic Catalysis “on Water”.
  \emph{J. Am. Chem. Soc.} \textbf{2007}, \emph{129}, 5492--5502, PMID:
  17388592\relax
\mciteBstWouldAddEndPuncttrue
\mciteSetBstMidEndSepPunct{\mcitedefaultmidpunct}
{\mcitedefaultendpunct}{\mcitedefaultseppunct}\relax
\EndOfBibitem
\bibitem[White and Wolfarth(1970)White, and Wolfarth]{White1970}
White,~W.~N.; Wolfarth,~E.~F. The O-claisen Rearrangement. Viii. Solvent
  Effects. \emph{J. Org. Chem.} \textbf{1970}, \emph{35}, 2196\relax
\mciteBstWouldAddEndPuncttrue
\mciteSetBstMidEndSepPunct{\mcitedefaultmidpunct}
{\mcitedefaultendpunct}{\mcitedefaultseppunct}\relax
\EndOfBibitem
\bibitem[Acevedo and Armacost(2010)Acevedo, and Armacost]{Acevedo2010}
Acevedo,~O.; Armacost,~K. Claisen Rearrangements: Insight into Solvent Effects
  and “on Water” Reactivity from QM/MM Simulations. \emph{J. Am. Chem.
  Soc.} \textbf{2010}, \emph{132}, 1966\relax
\mciteBstWouldAddEndPuncttrue
\mciteSetBstMidEndSepPunct{\mcitedefaultmidpunct}
{\mcitedefaultendpunct}{\mcitedefaultseppunct}\relax
\EndOfBibitem
\bibitem[van Keulen \latin{et~al.}(2017)van Keulen, Solano, and
  Rothlisberger]{Rothlisberger2017}
van Keulen,~S.~C.; Solano,~A.; Rothlisberger,~U. How Rhodopsin Tunes the
  Equilibrium between Protonated and Deprotonated Forms of the Retinal
  Chromophore. \emph{J. Chem. Theo. Comp.} \textbf{2017}, \emph{13},
  4524--4534, PMID: 28731695\relax
\mciteBstWouldAddEndPuncttrue
\mciteSetBstMidEndSepPunct{\mcitedefaultmidpunct}
{\mcitedefaultendpunct}{\mcitedefaultseppunct}\relax
\EndOfBibitem
\bibitem[El~Hage \latin{et~al.}(2017)El~Hage, Brickel, Hermelin, Gaulier,
  Schmidt, Bonacina, van Keulen, Bhattacharyya, Chergui, Hamm, Rothlisberger,
  Wolf, and Meuwly]{Hage2017}
El~Hage,~K.; Brickel,~S.; Hermelin,~S.; Gaulier,~G.; Schmidt,~C.; Bonacina,~L.;
  van Keulen,~S.~C.; Bhattacharyya,~S.; Chergui,~M.; Hamm,~P. \latin{et~al.}
  Implications of Short Time Scale Dynamics on Long Time Processes.
  \emph{Struct. Dyn.} \textbf{2017}, \emph{4}, 061507\relax
\mciteBstWouldAddEndPuncttrue
\mciteSetBstMidEndSepPunct{\mcitedefaultmidpunct}
{\mcitedefaultendpunct}{\mcitedefaultseppunct}\relax
\EndOfBibitem
\bibitem[Mulholland \latin{et~al.}(2000)Mulholland, Lyne, and
  Karplus]{Karplus2000}
Mulholland,~A.~J.; Lyne,~P.~D.; Karplus,~M. Ab Initio QM/MM Study of the
  Citrate Synthase Mechanism. A Low-barrier Hydrogen Bond Is Not Involved.
  \emph{J. Am. Chem. Soc.} \textbf{2000}, \emph{122}, 534--535\relax
\mciteBstWouldAddEndPuncttrue
\mciteSetBstMidEndSepPunct{\mcitedefaultmidpunct}
{\mcitedefaultendpunct}{\mcitedefaultseppunct}\relax
\EndOfBibitem
\bibitem[Senn and Thiel(2009)Senn, and Thiel]{Senn2009}
Senn,~H.~M.; Thiel,~W. QM/MM Methods for Biomolecular Systems. \emph{Angew.
  Chem. Intern. Ed.} \textbf{2009}, \emph{48}, 1198--1229\relax
\mciteBstWouldAddEndPuncttrue
\mciteSetBstMidEndSepPunct{\mcitedefaultmidpunct}
{\mcitedefaultendpunct}{\mcitedefaultseppunct}\relax
\EndOfBibitem
\bibitem[Groenhof(2013)]{Groenhof2013}
Groenhof,~G. In \emph{Biomolecular Simulations: Methods and Protocols};
  Monticelli,~L., Salonen,~E., Eds.; Humana Press: Totowa, NJ, 2013; pp
  43--66\relax
\mciteBstWouldAddEndPuncttrue
\mciteSetBstMidEndSepPunct{\mcitedefaultmidpunct}
{\mcitedefaultendpunct}{\mcitedefaultseppunct}\relax
\EndOfBibitem
\bibitem[{van Duin} \latin{et~al.}(2001){van Duin}, Dasgupta, Lorant, and
  {Goddard III}]{goddard01reaxff}
{van Duin},~A. C.~T.; Dasgupta,~S.; Lorant,~F.; {Goddard III},~W. A.~. ReaxFF:
  A Reactive Force Field for Hydrocarbons. \emph{J. Phys. Chem. A}
  \textbf{2001}, \emph{105}, 9396--9409\relax
\mciteBstWouldAddEndPuncttrue
\mciteSetBstMidEndSepPunct{\mcitedefaultmidpunct}
{\mcitedefaultendpunct}{\mcitedefaultseppunct}\relax
\EndOfBibitem
\bibitem[Nagy \latin{et~al.}(2014)Nagy, Yosa~Reyes, and Meuwly]{MM.armd:2014}
Nagy,~T.; Yosa~Reyes,~J.; Meuwly,~M. Multisurface Adiabatic Reactive Molecular
  Dynamics. \emph{J. Chem. Theo. Comp.} \textbf{2014}, \emph{10},
  1366--1375\relax
\mciteBstWouldAddEndPuncttrue
\mciteSetBstMidEndSepPunct{\mcitedefaultmidpunct}
{\mcitedefaultendpunct}{\mcitedefaultseppunct}\relax
\EndOfBibitem
\bibitem[Ang \latin{et~al.}(2021)Ang, Wang, Schwalbe-Koda, Axelrod, and
  G\'omez-Bombarelli]{Ang2021}
Ang,~S.~J.; Wang,~W.; Schwalbe-Koda,~D.; Axelrod,~S.; G\'omez-Bombarelli,~R.
  Active Learning Accelerates Ab Initio Molecular Dynamics on Reactive Energy
  Surfaces. \emph{Chem} \textbf{2021}, \emph{7}, 738--751\relax
\mciteBstWouldAddEndPuncttrue
\mciteSetBstMidEndSepPunct{\mcitedefaultmidpunct}
{\mcitedefaultendpunct}{\mcitedefaultseppunct}\relax
\EndOfBibitem
\bibitem[B\"oselt \latin{et~al.}(2021)B\"oselt, Th\"urlemann, and
  Riniker]{Riniker2021}
B\"oselt,~L.; Th\"urlemann,~M.; Riniker,~S. Machine Learning in QM/MM Molecular
  Dynamics Simulations of Condensed-phase Systems. \emph{J. Chem. Theo. Comp.}
  \textbf{2021}, \emph{17}, 2641--2658\relax
\mciteBstWouldAddEndPuncttrue
\mciteSetBstMidEndSepPunct{\mcitedefaultmidpunct}
{\mcitedefaultendpunct}{\mcitedefaultseppunct}\relax
\EndOfBibitem
\bibitem[Pan and McAllister(1997)Pan, and McAllister]{Pan1997}
Pan,~Y.; McAllister,~M.~A. Characterization of Low-barrier Hydrogen Bonds. 1.
  Microsolvation Effects. An Ab Initio and DFT Investigation. \emph{J. Am.
  Chem. Soc.} \textbf{1997}, \emph{119}, 7561--7566\relax
\mciteBstWouldAddEndPuncttrue
\mciteSetBstMidEndSepPunct{\mcitedefaultmidpunct}
{\mcitedefaultendpunct}{\mcitedefaultseppunct}\relax
\EndOfBibitem
\bibitem[Lim \latin{et~al.}(1997)Lim, Lee, and Kim]{Lim1997}
Lim,~J.-H.; Lee,~E.~K.; Kim,~Y. Theoretical Study for Solvent Effect on the
  Potential Energy Surface for the Double Proton Transfer in Formic Acid Dimer
  and Formamidine Dimer. \emph{J. Phys. Chem. A} \textbf{1997}, \emph{101},
  2233--2239\relax
\mciteBstWouldAddEndPuncttrue
\mciteSetBstMidEndSepPunct{\mcitedefaultmidpunct}
{\mcitedefaultendpunct}{\mcitedefaultseppunct}\relax
\EndOfBibitem
\bibitem[Miura \latin{et~al.}(1998)Miura, Tuckerman, and Klein]{Klein1998}
Miura,~S.; Tuckerman,~M.~E.; Klein,~M.~L. An Ab Initio Path Integral Molecular
  Dynamics Study of Double Proton Transfer in the Formic Acid Dimer. \emph{J.
  Chem. Phys.} \textbf{1998}, \emph{109}, 5290--5299\relax
\mciteBstWouldAddEndPuncttrue
\mciteSetBstMidEndSepPunct{\mcitedefaultmidpunct}
{\mcitedefaultendpunct}{\mcitedefaultseppunct}\relax
\EndOfBibitem
\bibitem[Kohanoff \latin{et~al.}(2000)Kohanoff, Koval, Estrin, Laria, and
  Abashkin]{Kohanoff2000}
Kohanoff,~J.; Koval,~S.; Estrin,~D.~A.; Laria,~D.; Abashkin,~Y. Concertedness
  and Solvent Effects in Multiple Proton Transfer Reactions: The Formic Acid
  Dimer in Solution. \emph{J. Chem. Phys.} \textbf{2000}, \emph{112},
  9498--9508\relax
\mciteBstWouldAddEndPuncttrue
\mciteSetBstMidEndSepPunct{\mcitedefaultmidpunct}
{\mcitedefaultendpunct}{\mcitedefaultseppunct}\relax
\EndOfBibitem
\bibitem[Ushiyama and Takatsuka(2001)Ushiyama, and Takatsuka]{Ushiyama2001}
Ushiyama,~H.; Takatsuka,~K. Successive Mechanism of Double-proton Transfer in
  Formic Acid Dimer: A Classical Study. \emph{J. Chem. Phys.} \textbf{2001},
  \emph{115}, 5903--5912\relax
\mciteBstWouldAddEndPuncttrue
\mciteSetBstMidEndSepPunct{\mcitedefaultmidpunct}
{\mcitedefaultendpunct}{\mcitedefaultseppunct}\relax
\EndOfBibitem
\bibitem[Kalescky \latin{et~al.}(2013)Kalescky, Kraka, and
  Cremer]{kalescky2013local}
Kalescky,~R.; Kraka,~E.; Cremer,~D. Local Vibrational Modes of the Formic Acid
  Dimer--the Strength of the Double Hydrogen Bond. \emph{Mol. Phys.}
  \textbf{2013}, \emph{111}, 1497--1510\relax
\mciteBstWouldAddEndPuncttrue
\mciteSetBstMidEndSepPunct{\mcitedefaultmidpunct}
{\mcitedefaultendpunct}{\mcitedefaultseppunct}\relax
\EndOfBibitem
\bibitem[Ivanov \latin{et~al.}(2015)Ivanov, Grant, and Marx]{marx:2015}
Ivanov,~S.~D.; Grant,~I.~M.; Marx,~D. Quantum Free Energy Landscapes from Ab
  Initio Path Integral Metadynamics: Double Proton Transfer in the Formic Acid
  Dimer Is Concerted but Not Correlated. \emph{J. Chem. Phys.} \textbf{2015},
  \emph{143}, 124304\relax
\mciteBstWouldAddEndPuncttrue
\mciteSetBstMidEndSepPunct{\mcitedefaultmidpunct}
{\mcitedefaultendpunct}{\mcitedefaultseppunct}\relax
\EndOfBibitem
\bibitem[Miliordos and Xantheas(2015)Miliordos, and Xantheas]{Miliordos2015}
Miliordos,~E.; Xantheas,~S.~S. On the Validity of the Basis Set Superposition
  Error and Complete Basis Set Limit Extrapolations for the Binding Energy of
  the Formic Acid Dimer. \emph{J. Chem. Phys.} \textbf{2015}, \emph{142},
  094311\relax
\mciteBstWouldAddEndPuncttrue
\mciteSetBstMidEndSepPunct{\mcitedefaultmidpunct}
{\mcitedefaultendpunct}{\mcitedefaultseppunct}\relax
\EndOfBibitem
\bibitem[Tew and Mizukami(2016)Tew, and Mizukami]{tew2016ab}
Tew,~D.~P.; Mizukami,~W. Ab Initio Vibrational Spectroscopy of Cis-and
  Trans-formic Acid from a Global Potential Energy Surface. \emph{J. Phys.
  Chem. A} \textbf{2016}, \emph{120}, 9815--9828\relax
\mciteBstWouldAddEndPuncttrue
\mciteSetBstMidEndSepPunct{\mcitedefaultmidpunct}
{\mcitedefaultendpunct}{\mcitedefaultseppunct}\relax
\EndOfBibitem
\bibitem[Qu and Bowman(2016)Qu, and Bowman]{bowman.fad:2016}
Qu,~C.; Bowman,~J.~M. An Ab Initio Potential Energy Surface for the Formic Acid
  Dimer: Zero-point Energy, Selected Anharmonic Fundamental Energies, and
  Ground-State Tunneling Splitting Calculated in Relaxed 1--4-mode Subspaces.
  \emph{Phys. Chem. Chem. Phys.} \textbf{2016}, \emph{18}, 24835--24840\relax
\mciteBstWouldAddEndPuncttrue
\mciteSetBstMidEndSepPunct{\mcitedefaultmidpunct}
{\mcitedefaultendpunct}{\mcitedefaultseppunct}\relax
\EndOfBibitem
\bibitem[Mackeprang \latin{et~al.}(2016)Mackeprang, Xu, Maroun, Meuwly, and
  Kjaergaard]{MM.fad:2016}
Mackeprang,~K.; Xu,~Z.-H.; Maroun,~Z.; Meuwly,~M.; Kjaergaard,~H.~G.
  Spectroscopy and Dynamics of Double Proton Transfer in Formic Acid Dimer.
  \emph{Phys. Chem. Chem. Phys.} \textbf{2016}, \emph{18}, 24654--24662\relax
\mciteBstWouldAddEndPuncttrue
\mciteSetBstMidEndSepPunct{\mcitedefaultmidpunct}
{\mcitedefaultendpunct}{\mcitedefaultseppunct}\relax
\EndOfBibitem
\bibitem[Richardson(2017)]{richardson:2017}
Richardson,~J.~O. Full-and Reduced-dimensionality Instanton Calculations of the
  Tunnelling Splitting in the Formic Acid Dimer. \emph{Phys. Chem. Chem. Phys.}
  \textbf{2017}, \emph{19}, 966--970\relax
\mciteBstWouldAddEndPuncttrue
\mciteSetBstMidEndSepPunct{\mcitedefaultmidpunct}
{\mcitedefaultendpunct}{\mcitedefaultseppunct}\relax
\EndOfBibitem
\bibitem[Qu and Bowman(2018)Qu, and Bowman]{qu2018high}
Qu,~C.; Bowman,~J.~M. High-dimensional Fitting of Sparse Datasets of CCSD(T)
  Electronic Energies and MP2 Dipole Moments, Illustrated for the Formic Acid
  Dimer and Its Complex IR Spectrum. \emph{J. Chem. Phys.} \textbf{2018},
  \emph{148}, 241713\relax
\mciteBstWouldAddEndPuncttrue
\mciteSetBstMidEndSepPunct{\mcitedefaultmidpunct}
{\mcitedefaultendpunct}{\mcitedefaultseppunct}\relax
\EndOfBibitem
\bibitem[Qu and Bowman(2018)Qu, and Bowman]{qu2018quantum}
Qu,~C.; Bowman,~J.~M. Quantum and Classical Ir Spectra of (HCOOH)$_2$,
  (DCOOH)$_2$ and (DCOOD)$_2$ Using Ab Initio Potential Energy and Dipole
  Moment Surfaces. \emph{Faraday Discuss.} \textbf{2018}, \emph{212},
  33--49\relax
\mciteBstWouldAddEndPuncttrue
\mciteSetBstMidEndSepPunct{\mcitedefaultmidpunct}
{\mcitedefaultendpunct}{\mcitedefaultseppunct}\relax
\EndOfBibitem
\bibitem[Qu and Bowman(2018)Qu, and Bowman]{qu2018ir}
Qu,~C.; Bowman,~J.~M. Ir Spectra of (HCOOH)$_2$ and (DCOOH)$_2$: Experiment,
  VSCF/VCI, and Ab Initio Molecular Dynamics Calculations Using
  Full-dimensional Potential and Dipole Moment Surfaces. \emph{J. Phys. Chem.
  Lett.} \textbf{2018}, \emph{9}, 2604--2610\relax
\mciteBstWouldAddEndPuncttrue
\mciteSetBstMidEndSepPunct{\mcitedefaultmidpunct}
{\mcitedefaultendpunct}{\mcitedefaultseppunct}\relax
\EndOfBibitem
\bibitem[K{\"a}ser and Meuwly(2022)K{\"a}ser, and Meuwly]{MM.fad:2022}
K{\"a}ser,~S.; Meuwly,~M. Transfer Learned Potential Energy Surfaces: Accurate
  Anharmonic Vibrational Dynamics and Dissociation Energies for the Formic Acid
  Monomer and Dimer. \emph{Phys. Chem. Chem. Phys.} \textbf{2022}, \relax
\mciteBstWouldAddEndPunctfalse
\mciteSetBstMidEndSepPunct{\mcitedefaultmidpunct}
{}{\mcitedefaultseppunct}\relax
\EndOfBibitem
\bibitem[Ito and Nakanaga(2000)Ito, and Nakanaga]{ito2000jet}
Ito,~F.; Nakanaga,~T. A Jet-cooled Infrared Spectrum of the Formic Acid Dimer
  by Cavity Ring-down Spectroscopy. \emph{Chem. Phys. Lett.} \textbf{2000},
  \emph{318}, 571--577\relax
\mciteBstWouldAddEndPuncttrue
\mciteSetBstMidEndSepPunct{\mcitedefaultmidpunct}
{\mcitedefaultendpunct}{\mcitedefaultseppunct}\relax
\EndOfBibitem
\bibitem[Freytes \latin{et~al.}(2002)Freytes, Hurtmans, Kassi, Li{\'e}vin,
  Vander~Auwera, Campargue, and Herman]{freytes2002overtone}
Freytes,~M.; Hurtmans,~D.; Kassi,~S.; Li{\'e}vin,~J.; Vander~Auwera,~J.;
  Campargue,~A.; Herman,~M. Overtone Spectroscopy of Formic Acid. \emph{Chem.
  Phys.} \textbf{2002}, \emph{283}, 47--61\relax
\mciteBstWouldAddEndPuncttrue
\mciteSetBstMidEndSepPunct{\mcitedefaultmidpunct}
{\mcitedefaultendpunct}{\mcitedefaultseppunct}\relax
\EndOfBibitem
\bibitem[Georges \latin{et~al.}(2004)Georges, Freytes, Hurtmans, Kleiner,
  Vander~Auwera, and Herman]{georges2004jet}
Georges,~R.; Freytes,~M.; Hurtmans,~D.; Kleiner,~I.; Vander~Auwera,~J.;
  Herman,~M. Jet-cooled and Room Temperature FTIR Spectra of the Dimer of
  Formic Acid in the Gas Phase. \emph{Chem. Phys.} \textbf{2004}, \emph{305},
  187--196\relax
\mciteBstWouldAddEndPuncttrue
\mciteSetBstMidEndSepPunct{\mcitedefaultmidpunct}
{\mcitedefaultendpunct}{\mcitedefaultseppunct}\relax
\EndOfBibitem
\bibitem[Zielke and Suhm(2007)Zielke, and Suhm]{zielke:2007}
Zielke,~P.; Suhm,~M. Raman Jet Spectroscopy of Formic Acid Dimers: Low
  Frequency Vibrational Dynamics and Beyond. \emph{Phys. Chem. Chem. Phys.}
  \textbf{2007}, \emph{9}, 4528--4534\relax
\mciteBstWouldAddEndPuncttrue
\mciteSetBstMidEndSepPunct{\mcitedefaultmidpunct}
{\mcitedefaultendpunct}{\mcitedefaultseppunct}\relax
\EndOfBibitem
\bibitem[Xue and Suhm(2009)Xue, and Suhm]{xue2009}
Xue,~Z.; Suhm,~M.~A. Probing the Stiffness of the Simplest Double Hydrogen
  Bond: The Symmetric Hydrogen Bond Modes of Jet-cooled Formic Acid Dimer.
  \emph{J. Chem. Phys.} \textbf{2009}, \emph{131}, 054301\relax
\mciteBstWouldAddEndPuncttrue
\mciteSetBstMidEndSepPunct{\mcitedefaultmidpunct}
{\mcitedefaultendpunct}{\mcitedefaultseppunct}\relax
\EndOfBibitem
\bibitem[Kollipost \latin{et~al.}(2012)Kollipost, Larsen, Domanskaya,
  N{\"o}renberg, and Suhm]{suhm:2012}
Kollipost,~F.; Larsen,~R.~W.; Domanskaya,~A.; N{\"o}renberg,~M.; Suhm,~M.
  Communication: The Highest Frequency Hydrogen Bond Vibration and an
  Experimental Value for the Dissociation Energy of Formic Acid Dimer. \emph{J.
  Chem. Phys.} \textbf{2012}, \emph{136}, 151101\relax
\mciteBstWouldAddEndPuncttrue
\mciteSetBstMidEndSepPunct{\mcitedefaultmidpunct}
{\mcitedefaultendpunct}{\mcitedefaultseppunct}\relax
\EndOfBibitem
\bibitem[Nejad and Suhm(2020)Nejad, and Suhm]{suhm:2020}
Nejad,~A.; Suhm,~M.~A. Concerted Pair Motion Due to Double Hydrogen Bonding:
  The Formic Acid Dimer Case. \emph{J. Ind. Inst. Sci.} \textbf{2020},
  \emph{100}, 5--19\relax
\mciteBstWouldAddEndPuncttrue
\mciteSetBstMidEndSepPunct{\mcitedefaultmidpunct}
{\mcitedefaultendpunct}{\mcitedefaultseppunct}\relax
\EndOfBibitem
\bibitem[Reutemann and Kieczka(2011)Reutemann, and
  Kieczka]{reutemann2011formic}
Reutemann,~W.; Kieczka,~H. \emph{Ullmann's Encyclopedia of Industrial
  Chemistry}; American Cancer Society, 2011\relax
\mciteBstWouldAddEndPuncttrue
\mciteSetBstMidEndSepPunct{\mcitedefaultmidpunct}
{\mcitedefaultendpunct}{\mcitedefaultseppunct}\relax
\EndOfBibitem
\bibitem[Balabin(2009)]{balabin2009}
Balabin,~R.~M. Polar (acyclic) Isomer of Formic Acid Dimer: Gas-phase Raman
  Spectroscopy Study and Thermodynamic Parameters. \emph{J. Phys. Chem. A}
  \textbf{2009}, \emph{113}, 4910--4918\relax
\mciteBstWouldAddEndPuncttrue
\mciteSetBstMidEndSepPunct{\mcitedefaultmidpunct}
{\mcitedefaultendpunct}{\mcitedefaultseppunct}\relax
\EndOfBibitem
\bibitem[Li \latin{et~al.}(2019)Li, Evangelisti, Gou, Caminati, and
  Meyer]{caminati:2019}
Li,~W.; Evangelisti,~L.; Gou,~Q.; Caminati,~W.; Meyer,~R. The Barrier to Proton
  Transfer in the Dimer of Formic Acid: A Pure Rotational Study. \emph{Angew.
  Chem. Intern. Ed.} \textbf{2019}, \emph{58}, 859--865\relax
\mciteBstWouldAddEndPuncttrue
\mciteSetBstMidEndSepPunct{\mcitedefaultmidpunct}
{\mcitedefaultendpunct}{\mcitedefaultseppunct}\relax
\EndOfBibitem
\bibitem[Zhang \latin{et~al.}(2017)Zhang, Li, Luo, Zhu, and Duan]{duan:2017}
Zhang,~Y.; Li,~W.; Luo,~W.; Zhu,~Y.; Duan,~C. High Resolution Jet-cooled
  Infrared Absorption Spectra of (HCOOH)$_2$, (HCOOD)$_2$, and HCOOH-HCOOD
  Complexes in 7.2 $\mu$m Region. \emph{J. Chem. Phys.} \textbf{2017},
  \emph{146}, 244306\relax
\mciteBstWouldAddEndPuncttrue
\mciteSetBstMidEndSepPunct{\mcitedefaultmidpunct}
{\mcitedefaultendpunct}{\mcitedefaultseppunct}\relax
\EndOfBibitem
\bibitem[Ortlieb and Havenith(2007)Ortlieb, and Havenith]{ortlieb2007proton}
Ortlieb,~M.; Havenith,~M. Proton Transfer in (HCOOH)$_2$: An IR High-resolution
  Spectroscopic Study of the Antisymmetric C--O Stretch. \emph{J. Phys. Chem.
  A} \textbf{2007}, \emph{111}, 7355--7363\relax
\mciteBstWouldAddEndPuncttrue
\mciteSetBstMidEndSepPunct{\mcitedefaultmidpunct}
{\mcitedefaultendpunct}{\mcitedefaultseppunct}\relax
\EndOfBibitem
\bibitem[Goroya \latin{et~al.}(2014)Goroya, Zhu, Sun, and Duan]{goroya2014high}
Goroya,~K.~G.; Zhu,~Y.; Sun,~P.; Duan,~C. High Resolution Jet-cooled Infrared
  Absorption Spectra of the Formic Acid Dimer: A Reinvestigation of the C--O
  Stretch Region. \emph{J. Chem. Phys.} \textbf{2014}, \emph{140}, 164311\relax
\mciteBstWouldAddEndPuncttrue
\mciteSetBstMidEndSepPunct{\mcitedefaultmidpunct}
{\mcitedefaultendpunct}{\mcitedefaultseppunct}\relax
\EndOfBibitem
\bibitem[Zoete and Meuwly(2004)Zoete, and Meuwly]{MM.dna:2004}
Zoete,~V.; Meuwly,~M. Double Proton Transfer in the Isolated and DNA-embedded
  Guanine-cytosine Base Pair. \emph{J. Chem. Phys.} \textbf{2004}, \emph{121},
  4377--4388\relax
\mciteBstWouldAddEndPuncttrue
\mciteSetBstMidEndSepPunct{\mcitedefaultmidpunct}
{\mcitedefaultendpunct}{\mcitedefaultseppunct}\relax
\EndOfBibitem
\bibitem[Arabi and Matta(2018)Arabi, and Matta]{arabi:2018}
Arabi,~A.~A.; Matta,~C.~F. Effects of Intense Electric Fields on the Double
  Proton Transfer in the Watson--crick Guanine--Cytosine Base Pair. \emph{J.
  Phys. Chem. B} \textbf{2018}, \emph{122}, 8631--8641\relax
\mciteBstWouldAddEndPuncttrue
\mciteSetBstMidEndSepPunct{\mcitedefaultmidpunct}
{\mcitedefaultendpunct}{\mcitedefaultseppunct}\relax
\EndOfBibitem
\bibitem[Chen \latin{et~al.}(2008)Chen, Brooks, and Scheraga]{brooks:2008}
Chen,~J.; Brooks,~C.~L.; Scheraga,~H.~A. Revisiting the Carboxylic Acid Dimers
  in Aqueous Solution: Interplay of Hydrogen Bonding, Hydrophobic Interactions,
  and Entropy. \emph{J. Phys. Chem. B} \textbf{2008}, \emph{112},
  242--249\relax
\mciteBstWouldAddEndPuncttrue
\mciteSetBstMidEndSepPunct{\mcitedefaultmidpunct}
{\mcitedefaultendpunct}{\mcitedefaultseppunct}\relax
\EndOfBibitem
\bibitem[Katchalsky \latin{et~al.}(1951)Katchalsky, Eisenberg, and
  Lifson]{katchalsky:1951}
Katchalsky,~A.; Eisenberg,~H.; Lifson,~S. Hydrogen Bonding and Ionization of
  Carboxylic Acids in Aqueous Solutions. \emph{J. Am. Chem. Soc.}
  \textbf{1951}, \emph{73}, 5889--5890\relax
\mciteBstWouldAddEndPuncttrue
\mciteSetBstMidEndSepPunct{\mcitedefaultmidpunct}
{\mcitedefaultendpunct}{\mcitedefaultseppunct}\relax
\EndOfBibitem
\bibitem[Schrier \latin{et~al.}(1964)Schrier, Pottle, and
  Scheraga]{scheraga:1964}
Schrier,~E.~E.; Pottle,~M.; Scheraga,~H.~A. The Influence of Hydrogen and
  Hydrophobic Bonds on the Stability of the Carboxylic Acid Dimers in Aqueous
  Solution. \emph{J. Am. Chem. Soc.} \textbf{1964}, \emph{86}, 3444--3449\relax
\mciteBstWouldAddEndPuncttrue
\mciteSetBstMidEndSepPunct{\mcitedefaultmidpunct}
{\mcitedefaultendpunct}{\mcitedefaultseppunct}\relax
\EndOfBibitem
\bibitem[Soffientini \latin{et~al.}(2015)Soffientini, Bernasconi, and
  Imberti]{imberti:2015}
Soffientini,~S.; Bernasconi,~L.; Imberti,~S. The Hydration of Formic Acid and
  Acetic Acid. \emph{J. Mol. Liqu.} \textbf{2015}, \emph{205}, 85--92\relax
\mciteBstWouldAddEndPuncttrue
\mciteSetBstMidEndSepPunct{\mcitedefaultmidpunct}
{\mcitedefaultendpunct}{\mcitedefaultseppunct}\relax
\EndOfBibitem
\bibitem[Sobyra \latin{et~al.}(2017)Sobyra, Melvin, and Nathanson]{sobyra:2017}
Sobyra,~T.~B.; Melvin,~M.~P.; Nathanson,~G.~M. Liquid Microjet Measurements of
  the Entry of Organic Acids and Bases into Salty Water. \emph{J. Phys. Chem.
  C} \textbf{2017}, \emph{121}, 20911--20924\relax
\mciteBstWouldAddEndPuncttrue
\mciteSetBstMidEndSepPunct{\mcitedefaultmidpunct}
{\mcitedefaultendpunct}{\mcitedefaultseppunct}\relax
\EndOfBibitem
\bibitem[H{\"a}nninen \latin{et~al.}(2018)H{\"a}nninen, Murdachaew, Nathanson,
  Gerber, and Halonen]{hanninen:2018}
H{\"a}nninen,~V.; Murdachaew,~G.; Nathanson,~G.~M.; Gerber,~R.~B.; Halonen,~L.
  Ab Initio Molecular Dynamics Studies of Formic Acid Dimer Colliding with
  Liquid Water. \emph{Phys. Chem. Chem. Phys.} \textbf{2018}, \emph{20},
  23717--23725\relax
\mciteBstWouldAddEndPuncttrue
\mciteSetBstMidEndSepPunct{\mcitedefaultmidpunct}
{\mcitedefaultendpunct}{\mcitedefaultseppunct}\relax
\EndOfBibitem
\bibitem[Tarakanova \latin{et~al.}(2019)Tarakanova, Voloshenko, Kislina,
  Mayorov, Yukhnevich, and Lyashchenko]{tarakanova:2019}
Tarakanova,~E.; Voloshenko,~G.; Kislina,~I.; Mayorov,~V.; Yukhnevich,~G.;
  Lyashchenko,~A. Composition and Structure of Hydrates Formed in Aqueous
  Solutions of Formic Acid. \emph{J. Struct. Chem.} \textbf{2019}, \emph{60},
  255--267\relax
\mciteBstWouldAddEndPuncttrue
\mciteSetBstMidEndSepPunct{\mcitedefaultmidpunct}
{\mcitedefaultendpunct}{\mcitedefaultseppunct}\relax
\EndOfBibitem
\bibitem[Dou \latin{et~al.}(2020)Dou, Wang, Hu, Fang, Sun, and Men]{men:2020}
Dou,~Z.; Wang,~L.; Hu,~J.; Fang,~W.; Sun,~C.; Men,~Z. Hydrogen Bonding Effect
  on Raman Modes of Formic Acid-water Binary Solutions. \emph{J. Mol. Liq.}
  \textbf{2020}, \emph{313}, 113595\relax
\mciteBstWouldAddEndPuncttrue
\mciteSetBstMidEndSepPunct{\mcitedefaultmidpunct}
{\mcitedefaultendpunct}{\mcitedefaultseppunct}\relax
\EndOfBibitem
\bibitem[Larsen \latin{et~al.}(2017)Larsen, Mortensen, Blomqvist, Castelli,
  Christensen, Du{\l}ak, Friis, Groves, Hammer, Hargus, Hermes, Jennings,
  Jensen, Kermode, Kitchin, Kolsbjerg, Kubal, Kaasbjerg, Lysgaard, Maronsson,
  Maxson, Olsen, Pastewka, Peterson, Rostgaard, Schi{\o}tz, Sch\"{u}tt,
  Strange, Thygesen, Vegge, Vilhelmsen, Walter, Zeng, and
  Jacobsen]{HjorthLarsen2017}
Larsen,~A.~H.; Mortensen,~J.~J.; Blomqvist,~J.; Castelli,~I.~E.;
  Christensen,~R.; Du{\l}ak,~M.; Friis,~J.; Groves,~M.~N.; Hammer,~B.;
  Hargus,~C. \latin{et~al.}  The Atomic Simulation Environment -- a Python
  Library for Working with Atoms. \emph{J. Phys. Condens. Matter}
  \textbf{2017}, \emph{29}, 273002\relax
\mciteBstWouldAddEndPuncttrue
\mciteSetBstMidEndSepPunct{\mcitedefaultmidpunct}
{\mcitedefaultendpunct}{\mcitedefaultseppunct}\relax
\EndOfBibitem
\bibitem[Unke and Meuwly(2019)Unke, and Meuwly]{Unke2019}
Unke,~O.~T.; Meuwly,~M. Physnet: A Neural Network for Predicting Energies,
  Forces, Dipole Moments, and Partial Charges. \emph{J. Chem. Theo. Comp.}
  \textbf{2019}, \emph{15}, 3678--3693\relax
\mciteBstWouldAddEndPuncttrue
\mciteSetBstMidEndSepPunct{\mcitedefaultmidpunct}
{\mcitedefaultendpunct}{\mcitedefaultseppunct}\relax
\EndOfBibitem
\bibitem[Jorgensen \latin{et~al.}(1983)Jorgensen, Chandrasekhar, Madura, Impey,
  and Klein]{Jorgensen1983}
Jorgensen,~W.~L.; Chandrasekhar,~J.; Madura,~J.~D.; Impey,~R.~W.; Klein,~M.~L.
  Comparison of Simple Potential Functions for Simulating Liquid Water.
  \emph{J. Chem. Phys.} \textbf{1983}, \emph{79}, 926--935\relax
\mciteBstWouldAddEndPuncttrue
\mciteSetBstMidEndSepPunct{\mcitedefaultmidpunct}
{\mcitedefaultendpunct}{\mcitedefaultseppunct}\relax
\EndOfBibitem
\bibitem[Werner \latin{et~al.}(2020)Werner, Knowles, Manby, Black, Doll,
  Heßelmann, Kats, Köhn, Korona, Kreplin, Ma, Miller, Mitrushchenkov,
  Peterson, Polyak, Rauhut, and Sibaev]{Werner2020}
Werner,~H.-J.; Knowles,~P.~J.; Manby,~F.~R.; Black,~J.~A.; Doll,~K.;
  Heßelmann,~A.; Kats,~D.; Köhn,~A.; Korona,~T.; Kreplin,~D.~A.
  \latin{et~al.}  The Molpro Quantum Chemistry Package. \emph{J. Chem. Phys.}
  \textbf{2020}, \emph{152}, 144107\relax
\mciteBstWouldAddEndPuncttrue
\mciteSetBstMidEndSepPunct{\mcitedefaultmidpunct}
{\mcitedefaultendpunct}{\mcitedefaultseppunct}\relax
\EndOfBibitem
\bibitem[Huang and von Lilienfeld(2020)Huang, and von Lilienfeld]{Huang2020}
Huang,~B.; von Lilienfeld,~O.~A. Quantum Machine Learning Using
  Atom-in-molecule-based Fragments Selected on the Fly. \emph{Nat. Chem.}
  \textbf{2020}, \emph{12}, 945--951\relax
\mciteBstWouldAddEndPuncttrue
\mciteSetBstMidEndSepPunct{\mcitedefaultmidpunct}
{\mcitedefaultendpunct}{\mcitedefaultseppunct}\relax
\EndOfBibitem
\bibitem[Behler(2015)]{behler2015constructing}
Behler,~J. Constructing High-Dimensional Neural Network Potentials: A Tutorial
  Review. \emph{Int. J. Quantum. Chem.} \textbf{2015}, \emph{115},
  1032--1050\relax
\mciteBstWouldAddEndPuncttrue
\mciteSetBstMidEndSepPunct{\mcitedefaultmidpunct}
{\mcitedefaultendpunct}{\mcitedefaultseppunct}\relax
\EndOfBibitem
\bibitem[Unke and Meuwly(2018)Unke, and Meuwly]{Unke2018}
Unke,~O.~T.; Meuwly,~M. A Reactive, Scalable, and Transferable Model for
  Molecular Energies from a Neural Network Approach Based on Local Information.
  \emph{J. Chem. Phys.} \textbf{2018}, \emph{148}, 241708\relax
\mciteBstWouldAddEndPuncttrue
\mciteSetBstMidEndSepPunct{\mcitedefaultmidpunct}
{\mcitedefaultendpunct}{\mcitedefaultseppunct}\relax
\EndOfBibitem
\bibitem[Grimme \latin{et~al.}(2011)Grimme, Ehrlich, and Goerigk]{Grimme2011}
Grimme,~S.; Ehrlich,~S.; Goerigk,~L. Effect of the Damping Function in
  Dispersion Corrected Density Functional Theory. \emph{J. Comput. Phys.}
  \textbf{2011}, \emph{32}, 1456--1465\relax
\mciteBstWouldAddEndPuncttrue
\mciteSetBstMidEndSepPunct{\mcitedefaultmidpunct}
{\mcitedefaultendpunct}{\mcitedefaultseppunct}\relax
\EndOfBibitem
\bibitem[Abadi \latin{et~al.}(2015)Abadi, Agarwal, Barham, Brevdo, Chen, Citro,
  Corrado, Davis, Dean, Devin, Ghemawat, Goodfellow, Harp, Irving, Isard, Jia,
  Jozefowicz, Kaiser, Kudlur, Levenberg, Man\'{e}, Monga, Moore, Murray, Olah,
  Schuster, Shlens, Steiner, Sutskever, Talwar, Tucker, Vanhoucke, Vasudevan,
  Vi\'{e}gas, Vinyals, Warden, Wattenberg, Wicke, Yu, and
  Zheng]{tensorflow2015}
Abadi,~M.; Agarwal,~A.; Barham,~P.; Brevdo,~E.; Chen,~Z.; Citro,~C.;
  Corrado,~G.~S.; Davis,~A.; Dean,~J.; Devin,~M. \latin{et~al.}  {TensorFlow}:
  Large-scale Machine Learning on Heterogeneous Systems. 2015;
  \url{http://tensorflow.org/}, Software available from tensorflow.org\relax
\mciteBstWouldAddEndPuncttrue
\mciteSetBstMidEndSepPunct{\mcitedefaultmidpunct}
{\mcitedefaultendpunct}{\mcitedefaultseppunct}\relax
\EndOfBibitem
\bibitem[Baydin \latin{et~al.}(2018)Baydin, Pearlmutter, Radul, and
  Siskind]{Baydin2018}
Baydin,~A.~G.; Pearlmutter,~B.~A.; Radul,~A.~A.; Siskind,~J.~M. Automatic
  Differentiation in Machine Learning: A Survey. \emph{J. Mach. Learn. Res.}
  \textbf{2018}, \emph{18}, 1--43\relax
\mciteBstWouldAddEndPuncttrue
\mciteSetBstMidEndSepPunct{\mcitedefaultmidpunct}
{\mcitedefaultendpunct}{\mcitedefaultseppunct}\relax
\EndOfBibitem
\bibitem[Mortensen \latin{et~al.}(2005)Mortensen, Hansen, and
  Jacobsen]{Mortensen2005}
Mortensen,~J.~J.; Hansen,~L.~B.; Jacobsen,~K.~W. Real-space Grid Implementation
  of the Projector Augmented Wave Method. \emph{Phys. Rev. B} \textbf{2005},
  \emph{71}, 035109\relax
\mciteBstWouldAddEndPuncttrue
\mciteSetBstMidEndSepPunct{\mcitedefaultmidpunct}
{\mcitedefaultendpunct}{\mcitedefaultseppunct}\relax
\EndOfBibitem
\bibitem[Vanommeslaeghe \latin{et~al.}(2010)Vanommeslaeghe, Hatcher, Acharya,
  Kundu, Zhong, Shim, Darian, Guvench, Lopes, Vorobyov, and
  Mackerell]{cgenff:2010}
Vanommeslaeghe,~K.; Hatcher,~E.; Acharya,~C.; Kundu,~S.; Zhong,~S.; Shim,~J.;
  Darian,~E.; Guvench,~O.; Lopes,~P.; Vorobyov,~I. \latin{et~al.}  CHARMM
  General Force Field: A Force Field for Drug-like Molecules Compatible with
  the CHARMM All-Atom Additive Biological Force Fields. \emph{J. Comput. Phys.}
  \textbf{2010}, \emph{31}, 671--690\relax
\mciteBstWouldAddEndPuncttrue
\mciteSetBstMidEndSepPunct{\mcitedefaultmidpunct}
{\mcitedefaultendpunct}{\mcitedefaultseppunct}\relax
\EndOfBibitem
\bibitem[Spohr(1997)]{Spohr1997}
Spohr,~E. Effect of Electrostatic Boundary Conditions and System Size on the
  Interfacial Properties of Water and Aqueous Solutions. \emph{J. Chem. Phys.}
  \textbf{1997}, \emph{107}, 6342--6348\relax
\mciteBstWouldAddEndPuncttrue
\mciteSetBstMidEndSepPunct{\mcitedefaultmidpunct}
{\mcitedefaultendpunct}{\mcitedefaultseppunct}\relax
\EndOfBibitem
\bibitem[Torrie and Valleau(1977)Torrie, and Valleau]{torrie1977nonphysical}
Torrie,~G.~M.; Valleau,~J.~P. Nonphysical Sampling Distributions in Monte Carlo
  Free-energy Estimation: Umbrella Sampling. \emph{J. Chem. Phys.}
  \textbf{1977}, \emph{23}, 187--199\relax
\mciteBstWouldAddEndPuncttrue
\mciteSetBstMidEndSepPunct{\mcitedefaultmidpunct}
{\mcitedefaultendpunct}{\mcitedefaultseppunct}\relax
\EndOfBibitem
\bibitem[Andersen(1983)]{Andersen1983}
Andersen,~H.~C. Rattle: A “Velocity” Version of the Shake Algorithm for
  Molecular Dynamics Calculations. \emph{J. Comput. Phys.} \textbf{1983},
  \emph{52}, 24--34\relax
\mciteBstWouldAddEndPuncttrue
\mciteSetBstMidEndSepPunct{\mcitedefaultmidpunct}
{\mcitedefaultendpunct}{\mcitedefaultseppunct}\relax
\EndOfBibitem
\bibitem[Enkovaara \latin{et~al.}(2010)Enkovaara, Rostgaard, Mortensen, Chen,
  Du{\l}ak, Ferrighi, Gavnholt, Glinsvad, Haikola, Hansen, Kristoffersen,
  Kuisma, Larsen, Lehtovaara, Ljungberg, Lopez-Acevedo, Moses, Ojanen, Olsen,
  Petzold, Romero, Stausholm-M{\o}ller, Strange, Tritsaris, Vanin, Walter,
  Hammer, H\"akkinen, Madsen, Nieminen, N{\o}rskov, Puska, Rantala, Schi{\o}tz,
  Thygesen, and Jacobsen]{Enkovaara2010}
Enkovaara,~J.; Rostgaard,~C.; Mortensen,~J.~J.; Chen,~J.; Du{\l}ak,~M.;
  Ferrighi,~L.; Gavnholt,~J.; Glinsvad,~C.; Haikola,~V.; Hansen,~H.~A.
  \latin{et~al.}  Electronic Structure Calculations with {GPAW}: A Real-space
  Implementation of the Projector Augmented-wave Method. \emph{J. Phys.:
  Condens. Matter} \textbf{2010}, \emph{22}, 253202\relax
\mciteBstWouldAddEndPuncttrue
\mciteSetBstMidEndSepPunct{\mcitedefaultmidpunct}
{\mcitedefaultendpunct}{\mcitedefaultseppunct}\relax
\EndOfBibitem
\bibitem[Meuwly and Karplus(2002)Meuwly, and Karplus]{MM.amm:2002}
Meuwly,~M.; Karplus,~M. Simulation of Proton Transfer along Ammonia Wires: An
  “Ab Initio” and Semiempirical Density Functional Comparison of Potentials
  and Classical Molecular Dynamics. \emph{J. Chem. Phys.} \textbf{2002},
  \emph{116}, 2572--2585\relax
\mciteBstWouldAddEndPuncttrue
\mciteSetBstMidEndSepPunct{\mcitedefaultmidpunct}
{\mcitedefaultendpunct}{\mcitedefaultseppunct}\relax
\EndOfBibitem
\bibitem[Becke(1993)]{becke:1993}
Becke,~A.~D. Density‐functional Thermochemistry. Iii. The Role of Exact
  Exchange. \emph{J. Chem. Phys.} \textbf{1993}, \emph{98}, 5648--5652\relax
\mciteBstWouldAddEndPuncttrue
\mciteSetBstMidEndSepPunct{\mcitedefaultmidpunct}
{\mcitedefaultendpunct}{\mcitedefaultseppunct}\relax
\EndOfBibitem
\bibitem[Lee \latin{et~al.}(1988)Lee, Yang, and Parr]{lyp:1988}
Lee,~C.; Yang,~W.; Parr,~R.~G. Development of the Colle-Salvetti
  Correlation-energy Formula into a Functional of the Electron Density.
  \emph{Phys. Rev. B} \textbf{1988}, \emph{37}, 785--789\relax
\mciteBstWouldAddEndPuncttrue
\mciteSetBstMidEndSepPunct{\mcitedefaultmidpunct}
{\mcitedefaultendpunct}{\mcitedefaultseppunct}\relax
\EndOfBibitem
\bibitem[Grimme \latin{et~al.}(2010)Grimme, Antony, Ehrlich, and
  Krieg]{grimme:2010}
Grimme,~S.; Antony,~J.; Ehrlich,~S.; Krieg,~H. A Consistent and Accurate Ab
  Initio Parametrization of Density Functional Dispersion Correction (DFT-D)
  for the 94 Elements H-Pu. \emph{J. Chem. Phys.} \textbf{2010}, \emph{132},
  154104\relax
\mciteBstWouldAddEndPuncttrue
\mciteSetBstMidEndSepPunct{\mcitedefaultmidpunct}
{\mcitedefaultendpunct}{\mcitedefaultseppunct}\relax
\EndOfBibitem
\bibitem[Sidler \latin{et~al.}(2018)Sidler, Meuwly, and Hamm]{MM.water:2018}
Sidler,~D.; Meuwly,~M.; Hamm,~P. An Efficient Water Force Field Calibrated
  against Intermolecular THz and Raman Spectra. \emph{J. Chem. Phys.}
  \textbf{2018}, \emph{148}, 244504\relax
\mciteBstWouldAddEndPuncttrue
\mciteSetBstMidEndSepPunct{\mcitedefaultmidpunct}
{\mcitedefaultendpunct}{\mcitedefaultseppunct}\relax
\EndOfBibitem
\bibitem[Huang \latin{et~al.}(2019)Huang, Tang, Li, Chen, Zheng, Zhang, Le, Li,
  Li, Liu, \latin{et~al.} others]{huang:2019}
Huang,~X.; Tang,~C.; Li,~J.; Chen,~L.-C.; Zheng,~J.; Zhang,~P.; Le,~J.; Li,~R.;
  Li,~X.; Liu,~J. \latin{et~al.}  Electric Field--induced Selective Catalysis
  of Single-molecule Reaction. \emph{Science advances} \textbf{2019}, \emph{5},
  eaaw3072\relax
\mciteBstWouldAddEndPuncttrue
\mciteSetBstMidEndSepPunct{\mcitedefaultmidpunct}
{\mcitedefaultendpunct}{\mcitedefaultseppunct}\relax
\EndOfBibitem
\bibitem[Shaik \latin{et~al.}(2004)Shaik, De~Visser, and Kumar]{shaik:2004}
Shaik,~S.; De~Visser,~S.~P.; Kumar,~D. External Electric Field Will Control the
  Selectivity of Enzymatic-like Bond Activations. \emph{J. Am. Chem. Soc.}
  \textbf{2004}, \emph{126}, 11746--11749\relax
\mciteBstWouldAddEndPuncttrue
\mciteSetBstMidEndSepPunct{\mcitedefaultmidpunct}
{\mcitedefaultendpunct}{\mcitedefaultseppunct}\relax
\EndOfBibitem
\bibitem[Ropp \latin{et~al.}(2001)Ropp, Lawrence, Farrar, and
  Skinner]{skinner:2001}
Ropp,~J.; Lawrence,~C.; Farrar,~T.; Skinner,~J. Rotational Motion in Liquid
  Water Is Anisotropic: A Nuclear Magnetic Resonance and Molecular Dynamics
  Simulation Study. \emph{J. Am. Chem. Soc.} \textbf{2001}, \emph{123},
  8047--8052\relax
\mciteBstWouldAddEndPuncttrue
\mciteSetBstMidEndSepPunct{\mcitedefaultmidpunct}
{\mcitedefaultendpunct}{\mcitedefaultseppunct}\relax
\EndOfBibitem
\bibitem[Martins \latin{et~al.}(2009)Martins, Middlemiss, Pulham, Wilson,
  Weller, Henry, Shankland, Shankland, Marshall, Ibberson, Knight, Moggach,
  Brunelli, and Morrison]{martins:2009}
Martins,~D. M.~S.; Middlemiss,~D.~S.; Pulham,~C.~R.; Wilson,~C.~C.;
  Weller,~M.~T.; Henry,~P.~F.; Shankland,~N.; Shankland,~K.; Marshall,~W.~G.;
  Ibberson,~R.~M. \latin{et~al.}  Temperature- and Pressure-induced Proton
  Transfer in the 1:1 Adduct Formed between Squaric Acid and 4,4'-bipyridine.
  \emph{J. Am. Chem. Soc.} \textbf{2009}, \emph{131}, 3884--3893\relax
\mciteBstWouldAddEndPuncttrue
\mciteSetBstMidEndSepPunct{\mcitedefaultmidpunct}
{\mcitedefaultendpunct}{\mcitedefaultseppunct}\relax
\EndOfBibitem
\bibitem[Ma \latin{et~al.}(2017)Ma, Li, Sun, and Zhou]{ma:2017}
Ma,~Z.; Li,~J.; Sun,~C.; Zhou,~M. High Pressure Spectroscopic Investigation on
  Proton Transfer in Squaric Acid and 4,4'-bipyridine Co-crystal. \emph{Sci.
  Rep.} \textbf{2017}, \emph{7}, 4677\relax
\mciteBstWouldAddEndPuncttrue
\mciteSetBstMidEndSepPunct{\mcitedefaultmidpunct}
{\mcitedefaultendpunct}{\mcitedefaultseppunct}\relax
\EndOfBibitem
\bibitem[Xu and Meuwly(2017)Xu, and Meuwly]{MM.oxalate:2017}
Xu,~Z.-H.; Meuwly,~M. Vibrational Spectroscopy and Proton Transfer Dynamics in
  Protonated Oxalate. \emph{J. Phys. Chem. A} \textbf{2017}, \emph{121},
  5389--5398\relax
\mciteBstWouldAddEndPuncttrue
\mciteSetBstMidEndSepPunct{\mcitedefaultmidpunct}
{\mcitedefaultendpunct}{\mcitedefaultseppunct}\relax
\EndOfBibitem
\end{mcitethebibliography}


\providecommand{\latin}[1]{#1}
\makeatletter
\providecommand{\doi}
  {\begingroup\let\do\@makeother\dospecials
  \catcode`\{=1 \catcode`\}=2 \doi@aux}
\providecommand{\doi@aux}[1]{\endgroup\texttt{#1}}
\makeatother
\providecommand*\mcitethebibliography{\thebibliography}
\csname @ifundefined\endcsname{endmcitethebibliography}
  {\let\endmcitethebibliography\endthebibliography}{}
\begin{mcitethebibliography}{2}
\providecommand*\natexlab[1]{#1}
\providecommand*\mciteSetBstSublistMode[1]{}
\providecommand*\mciteSetBstMaxWidthForm[2]{}
\providecommand*\mciteBstWouldAddEndPuncttrue
  {\def\EndOfBibitem{\unskip.}}
\providecommand*\mciteBstWouldAddEndPunctfalse
  {\let\EndOfBibitem\relax}
\providecommand*\mciteSetBstMidEndSepPunct[3]{}
\providecommand*\mciteSetBstSublistLabelBeginEnd[3]{}
\providecommand*\EndOfBibitem{}
\mciteSetBstSublistMode{f}
\mciteSetBstMaxWidthForm{subitem}{(\alph{mcitesubitemcount})}
\mciteSetBstSublistLabelBeginEnd
  {\mcitemaxwidthsubitemform\space}
  {\relax}
  {\relax}

\bibitem[Vanommeslaeghe \latin{et~al.}(2010)Vanommeslaeghe, Hatcher, Acharya,
  Kundu, Zhong, Shim, Darian, Guvench, Lopes, Vorobyov, and
  Mackerell]{cgenff:2010}
Vanommeslaeghe,~K.; Hatcher,~E.; Acharya,~C.; Kundu,~S.; Zhong,~S.; Shim,~J.;
  Darian,~E.; Guvench,~O.; Lopes,~P.; Vorobyov,~I. \latin{et~al.}  CHARMM
  General Force Field: A Force Field for Drug-like Molecules Compatible with
  the CHARMM All-Atom Additive Biological Force Fields. \emph{J. Comput. Phys.}
  \textbf{2010}, \emph{31}, 671--690\relax
\mciteBstWouldAddEndPuncttrue
\mciteSetBstMidEndSepPunct{\mcitedefaultmidpunct}
{\mcitedefaultendpunct}{\mcitedefaultseppunct}\relax
\EndOfBibitem
\end{mcitethebibliography}

\end{document}


\begin{figure}[htb!]
\centering
\includegraphics[width=\linewidth]{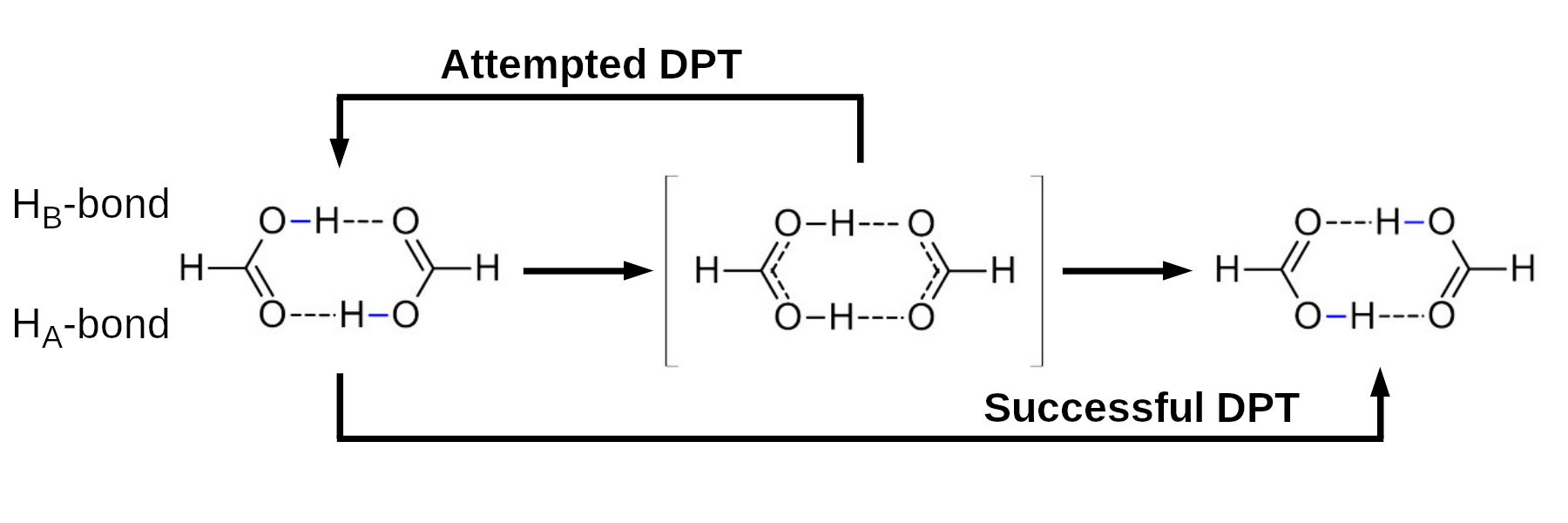}
\caption{Schematic reaction profile of DPT in FAD.  The
  intermediate ion-pair configuration after the first PT in the
  H$_\mathrm{A}$-bond is shown square brackets. The second PT occurs
  either in the H$_\mathrm{B}$-bond resulting in a successful DPT or
  in the H$_\mathrm{A}$-bond again labelled as attempted DPT.  Note
  the inversion of the covalent O--H bonds (blue line) in a successful
  DPT where the attempted DPT leads to the original configuration.  }
\label{sifig:1}
\end{figure}

\begin{figure}[htb!]
\centering
\includegraphics[width=0.8\linewidth]{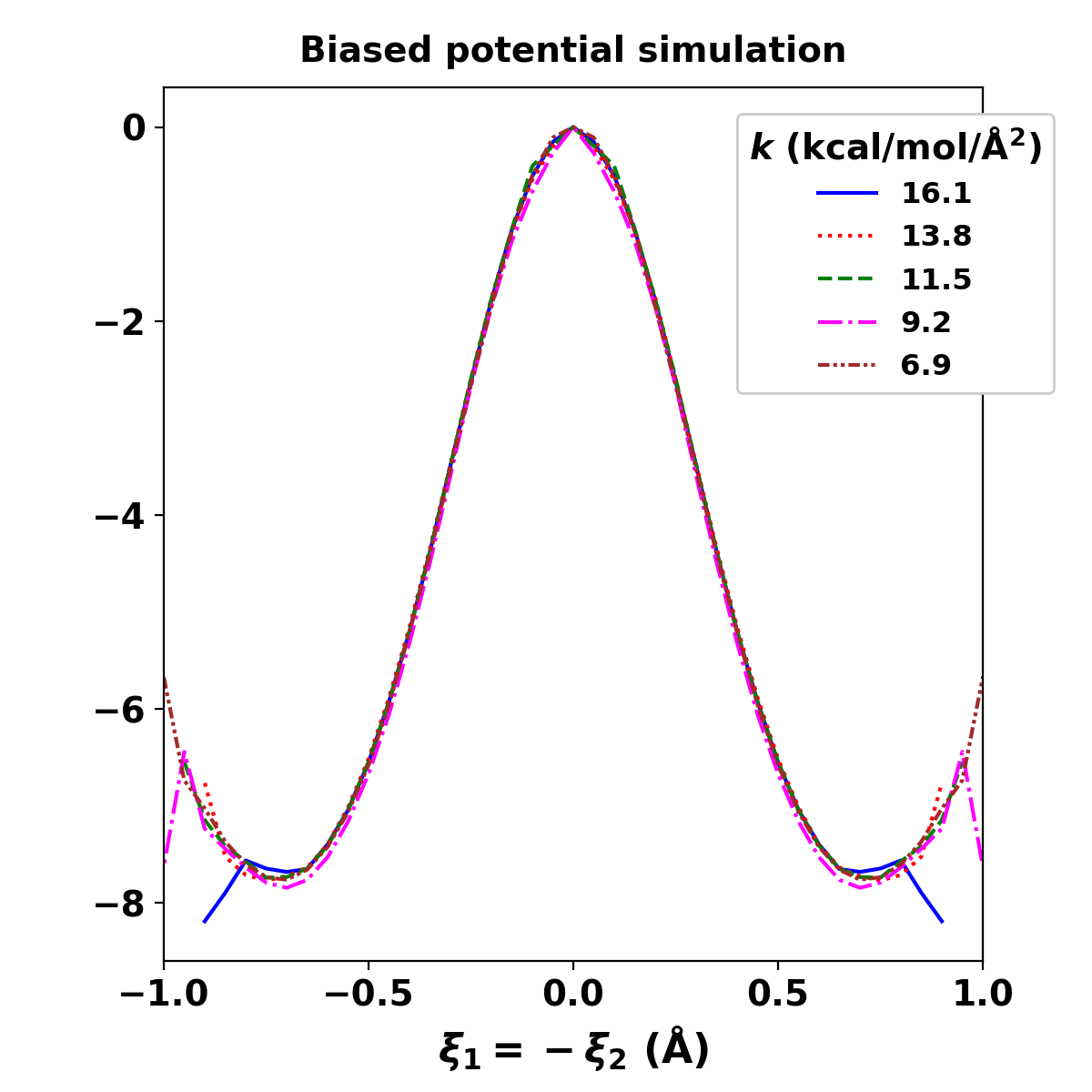}
\caption{1-dimensional free energy profiles from the 2-dimensional
  free energy surfaces at different force constants for the artificial
  harmonic potential in biased simulation of FAD in solution to
  constraint FAD around the TS conformation.}
\label{sifig:2}
\end{figure}

\begin{figure}[htb!]
\centering
\includegraphics[width=\linewidth]{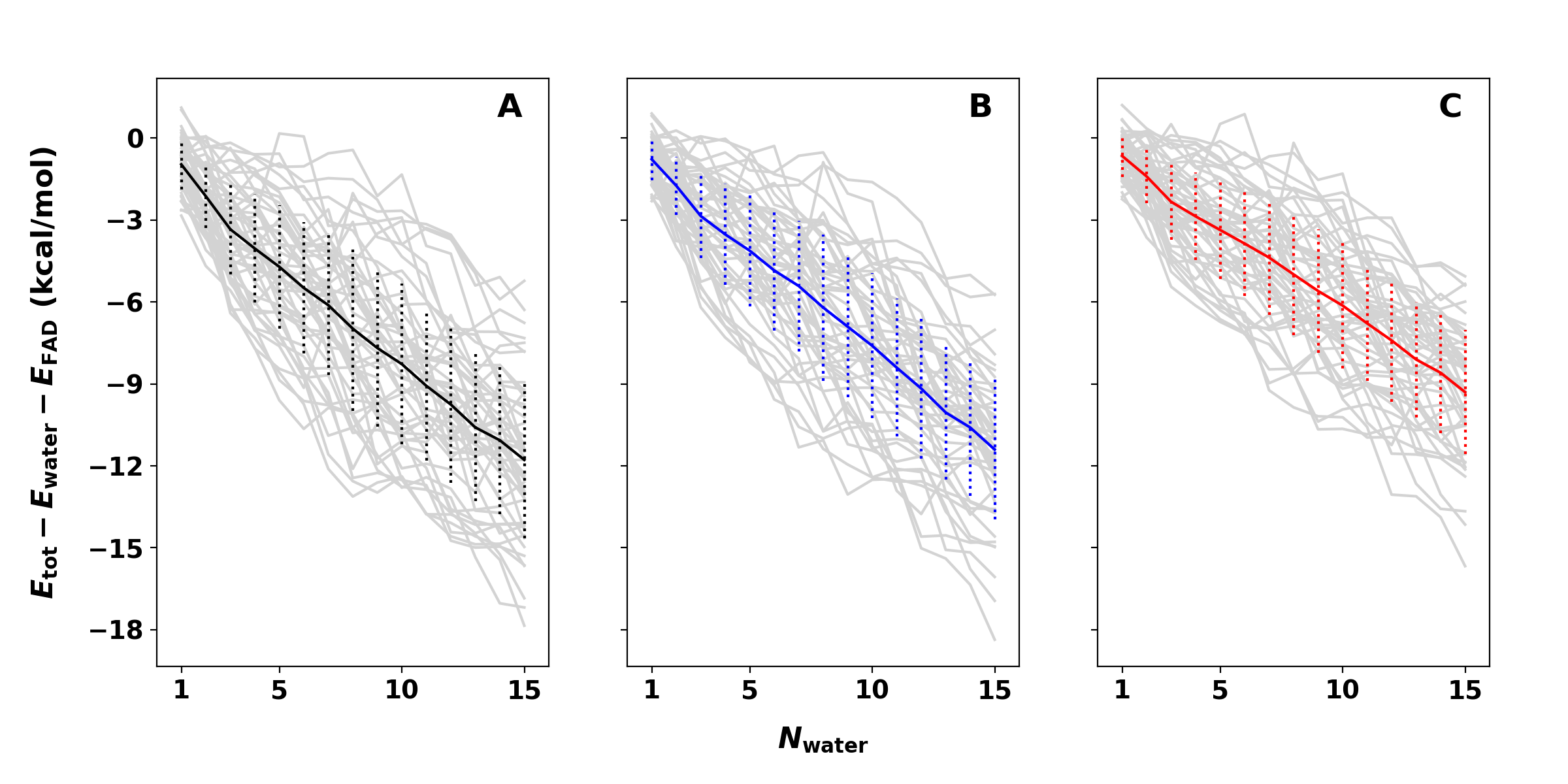}
\caption{Average interaction energy and fluctuation around it between
  FAD and increasing number of closest water molecules $N_{\rm water}$
  from 50 randomly chosen snapshots from a 2 ns simulation at 300
  K. The standard deviation is given by the dashed error bars and the
  light grey lines shows the hydration energy of the chosen
  snapshots. The interaction energies are determined by (left panel)
  DFT (B3LYP+D3/aug-cc-pVTZ, BSSE corrected), (center panel) the
  current ML/MM model and (right panel) with the CHARMM program
  package with the CGenFF force field.}
\label{sifig:3}
\end{figure}

\begin{figure}[htb!]
\centering
\includegraphics[width=\linewidth]{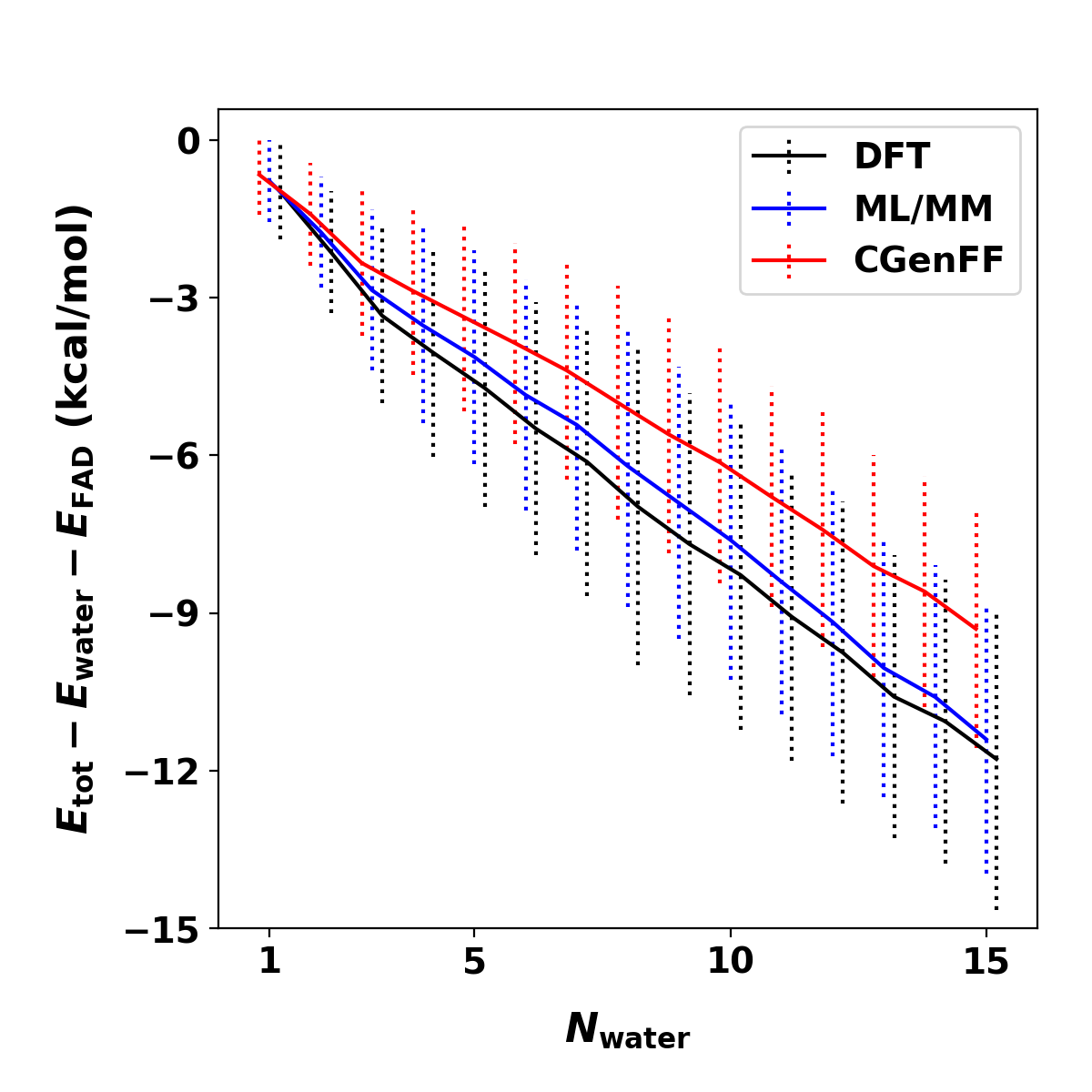}
\caption{Average interaction energy and fluctuation around it between
  FAD and increasing number of closest water molecules $N_{\rm water}$
  in 50 randomly chosen snapshots from 300 K simulations. The standard
  deviation is given by the dashed error bars which are slightly
  shifted for better visualization. The interaction energies are
  determined by DFT (B3LYP+D3/aug-cc-pVTZ, BSSE corrected), the
  current ML/MM model and with the CHARMM program package with the
  CGenFF force field.}
\label{sifig:4}
\end{figure}

\begin{figure}[htb!]
\centering
\includegraphics[width=\linewidth]{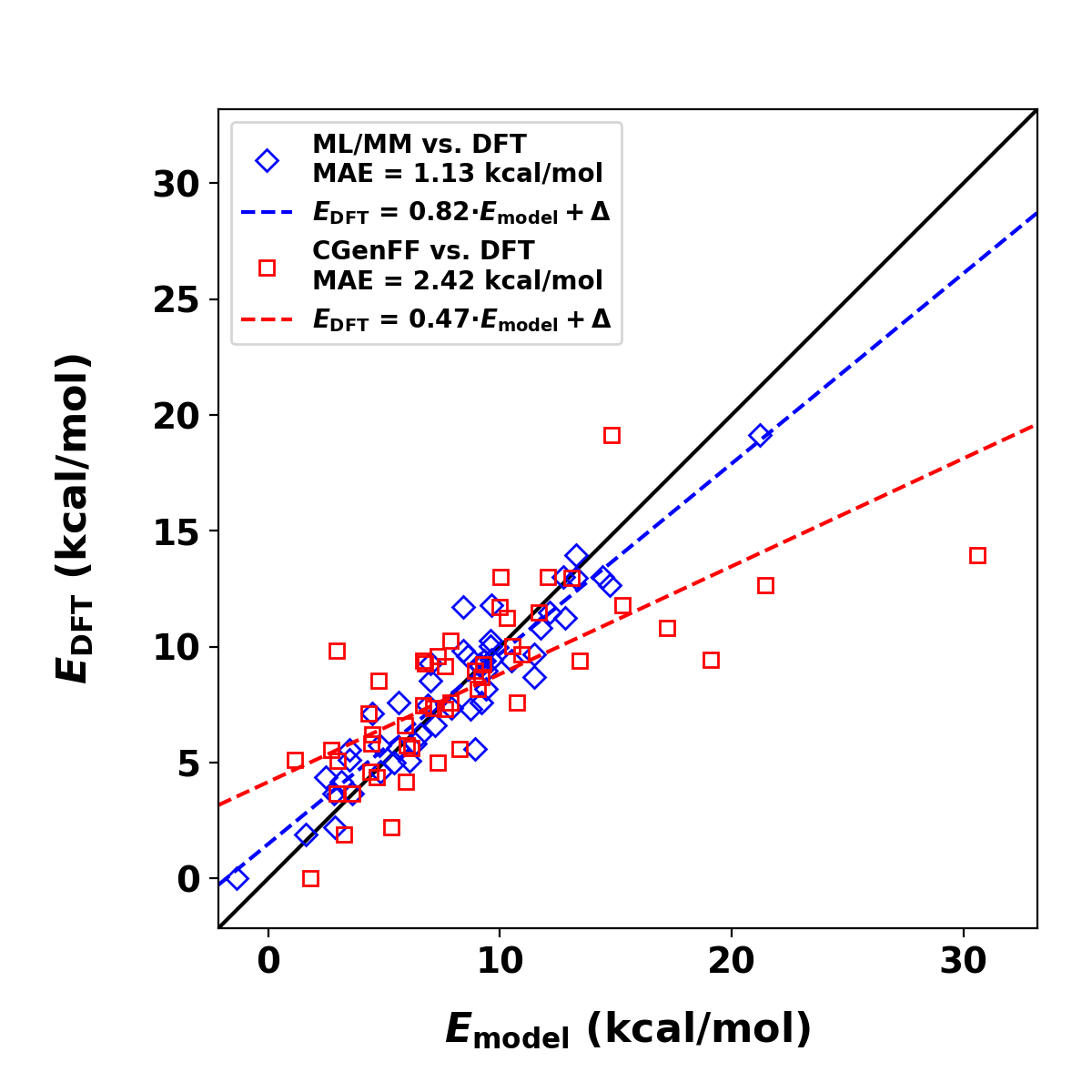}
\caption{Correlation between DFT and model potential energies for
  FAD surrounded by the 4 closest water molecules (relative to the FAD
  center of mass) for 50 randomly chosen snapshots from 300 K
  simulations. The DFT energy of the energetically lowest structure is
  shifted to zero and the model energies are shifted to minimize the
  mean absolute error (MAE) with the DFT energies. The linear
  regression for ML/MM vs. DFT and CGenFF vs. DFT are the dashed blue
  and red lines, respectively. The slope of the linear fit is given in
  the legend. The solid black line is the ideal 1:1 correlation and
  the MAE reported is with respect to the black solid line.}
\label{sifig:5}
\end{figure}

\begin{figure}[htb!]
\centering
\includegraphics[width=\linewidth]{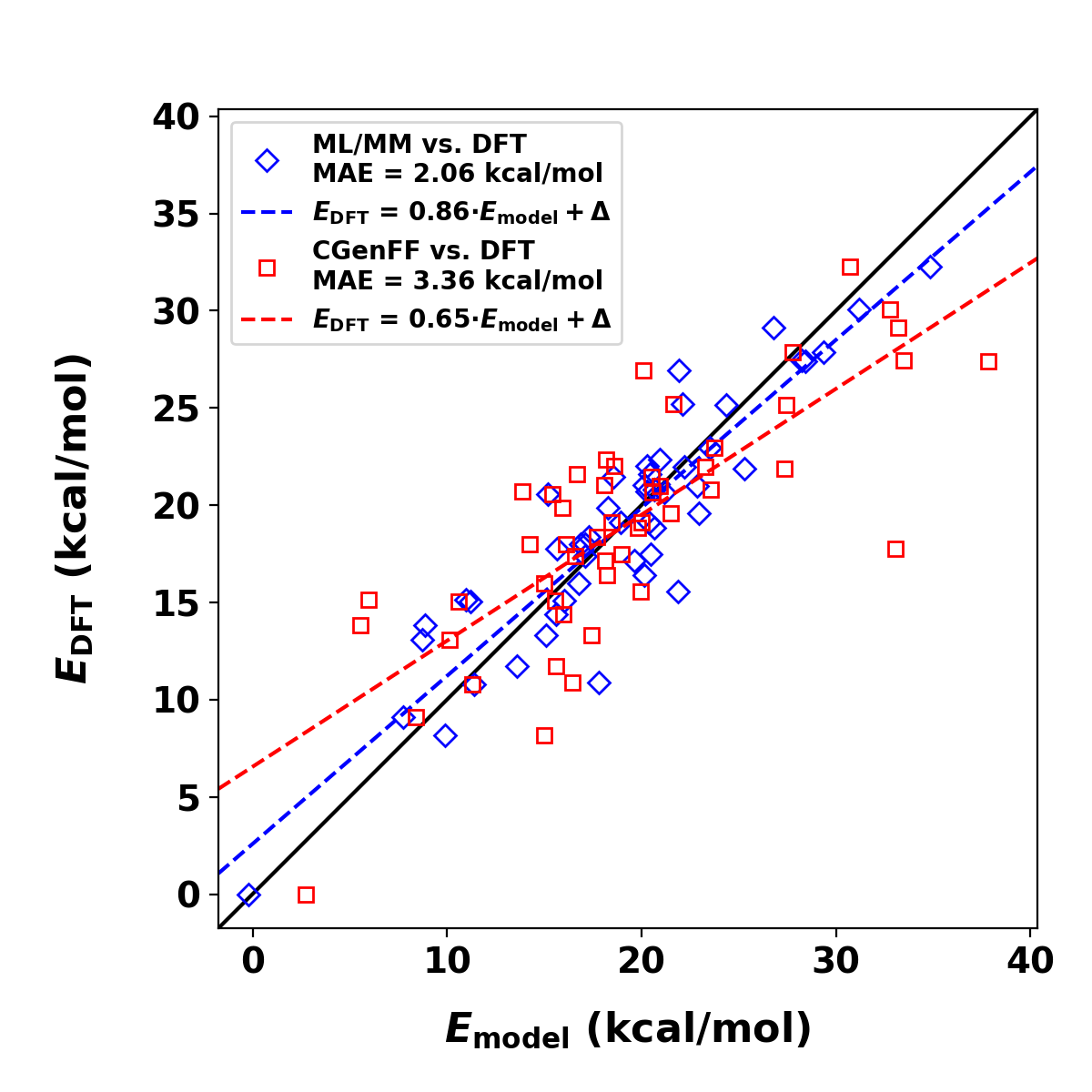}
\caption{Correlation between DFT and model potential energies for
  FAD surrounded by the 8 closest water molecules (relative to the FAD
  center of mass) for 50 randomly chosen snapshots from 300 K
  simulations. The DFT energy of the energetically lowest structure is
  shifted to zero and the model energies are shifted to minimize the
  mean absolute error (MAE) with the DFT energies. The linear
  regression for ML/MM vs. DFT and CGenFF vs. DFT are the dashed blue
  and red lines, respectively. The slope of the linear fit is given in
  the legend. The solid black line is the ideal 1:1 correlation and
  the MAE reported is with respect to the black solid line.}
\label{sifig:6}
\end{figure}

\begin{figure}[htb!]
\centering
\includegraphics[width=\linewidth]{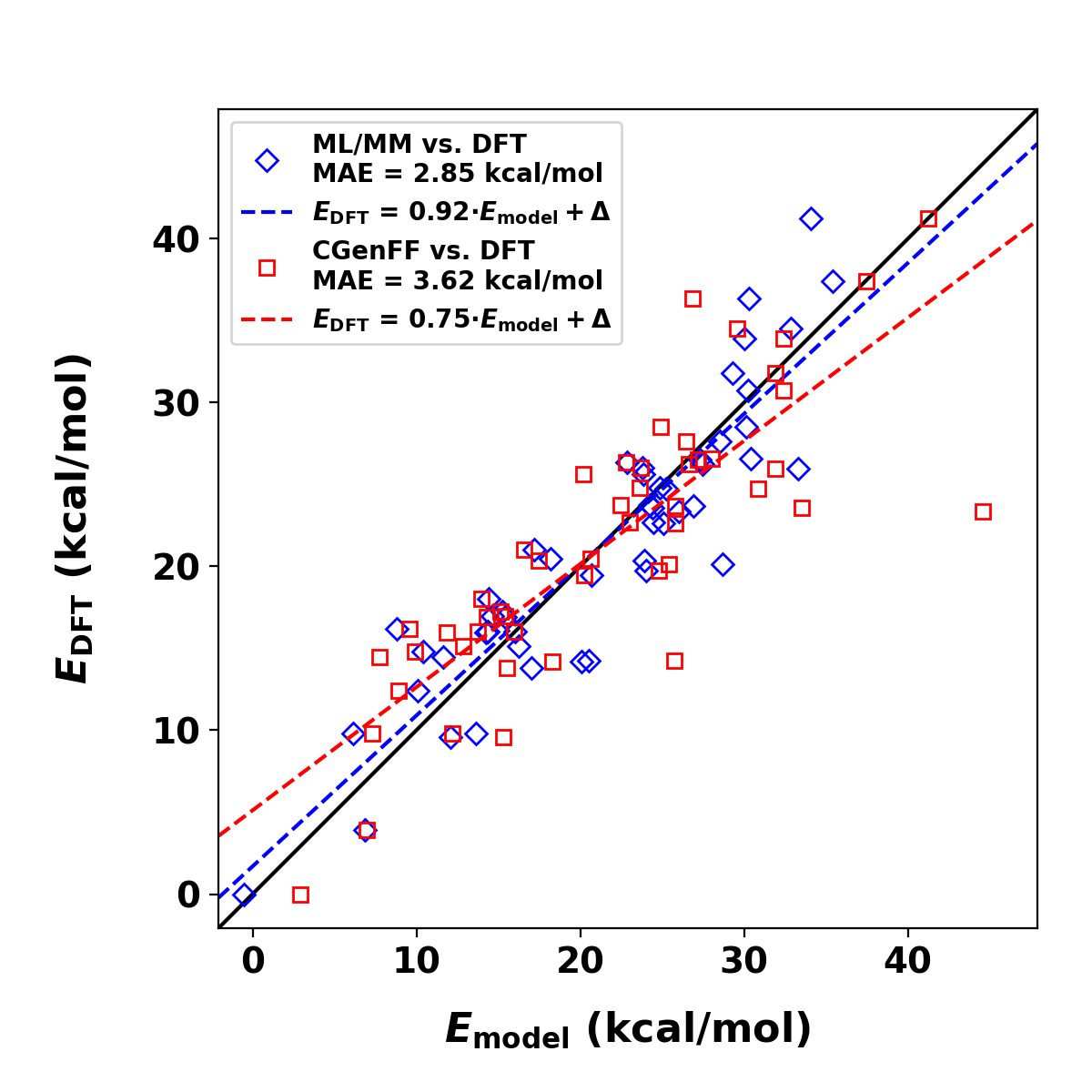}
\caption{Correlation between DFT and model potential energies for
  FAD surrounded by the 12 closest water molecules (relative to the FAD
  center of mass) for 50 randomly chosen snapshots from 300 K
  simulations. The DFT energy of the energetically lowest structure is
  shifted to zero and the model energies are shifted to minimize the
  mean absolute error (MAE) with the DFT energies. The linear
  regression for ML/MM vs. DFT and CGenFF vs. DFT are the dashed blue
  and red lines, respectively. The slope of the linear fit is given in
  the legend. The solid black line is the ideal 1:1 correlation and
  the MAE reported is with respect to the black solid line.}
\label{sifig:7}
\end{figure}

\begin{figure}[htb!]
\centering
\includegraphics[width=\linewidth]{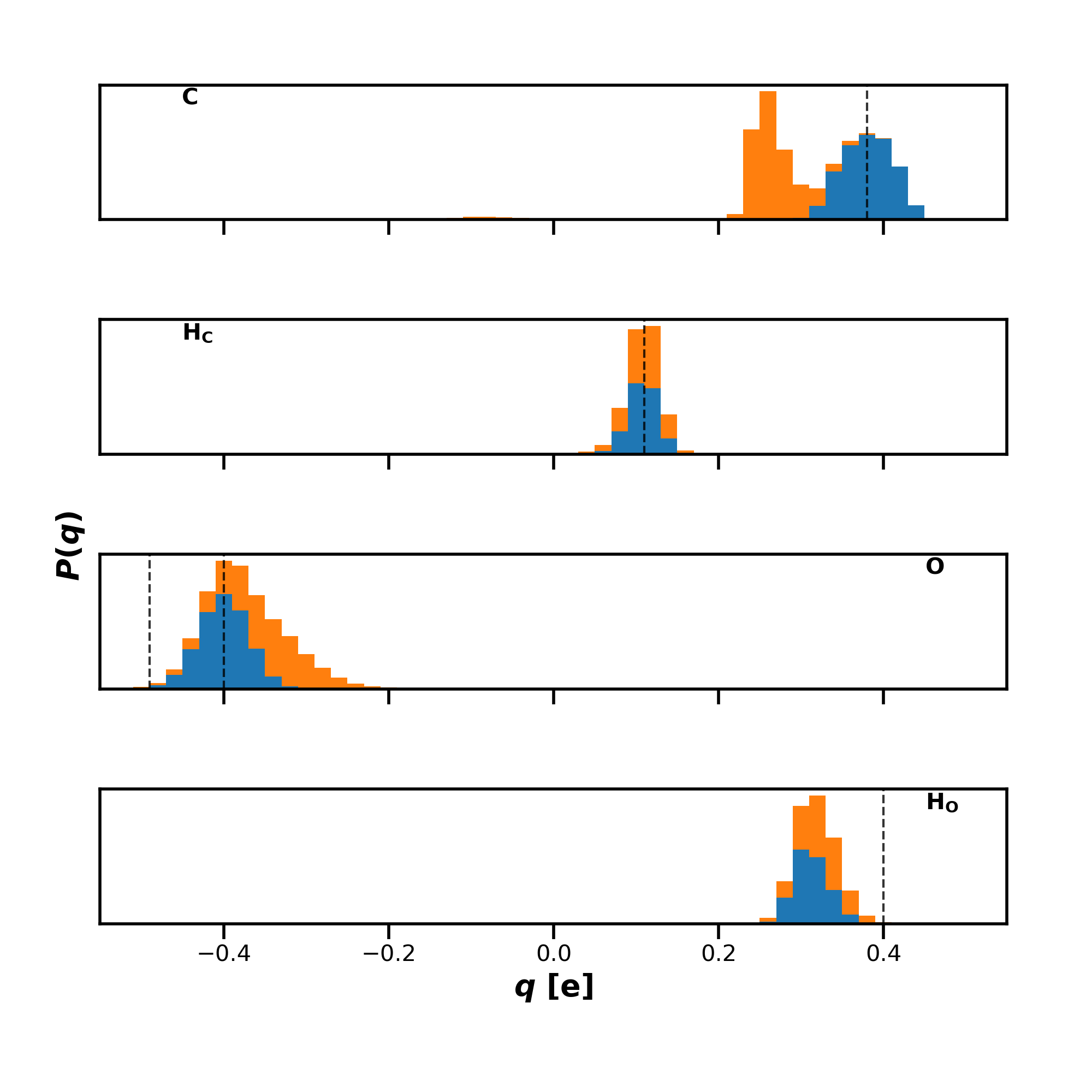}
\caption{Predicted atomic charge probability distributions $P(q)$ from
  the trained PhysNet model for FAD. The distributions are shown for
  the carbon atoms C, carbon-bonded hydrogen atoms H$_\mathrm{C}$,
  oxygen atoms O and oxygen-bonded hydrogen atoms
  H$_\mathrm{O}$. Blue: $P(q)$ for 10000 conformations of cyclic-FAD
  from gas phase MD simulations; orange: $P(q)$ for 10000
  conformations of non-cyclic FAD. The vertical black dashed lines
  mark the atomic charges for the corresponding atom types from the
  CGenFF force field.\cite{cgenff:2010}}
\label{sifig:8}
\end{figure}

\begin{figure}[htb!]
\centering
\includegraphics[width=0.8\linewidth]{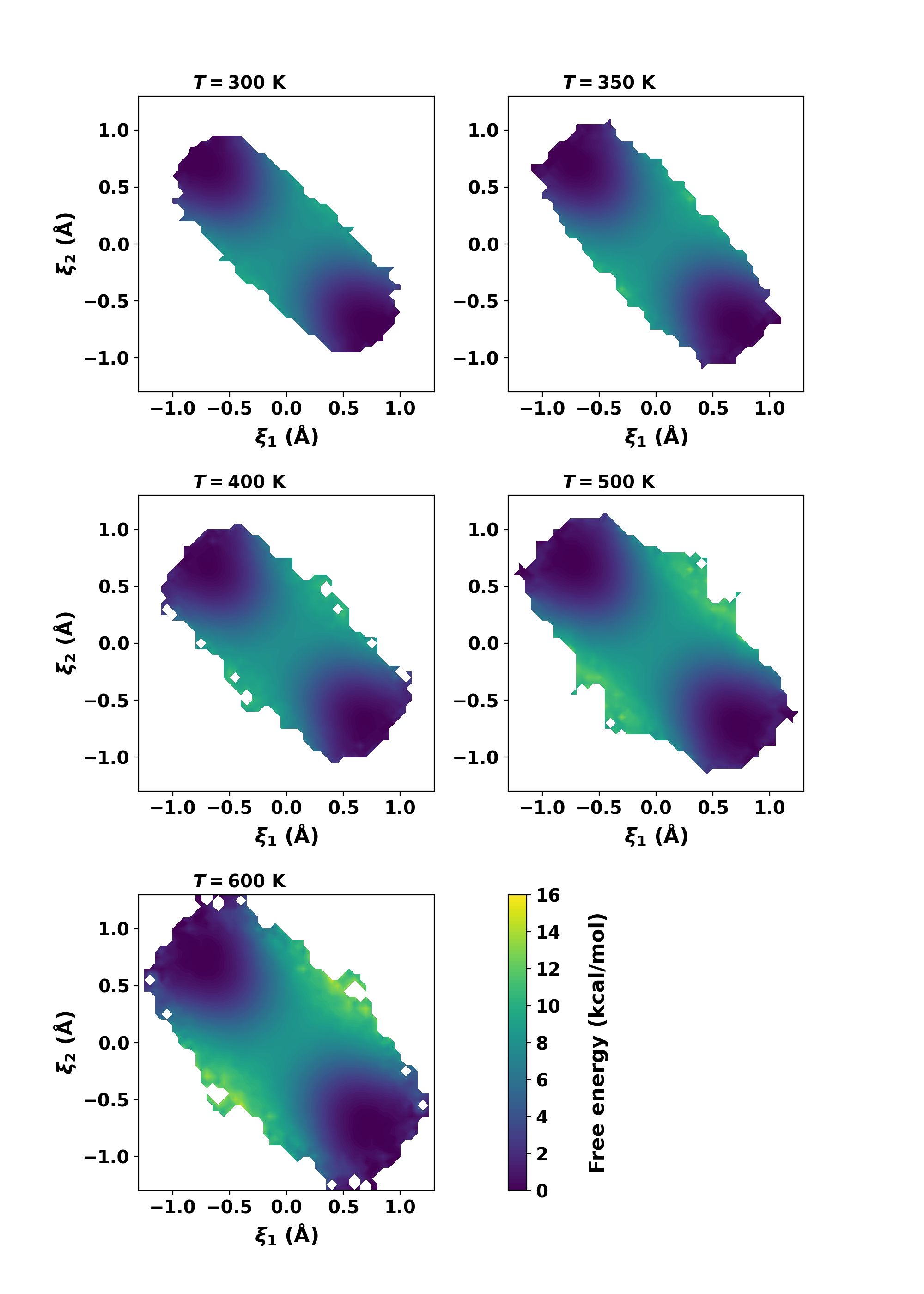}
\caption{2-dimensional free energy surfaces $G(\xi_1,\xi_2)$ for DPT
  in FAD in solution with fluctuating charges for different
  temperatures.}
\label{sifig:9}
\end{figure}

\begin{figure}[htb!]
\centering
\includegraphics[width=\linewidth]{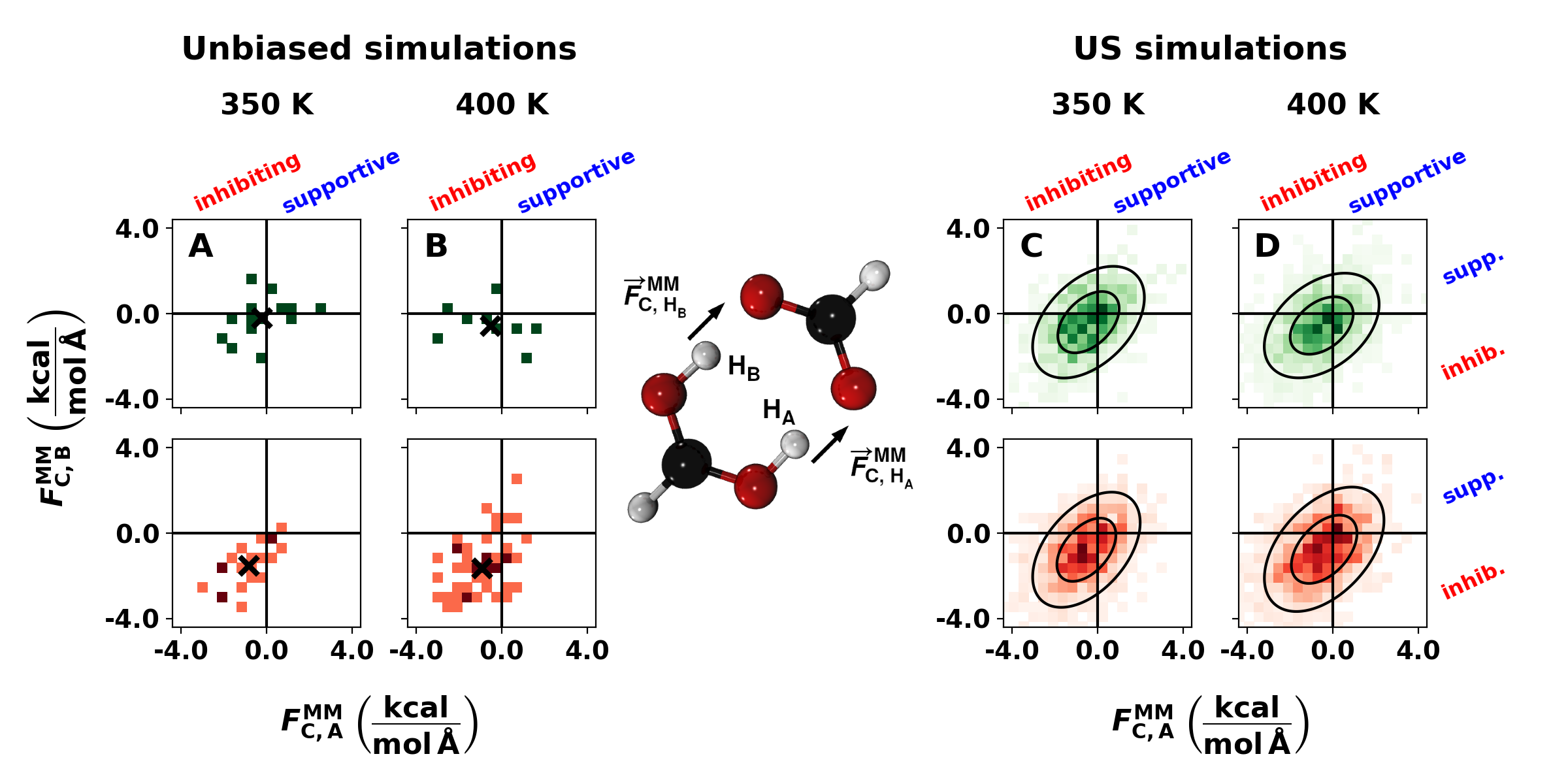}
\caption{Correlation between the magnitude and direction of the
  electrostatic component of the solvent force $F_{\rm C}^{\rm MM}$ at
  the position of the transferring hydrogen atom during the second PT
  for successful (green) and attempted (red) DPT. For definitions see
  Analysis section. Results from unbiased (A and B) and US simulations
  (C and D) are shown separately from simulations at (A and C) 350\,K
  and (B and D) 400\,K. The forces on the hydrogen atom H$_\mathrm{A}$
  involved in the first PT are given on the horizontal axis. In the
  vertical axis the forces on the hydrogen atom H$_\mathrm{B}$ are
  shown performing either the second PT or no PT for successful or
  attempted DPT, respectively. By definition, the force vector always
  points towards O$_{\rm acc}$ in the H-bond after the first PT.  An
  electrostatic force on the hydrogen atom towards O$_{\rm acc}$ in
  the respective H-bond is labelled as ``supportive'', and
  ``inhibiting'' otherwise.}
\label{sifig:10}
\end{figure}

\begin{figure}[htb!]
\centering
\includegraphics[width=\linewidth]{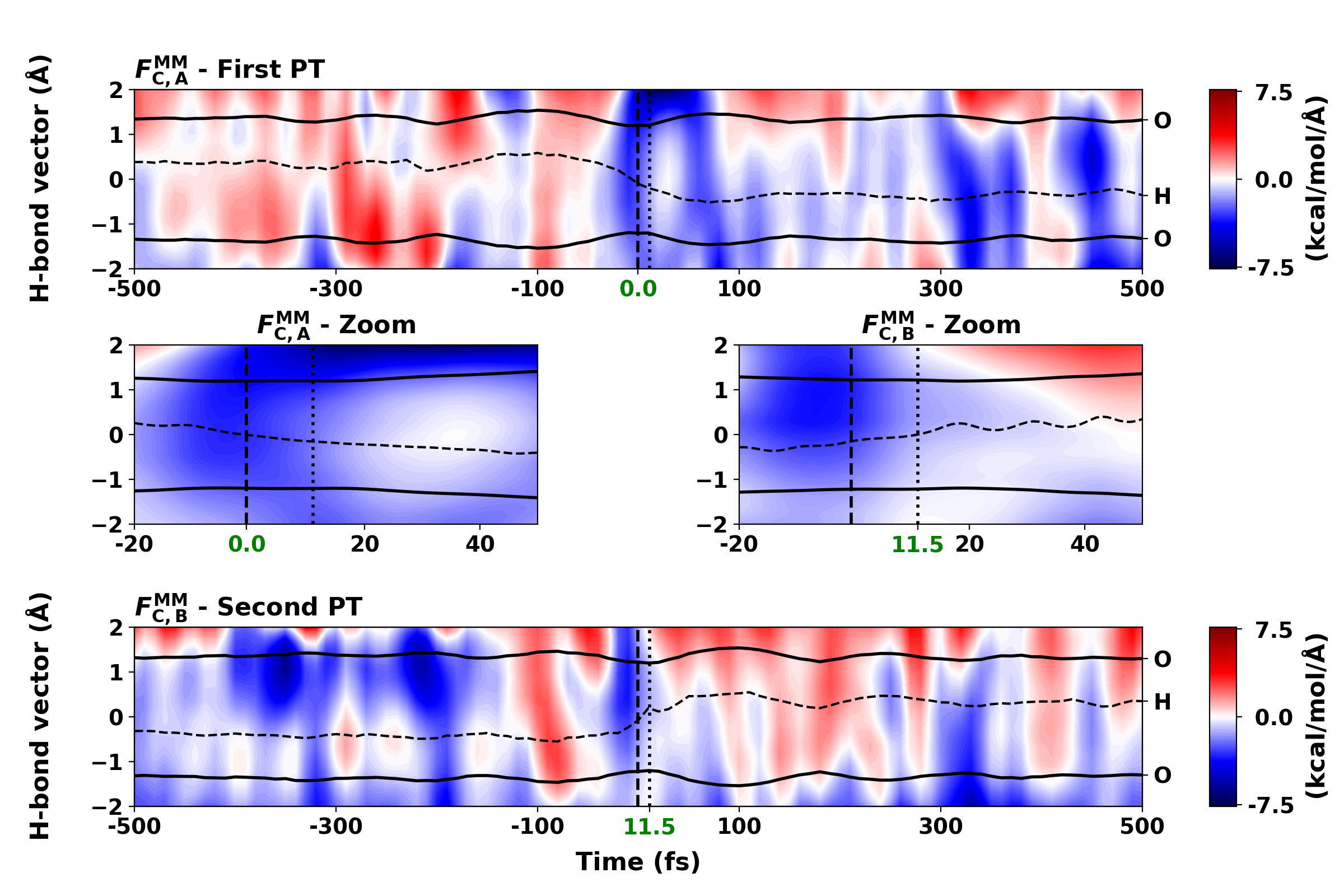}
\caption{Time sequence covering 1\,ps of reporting the
  solvent-generated Coulomb force $F_{\rm C}^{\rm MM}$ on the hydrogen
  atoms in both H-bonds before and after the first (top panel) and
  second PT (bottom panel) for a successful DPT at 350\,K. The black
  solid and dashed lines follow the position of the O and H atoms in
  the respective H-bond. The two center panels provide the $F_{\rm
    C}^{\rm MM}$ sequence in both H-bonds across a 70\,fs window
  around the DPT. The red colormap denotes a Coulomb force on the
  respective H atom parallel and the blue colormap antiparallel along
  the H-bond vector defined between both oxygen atoms. The horizontal
  solid and dashed lines follow the position of the O and H atoms in
  the respective H-bond. The vertical dashed line at 0\,fs is the time
  point of the first PT in the H$_{\rm A}$-bond and dotted line of the
  second PT in the H$_{\rm B}$-bond.}
\label{sifig:11}
\end{figure}

\begin{figure}[htb!]
\centering
\includegraphics[width=\linewidth]{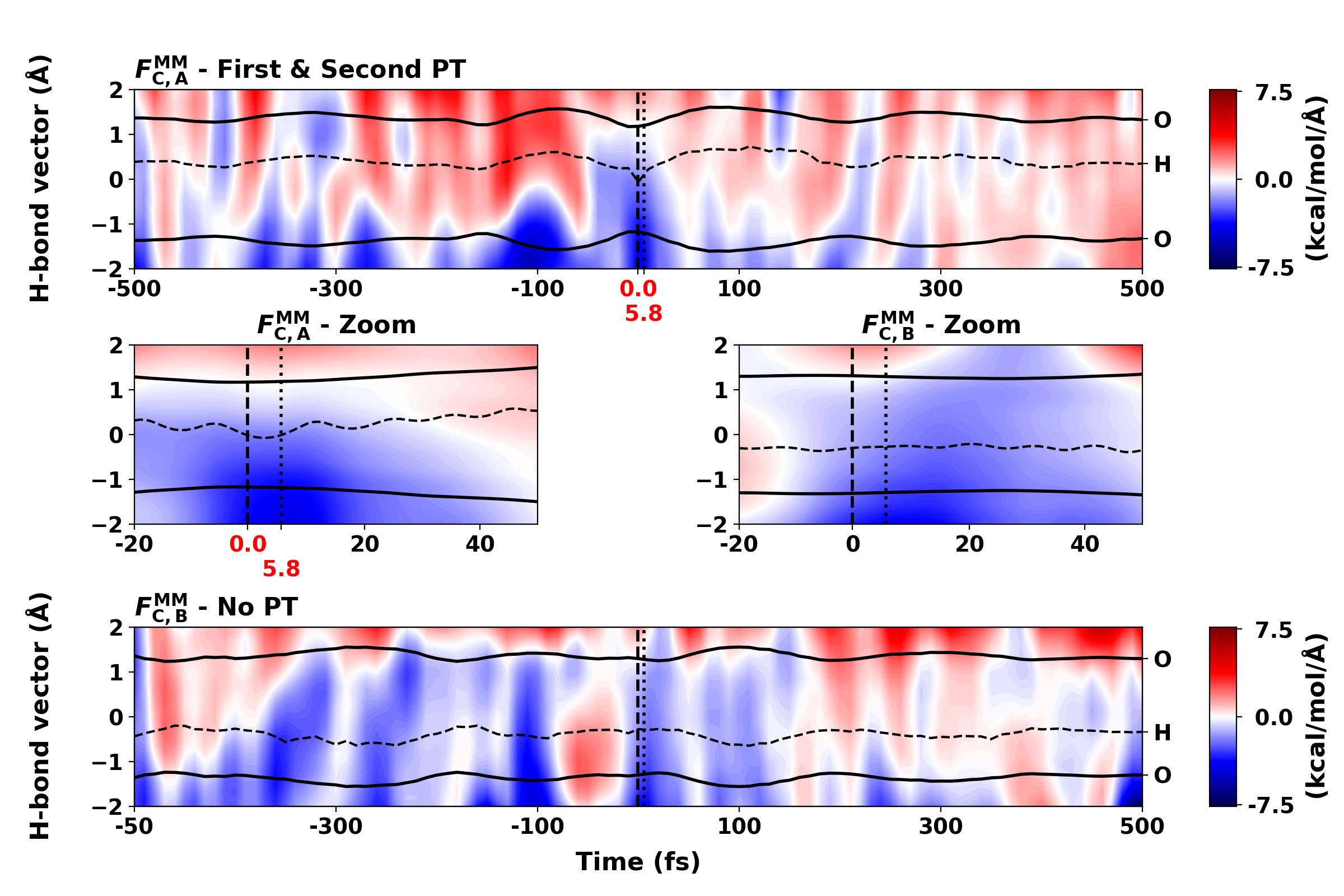}
\caption{Time sequence covering 1\,ps of reporting the
  solvent-generated Coulomb force $F_{\rm C}^{\rm MM}$ on the hydrogen
  atoms in both H-bonds before and after the first (top panel) and
  second PT (bottom panel) for an attempted DPT event at 350\,K. The
  black solid and dashed lines follow the position of the O and H
  atoms in the respective H-bond. The two center panels provide the
  $F_{\rm C}^{\rm MM}$ sequence in both H-bonds across a 70\,fs window
  around the DPT. The red colormap denotes a Coulomb force on the
  respective H atom parallel and the blue colormap antiparallel along
  the H-bond vector defined between both oxygen atoms. The horizontal
  passing solid and dashed line shows the position of the O and H
  atoms in the respective H-bond. The vertical dashed line at 0\,fs is
  the time point of the first PT in the H$_{\rm A}$-bond and dotted
  line of the second PT in the H$_{\rm B}$-bond.}
\label{sifig:12}
\end{figure}

\begin{figure}[htb!]
\centering
\includegraphics[width=\linewidth]{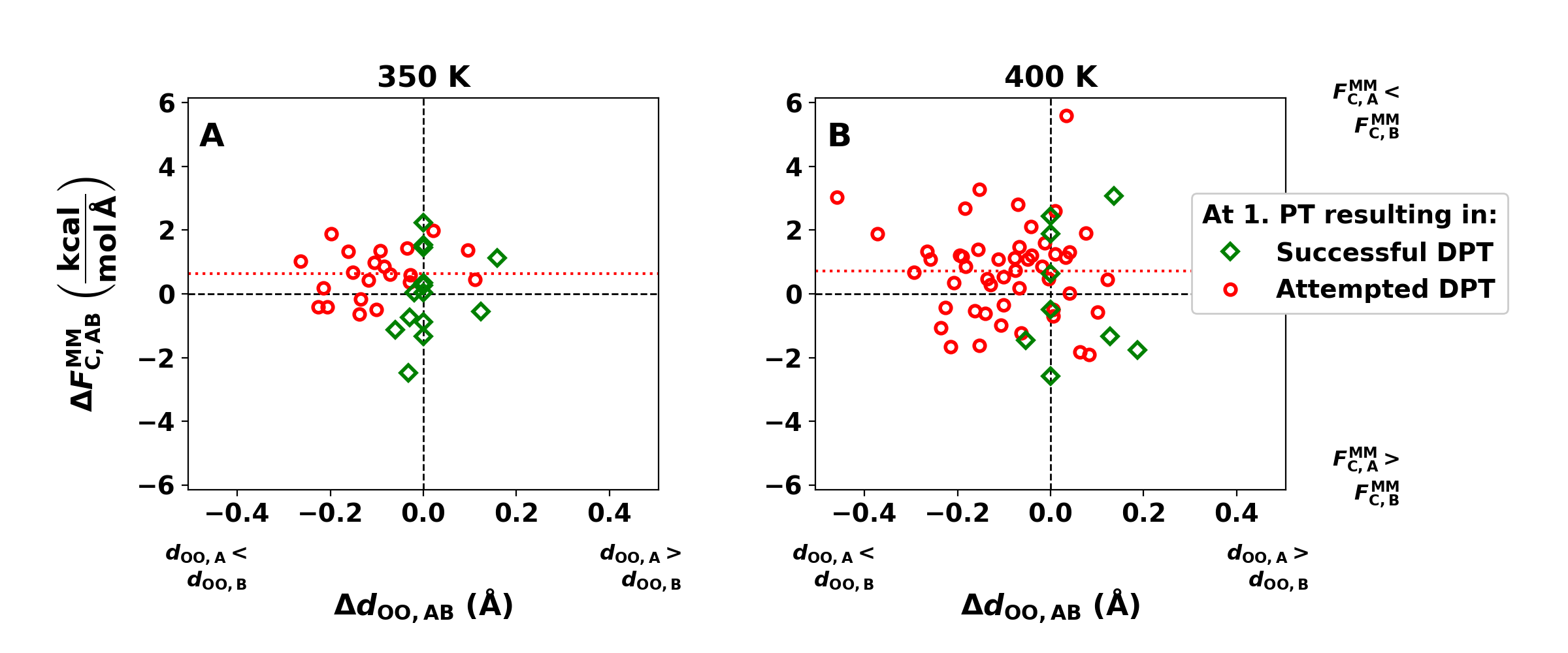}
\caption{Correlation between difference in O--O separation $\Delta
  d_{\rm OO,AB} = d_{\rm OO,A}-d_{\rm OO,B}$ and difference in
  solvent-generated Coulomb force $\Delta F_{\rm C,AB}^{\rm MM} =
  F^\mathrm{MM}_\mathrm{C,A} - F^\mathrm{MM}_\mathrm{C,B}$ of both
  H-bonds for the first PT at 350 K and 400 K of unbiased simulations,
  respectively. The average force difference for attempted DPT is
  indicated as the dashed red line. Manifestly, the average $\Delta
  d_{\rm OO,AB} < 0$ for attempted DPT whereas $\Delta d_{\rm OO,AB}
  \sim 0$, i.e. the structure is symmetric, for successful DPT.}
\label{sifig:13}
\end{figure}

\begin{figure}[htb!]
\centering
\includegraphics[width=\linewidth]{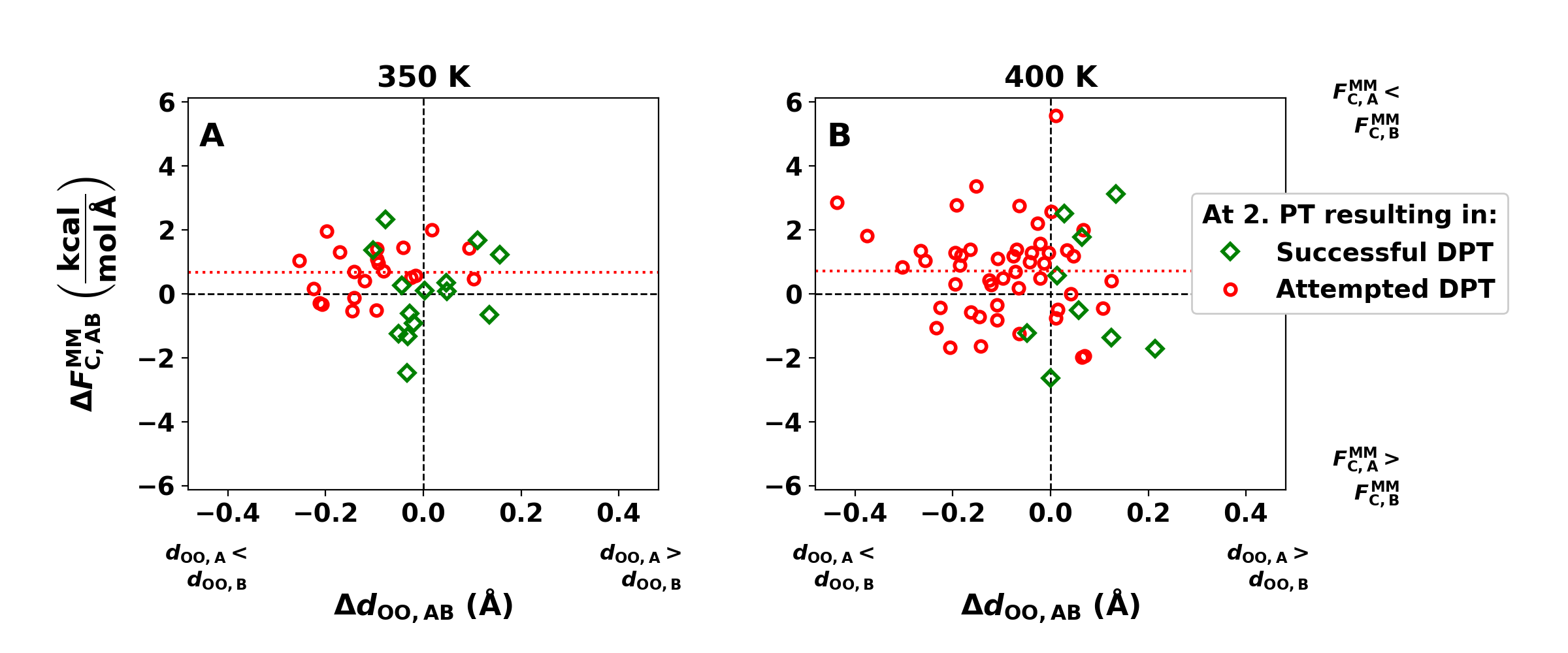}
\caption{Correlation between difference in O--O separation $\Delta
  d_{\rm OO,AB} = d_{\rm OO,A}-d_{\rm OO,B}$ and difference in
  solvent-generated Coulomb force $\Delta F_{\rm C,AB}^{\rm MM} =
  F^\mathrm{MM}_\mathrm{C,A} - F^\mathrm{MM}_\mathrm{C,B}$ of both
  H-bonds for the second PT at 350 K and 400 K of unbiased
  simulations, respectively. The average force difference for
  attempted DPT is indicated as the dashed red line. Manifestly, the
  average $\Delta d_{\rm OO,AB} < 0$ for attempted DPT whereas $\Delta
  d_{\rm OO,AB} \sim 0$, i.e. the structure is symmetric, for
  successful DPT. Note the slight displacements in position and forces
  compared with Figure \ref{sifig:13} as the first and second PT are
  separated by a few femtoseconds.}
\label{sifig:14}
\end{figure}

\clearpage
\bibliography{refs.tidy}